\documentclass[leqno,a4paper,12pt,oneside]{report}
\usepackage[latin1]{inputenc}
\usepackage[english]{babel}
\usepackage[T1]{fontenc}
\usepackage{amsfonts,amssymb,amsmath,amsthm,latexsym,amscd,mathtools,exscale}
\usepackage{makeidx}
\title{PATHWAYS OF HISTORY OF ELEMENTARY PARTICLE PHYSICS}
\author{\small Giuseppe Iurato\\\footnotesize \it University of Palermo, IT}
\date{}
\begin{document}\setcounter{tocdepth}{22}\maketitle\titlepage\tableofcontents\newpage

\section*{Preface}\addcontentsline{toc}{section}{Preface}

Most of the work of leading scientists has always been characterized both by an initial theoretical setting and
analysis of the given problem under examination and by the related experimental arrangement, and vice versa,
taking into account the main Galileian paradigm of scientific knowledge, essentially given by the dialectic and
inseparable relationships between experimental bases and theoretical-formal structures from which arise the
rational thought. These scientists have always been interested both to theoretical aspects and experimental
data, like Jun John Sakuari (1933-1982) as remembered by John S. Bell in (Sakurai 1985, Foreword). On the other
hand, just due to its Galileian nature, no history of theoretical physics can be disjoined from experimental
context, and vice versa. We have herein tried to adopt a new way of doing history of science: namely, trying to
delineate a technical (or internal) history of a certain field of knowledge through the life and the work of
those people who have, at international level, significantly and permanently contributed to it, along their
life. Amongst them, we shall consider, for example, some of the main works of A. Zichichi and R. Garwin, namely
those which have led to the first exact measurements of the anomalous magnetic moment of the muon, one of the
first precise test of QED.

To be precise, in drawing up this work, we adopt that unique possible historiographical methodology which has to
be followed to pursue a correct and objective historic-biographical report of the work of a given author under
examination, that is to say, the one consisting in giving primary and absolute priority to the study and to the
analysis of the original papers and works of the author under examination (primary literature). Only
subsequently it will be then possible also to take into account the related already existent secondary
literature. This for trying to minimize, as much as possible, the distortions and mystifications due to the
unavoidable \it personal equation\footnote{This concept has its own history which starts from Astronomy to
Freudian Psychoanalysis and Jungian Analytical Psychology. Here, we shall mean such a term in the latter wider
psychological meaning (see (Galimberti 2006) and (Thom\"{a} and K\"{a}chele 1990, Vol. 1., Chap. 3, § 3.1)).
Following (Carotenuto 1991, Chapter X), the \it personal equation \rm is an unavoidable subjective factor which
influences on the evaluation of objective data, leading to different visions of the phenomenological fields
under examination. It is determined by the individual history, by constitutional and typological elements and by
social-cultural factors. It acts as a perceptive filter, or rather as an internal transducer, which redefines,
according to personal parameters, the reality, shaping the knowledge's act. For instance, the various
mythological deformations and biases are mainly due to its action, so that, as regards historical sciences, we
agree with that historiographical method which gives priority to the study of the primary literature of a given
author, like her or his works.} \rm (biases, complexes, mythicizations, etc) which is implicitly presents in
everyone of us. As picturesquely recalled by Vittorio de Alfaro in (De Alfaro 1993, Introduction, p. 3), <<the
historical reconstruction is everything except a 'fractal': indeed, the level of enlargement with which we treat
a historical process, greatly shall influence the conclusions that we deduce>>. A similar case is besides also
recalled by Bruno Rossi in (Rossi 1964, Preface), in which he warns on the impartiality with which himself has
written the history of cosmic rays, since he was directly involved, as a leading actor, in the international
research on this field, so that he does not exclude to have given a major load to the work made by his research
group. On the other hand, this historiographical methodology is just that advocated by Benedetto Croce himself
(see (Croce 1938)) to obtain, through a rational analysis of the sources, an impartial historical judgement
devoid of any biased or partial mystification.

There are non-negligible historiographical questions about the general history of science, which we wish to
outline too as an apology for the method used in carrying on this work; in exposing such a historiographical
problematic, we mainly follow (Piattelli Palmarini 1992, Chap. I, §§ 2.I, 2.II and 2.III). Let us say it
immediately: such a problematic derives from the far from being trivial questions existing between myth and
science, which embed their roots in the crucial historical passage from mythological to philosophical thought.
The relationships between myth and science are far from being ancient and negligible: in this regards, it is
enough to remember as Wolfgang Pauli himself, after a long period of collaboration with Carl Gustav Jung, put
much attention to the possible links and intersections between epistemology and analytical psychology, writing
many interesting works on these arguments (see (Jacob 2000), (Tagliagambe and Malinconico 2011) and references
therein). The French anthropologist Pierre Smith claims that the myth is always \it in nuce \rm(that is to say,
implicitly) presents in the way in which each of us tells of herself or himself, above all as regards her or him
own past, this, in turn, implying unavoidable distortions and mystifications which give rise to a mythic
production. In short, the myth is an efficient way to organize and to coordinate the individual and collective
memory. The Smith's schoolmaster, Claude Lévi-Strauss, said that the myth is a story continuously transformed by
who believe only of repeating it and to which, instead, he or she gives ''an excess of meaning'' whenever it is
re-evoked.

Also in science there is an organization of collective and individual memory where the mythical element may
appear. For instance, Thomas S. Kuhn, collecting a great number of interviews carried out with the founders of
modern physics, frequently noticed many inaccuracies and inconsistences as concerns their biographies which
resulted to be strangely logical, linear and educationally edifying, but contrasting with the real facts and the
original sources; hence, Kuhn finished to conclude that the real history of science is not so perfectly
constructed and have not the direct pedagogical function of those a little mythical stories told by handbooks
and protagonists. Also Gerald Holton has experienced an emblematic case of the same type: precisely,
interviewing in old age, Einstein convinced himself to have developed his special theory of relativity on the
basis of the results of the famous Michelson-Morley experiment, building up, in such a manner, a logical
motivation to the birth of his theory\footnote{Also (Brown and Hoddeson 1983) confirm that \it <<people cannot
be totally objective about the events in which they participate; we tend unconsciously to reinterpret history in
terms of present-day values>>. \rm But, in doing so, often the historical reality may go lost.}. Indeed, such an
experiment was yes carried out few years before the Einsteinian publications, but Einstein did not know it when
he formulated his ideas. The Einstein reconstruction was logic and educationally efficacy even if historically
false; bona fide, he self-convinced himself that the things were just gone so. In these cases, we should
consider, according to Claude Lévi-Strauss and Fran\c{c}ois Jacob, the myth as an \it excess of meaning \rm
needed for organizing the memory and for giving a logical and instructive meaning to its contents, often to
detriment of the real historical and chronological truth. All this makes particularly difficult to do history of
science; it may be included in the wider unavoidable problematic concerning the already mentioned personal
equation, which would shed a certain shadow of discredit on the history of science if it weren't taken into the
right account. For these reasons, we think that the more correct historiographical method for carrying out a
scientific biography is that consisting at first in analyzing directly the primary related literature, trying to
prevent the non-objective deformations given by the effects of this personal equation, almost desiring to aspire
to emulate the coldness or indifference of a psychoanalyst which must simply reflect like a clean mirror (see
(Thom\"{a} and K\"{a}chele 1990, Vol. 1., Chap. 3, § 3.1)) with the highest objectivity degree\footnote{The
psychoanalysts try to attain this by means of the so-called \it didactic analysis, \rm which is strictly
correlated to the dualistic and dialectic interaction between transfer and countertransfer phenomena (see
(Thom\"{a} and K\"{a}chele 1990, Vol. 1., Chap. 3, § 3.1)).}.

This methodology, moreover, is the only way which permits us to may infer the formation and evolution of the
thought of a given author, undergoing his creative process (as in this case), along her or his social-cultural
and scientific career. Such a historiographical method resembles, in a certain sense, that already adopted by
Francesco Giacomo Tricomi in (Tricomi 1967) from the mathematics history side. On the other hand, all this had
already been highlighted by Sir Patrick M. S. Blackett (1897-1974) who remembered to what distortions may lead
the above recalled mythical production, by the so-called \it scientific divulgation, \rm if one does not take
the right position respect to the author under examination. The first and the most frequent methodological error
made by the historians of science concerns the location of the own \it Ego, \rm in the sense that often he puts
herself or himself as main subject rather than the author under examination. This decentralization of the Ego is
a primary epistemological process whose importance has been highlighted by Sigmund Freud himself: indeed,
following (Vegetti Finzi 1986, Chap. I, § 2), the scientific knowledge has reached its highest levels in
concomitance to real \it narcissistic wounds, \rm as those occurred with the Copernican revolution, the
Darwinism and the Freudian psychoanalysis, each of these having just reappraised the human Ego, self-limiting
this. Such a self-limitation of the Ego therefore corresponds to a general criterion of further improvement and
completion of knowledge, as highlighted too by Max Planck himself in (Planck 1964) (see also (Straneo 1947,
Introduction)).

This Ego decentralization also plays a non-trivial role in historiography as regards the position of the
historian compared to the object under attention. This fact, for instance, has been emblematically recalled by
one of the most important Italian mathematicians of the last century, Bruno Pini\footnote{For some brief
biobibliographical notes on the life and works of Bruno Pini (1918-2007), see (Cavallucci and Lanconelli 2011)
and (Lanconelli 2012).}, who had to say that <<sometimes, when one is called to commemorate someone, it goes end
up to overly speak of herself or himself>> (see (Lanconelli 2008)); this simple consideration may be extended to
the general biographical studies\footnote{The opposite case to this is that related to the deifications, like
those present in many hagiographies.}. This is simply due to the human weakness, scientist or not who he or she
be, even turned toward the own egocentric accomplishments (from which it follows, for some respects, the
well-known Latin maxim \it <<tot capita, tot sententi\ae>>\rm). Therefore, it turns out clear what
methodological importance has the examination of the original scientific production of every author under
examination, as we just hope to do in any case herein analyzed, to avoid any possible mystification.

Finally, since, according to Chen Ning Yang (see (Yang 1961, Preface)), <<a concept, especially a scientific
one, have not full meaning if it is not defined respect to that knowledge context from which it derived and has
developed>>, each examined original work or paper of the author under consideration shall be even contextually
laid out into the related theoretical framework of the time, so that, where possible, a brief historical
recognition should be mentioned as a contextual story meant as follows. In a certain sense, we might say to
follow an epistemological path analogous to that outlined by Stephan Hartmann in (Hartmann 1999) where he claims
on the importance, above all in hadron physics, to consider a theoretical model as the result of an interpreted
formalism plus a story, this last being meant both as a narrative but rigorous told around the formalism of the
given model and as a complement of it, hence an integral part of the model; the relationships between formalism
and story are then placeable out into the wider class of relationships subsisting between the syntactic and
semantic parts of a general physical model which are unavoidable just in Physics. Therefore, our work might be
considered as a sort of making a story to certain groups of works of the author under examination in order to
get an overall historical view of the subject matter in which her or him worked. For these reasons, it is also
indispensable refers us to the general technical-scientific literature to support what said. Only doing so, it
will be possible to pursue the highest objectivity degree and historical correctness in descrying the scientific
figure of an author, trying to avoid the above mentioned irrationality elements. At the same time, in dependence
on the scientific level of the treated author, with this method it shall be possible to outline a history of the
related work area.

Another confirmation of the validity of the above mentioned work program follows from some epistemological
considerations about the foundations of science, due to the modern French school which goes from G. Bachelard,
A. Koiré and G. Canguilhem to the structuralists J. Lacan, C. Lévi-Strauss, L. Althusser, M. Foucault, F.
Regnault, A. Badiou and F. Wahl. Indeed, following (Cressant 1971, Introduction), the scientific activity should
be looked at as a constructive process which pulls out the truth or the essence of the \it real objects \rm that
will constitute the central core upon which building up the corresponding \it knowledge object, \rm trying to
separate ideological questions from the mere scientific contents. Read an arbitrary work just means try to
separate the general ideological and philosophical context from the scientific one; it means to analyze the
problematic frame within which this work has been conceived, rebuilding up the prime structural causes from
which it shall develop. In doing so, a passive and sterile lecture will be replaced by an active and productive
re-enact (see (Wahl 1971)), almost analogously to what foreseen both by the Robin Collingwood historicism (see
(Kragh 1990) and (Iurato 2013)) and by Wilhelm Dilthey methodological hermeneutics, according to which any
written source should be laid out into the proper original historical context, according to the \it Zeitgeist
\rm of the time. Following (Schultz 1969, Chapter I) and (Wertheimer 1979, Chapter 1), the ideology is always an
unavoidable judgment component of human being, hence also of every historian, since it is a common perspective
to conceive the history as chiefly due to the subjective idiosyncrasies and to the preconceptions which will
play the role of selective mental grid of what to consider or not and of how to interpret this. Contrarily to
what one could thought, the ideology is also an unavoidable component of the normal scientific context: in this
regards, see (Boudon 1991, Chapter VIII).

\section*{1. Historical introduction: I}\addcontentsline{toc}{section}{1. Historical introduction: I}

Mario Gliozzi, in\footnote{The \it Enciclopedia delle Matematiche Elementari e Complementi \rm has been the most
important and notable Italian encyclopedic handbook on mathematical sciences and their applications, reviewed
abroad as one of the main encyclopedic work made in this context, as valuably remarked by (Archibald 1950) and
(Miller 1932). This article of Mario Gliozzi was the first systematic attempt to outline a brief history of
physics. It was later retaken as a first core for drawing up another more extended article published in the 1962
Nicola Abbagnano treatise on the history of science, in turn posthumously enlarged and revised by the sons of
Mario Gliozzi, in the new and definitive 2005 edition (Gliozzi 2005), which is one of the most complete textbook
on the history of physics. Herein, we have mainly followed (Gliozzi 1949) because of its conciseness which is
functional to the aims of this section, referring to (Gliozzi 2005) for a more complete and in-depth view.}
(Gliozzi 1949, § 30), outlines the main features of the experimental physics through the last 19th Century
decades to the 1940s. This was an almost unique period for the history of physics since, from the new results of
atomic physics of the 19th Century end, appears, in all its complexity, the new \it submicroscopic Weltbild \rm
to whose knowledge inextricably taken part philosophical, theoretical and experimental physical questions above
all characterized by the crucial passage from the classical determinism to the modern probabilism as recalled by
(Pignedoli 1968, Chap. I) which gives a clear and synthetic historical summary of this critical epistemological
step. Above all, the experimental physics had needed for new methods, techniques and tools to approach and to
examine this unexpected world so closed to our direct perceptions, this, in turn, implying the formulation of
new theories to explain it at the light of these experimental results which arose from the discovery of cathodic
and anodic emissions, channel and X rays, and radioactivity\footnote{The spontaneous radioactivity was
discovered by H. Becquerel in 1896 under advice of H.J. Poincaré. Indeed, the latter suggested to the former to
investigate on the possible relationships between optical fluorescence and X rays, which revealed to be fake,
but that led, for serendipity, to the discovery of radioactivity (see (Segrè 1999, Chap. 1)).} (see (Born 1976,
Chap. 2)). In this regards, in 1897, Charles T.R. Wilson discovered that the ions produced in air by ultraviolet
and X rays as well as by radioactive radiations, acted as condensation nuclei of water steam suitably
supersaturated by rapid adiabatic expansion. This notable discovery was at the basis of the so-called \it cloud
chamber, \rm one of the first valuable displaying particle detector, first set up at the Cambridge Cavendish
Laboratory in 1896 and subsequently improved by Wilson (see (Wilson and Littauer 1965, Chap. 3) and (Yang 1969,
Chap. 1)), so that it is often called too \it Wilson chamber\rm; it will play a fundamental role in experimental
atomic physics, even to be said ''an open window on the world'' (E. Persico). The particle detectors may be
classified into two main categories, namely the \it displaying \rm detectors and the \it optical \rm(or \it
electronic\rm) detectors; the first ones comprise the \v{C}erenkov and scintillation counters, the Wilson (or
cloud), bubble, spark and photographic emulsion chambers, whereas the second ones include the ionization
chamber, the Geiger-M\"{u}ller, the proportional and solid-state counters (see (Segrè 1999, Chap. 3), (Tolansky
1966, Chap. 17) and (Chiavassa, Ramello and Vercellin 1991, Chap. 2)).

Following (Segrè 1999, Chap. 1), after the discovery of electron in 1897, the first atomic models due to J.J.
Thomson, E. Rutherford and N.H. Bohr at the beginnings of 20th Century, together with the introduction of \it
quanta \rm by M. Planck and A. Einstein as regards the electromagnetic radiation, led to the formulation of
quantum mechanics which succeeded to explain many atomic phenomena. At the same time, after the discovery of
atomic nucleus in 1911, the new quantum theories gradually opened the way to nuclear physics with the first
$\alpha$ particle bombardment phenomena which led, after the 1919 pioneering Rutherford discovery of the \it
proton, \rm to the definitive 1924-25 experimental ascertainment of such a particle by P.M.S. Blackett, who was
a Rutherford's pupil (see (Gliozzi 1949, § 30, footnote $^{239)}$) and (Gamow 1963, Chap. VIII)). Thereafter, on
the basis of the previous works made by R.J. Van de Graaff\footnote{Nevertheless, the principle of the method
upon which relies the running of such machines is quite similar to one already studied by A. Righi at the end of
19th Century (see (Gliozzi 1949, § 30, footnote $^{241)}$)).}, J.D. Cockcroft, E.T.S. Walton, H. Greinacher and
R. Wilder\"{o}e, in 1933 E.O. Lawrence and S. Livingston built up the first particle accelerator, the so-called
\it cyclotron \rm (see (Wilson and Littauer 1965) and (Segrè 1976, Chap. XI)), based on the resonant
acceleration method. Independently of each other, in 1944 the Russian physicist V.I. Veksler proposed a new
particle accelerator based on the phase stability method, while in 1945 E.M. MacMillan proposed an analogous
particle accelerator which will be called \it synchrotron. \rm See (Lee 2004) for a complete and masterful
updated knowledge on accelerator physics, in which there are also interesting historical notes.

Retaking into account some previous experiences made by W. Bethe and H. Becker, the spouses I. Curie and F.
Joliot discovered a new particle, already suggested by Rutherford in 1920 and whose exact nature was
subsequently experimentally ascertained by J. Chadwick who called it \it neutron \rm (see (Hughes 1960)); the
Curie's experiences given rise to the first artificial radioactivity phenomena. In the years 1932-1934, a new
particle was observed, almost at the same time, by many scientists: amongst them, by I. Curie and F. Joliot in
collision phenomena with $\alpha$ particles, by C.D. Anderson in the United States and by P.M.S. Blackett with
G. Occhialini in England, in experiences concerning cosmic rays (see (Gliozzi 1949, § 30, footnote $^{243)}$)),
which was called, by C.D. Anderson, \it positive electron, \rm or \it positron. \rm Such a particle had already
been theoretically provided by P.A.M. Dirac with his elegant 1930s \it electron theory, \rm which, inter alia,
established too the so-called \it charge conjugation \rm invariance principle; this new particle was
experimentally determined having mass almost equal to the electron one but with positive charge. The discovery
of positron was a celebrated experimental confirmation of Dirac's electron theory, which was besides unknown to
Anderson but not to Blackett and Occhialini which made their above researches at the Cavendish Laboratory of
which Dirac was a member, at that time (see (Rossi 1964, Chap. VI)).

In the years 1933-1934, taking into account the previous works of the Curie-Joliot spouses, E. Fermi was the
first to use neutrons as collision particles, in place of $\alpha$ particles: indeed, he rightly argued that
neutrons were more suitable to this, due to the lack of electrostatic repulsion respect to an atomic nucleus;
slow neutrons turned out to be very efficient in breaking the atomic nucleus. Such ingenious intuitions were put
in practice in Rome, by E. Amaldi, O. D'Agostino, B. Pontecorvo, F. Rasetti and E. Segrè, where it was carried
out the celebrated experiences with slow neutrons (see (Gliozzi 1949, § 30, footnotes $^{245),\ 246)}$) which
will lead to the discovery of \it nuclear fission \rm and to the subsequent \it chain reactions, \rm all this at
the World War II eve (see (Gliozzi 1949, § 30, footnotes $^{247)-251)}$)). It was the beginning of the nuclear
physics with the use and applications of the nuclear energy by E. Fermi in 1942, in this, the Italian school
having been leader in the international research framework of the time. In this regards, from a historical
viewpoint, it is enough to give a glance to the fundamental works (Wick 1945; 1946) to witness all this, which
represented the first treatise on the new neutron physics; this unique two-volume treatise is the most valuable
historical source which exposes the ''state of the art'' of that time as regards this new chapter of nuclear
physics.

In the decade 1920s to 1930s, the building of quantum mechanics was achieved, with the elegant and rigorous
formulation given by P.A.M. Dirac in his celebrated textbook (Dirac 1958), whose first edition date back to 1930
and that is still the classical and definitive treatise on the subject with its last 1958 fourth edition; in it,
the chapters on the new quantum electrodynamics were updated till the results of 1950s. Once discovered the
neutron, one of the main problem of the new nuclear physics was to determine the interaction forces among the
constituents of the atomic nucleus, that is to say (d'après W. Heisenberg, D. Ivanenko and E. Majorana) protons
and neutrons, which have been interpreted as two states of the same particle, called \it nucleons, \rm having
different values of a well-determined numerical parameter called \it isospin. \rm This last quantum number is
related to the formal description of the notion of \it isotopic \rm (or \it isobaric\rm) \it invariance, \rm
that was first introduced by W. Heisenberg in 1932, then used by B. Cassen and E.U. Condon in 1936 and by E.P.
Wigner in 1937 (see (Landau and Lif\v{s}its 1982, Chap. XVI, § 116)) and subsequently applied to the
classification of other subnuclear particles, as we will see later. The next twenty years will see the birth of
the so-called \it quantum field theory \rm (QFT), before all with the new \it quantum
electrodynamics\footnote{For this fascinating story, see (Schweber 1983; 1994), (De Alfaro 1993) and (Weinberg
1999, Vol. I, Chap. 1). In (Schweber 1983; 1994) there is also an extensive history of quantum field theory of
20th Century, both from an internal and external historical standpoint.} \rm (QED), which develops, according to
the Galileian scientific method, in close concomitance with the related experimental physics contexts, above all
those concerning the radioactive emissions and the cosmic radiation, which will play a fundamental role in
developing the nuclear and subnuclear physics; within the theoretical framework given by the incoming QFT, they
will flow into the dawning of particle physics. The quantum electrodynamics started with the works of W.
Heisenberg, W. Pauli and P.A.M. Dirac, culminating in the Dirac's \it radiation theory \rm in which the photons
(already experimentally determined by E. Mayer and W. Gerlach in 1914 - see (Born 1976, Chap. 4, § 24)) are the
quanta of the electromagnetic field, this theory having been taken as main model for building up any further
quantum field theory, like the electronic-positronic field, the nucleonic and the mesonic ones, and so on (see
(Fermi 1963, Chap. 1, § 1)). In the decade from 1940s to 1950s, the electromagnetic field has been successfully
quantized starting from the Maxwell's equations, while the electronic-positronic field has been treated starting
from the Dirac's electron theory with a new formal process introduced by E.P. Wigner and W. Pauli in 1928,
called \it second quantization, \rm which is a modification of the previous quantization procedures to account
for supervened spin statistic problems. The situation concerning the electronic-positronic and nucleonic fields
was instead much more complex (see also (Weinberg 1999, Chap. 1, § 1.2)).

For our historical ends, we are more interested towards those aspects of particle physics history regarding both
radioactive decays and cosmic rays, which, as already said, have played a very fundamental role in the dawning
of particle physics and whose historical paths often have intertwined each other. Indeed, following (Weinberg
1999, Chap. 1, § 1.2), despite significant successes achieved by QFT (in primis, the Dirac's ones), a certain
dissatisfaction held towards it for all the 1930s, above all due to its apparent incapacity to explain many new
phenomena coming from the cosmic radiation as well as all the new type of particles contained in it. On the
other hand, as regards the various proposed theories explaining the radioactive emissions, above all that
regarding the $\beta$ decay to have played a fundamental role both in understanding the nuclear structure and in
developing QFT. Following (Persico 1959, Chaps. XI, XII and XIII), (Segrè 1999, Chap. 8), (Castagnoli 1975,
Chap. 3, § 3.7), (Tolansky 1966, Chap. XV), (Born 1976, Chap. 7, § 53) and (Friedlander and Kennedy 1965, Chaps.
6 and 7), the theory of $\alpha$ emission was successfully achieved by G. Gamow, R. Gurney and E. Condon in
1928-29, as the first attempt to apply the new quantum theories to nuclear structure, while the development of
the theory of $\gamma$ emission was parallel to that of quantum theory of radiation which started with an
interpretative theory analogous to the first atomic radiative emissions and continued, through 1920s to 1950s,
with the works of E. Rutherford, H. Robinsson, W.F. Rowlison, E.N. Da Costa Andrade, L. Meitner, H.J. Von
Baeyer, O. Hahn, C.F. Von Weizs\"{a}cker, H.A. Bethe, W. Heitler, O. Klein, Y. Nishina, J.P. Thibaud, E.
Feenberg, H. Primakoff, E.P. Wigner, M. Goldhaber, J.M. Blatt, V.F. Weisskopf, E. Wilson, K.T. Bainbrige, A.W.
Sunyar, P.B. Moon, R.L. M\"{o}ssbauer, and others (see (Heitler 1953)).

Instead, the $\beta$ emission, in both its $\beta^-$ and $\beta^+$ components, shown to have particular
difficulties to be laid out into a coherent theoretical description, above all in relation to the interpretation
of the related continuum electron velocity spectrum which was one of the most serious theoretical nuclear
physics problems of the time. The main theoretical problem concerned an apparent non-validity of the energy and
spin conservation laws at every elementary emission act, that Bohr attempted to justify invoking a sort of mean
validity of it. Nevertheless, in analogy with the case of $\gamma$ emission (which was experimentally excluded
to be associated with a $\beta$ emission), E. Fermi proposed an alternative and more valid quantitative
interpretation based on the possible contemporaneous emission of a new particle, called \it neutrino, \rm
together the electron involved in each elementary $\beta$ emission act. The neutrino was a particle first
theoretically predicted by H. Weyl in 1929 (see (Weyl 1931) and references therein) on the basis of Dirac's
electron theory, but his hypothesis was at once refused since it did not verify the parity symmetry. It however
will be reconsidered later in 1957 after the work of T.D. Lee and C.N. Yang on parity violation. Thereafter, in
1930, W. Pauli proposed to consider such a new particle to explain the lack of validity of the above mentioned
conservation laws as regards $\beta$ decay, which was later so named by E. Fermi in 1934. It was supposed to
have zero mass and spin one-half. The quantitative theory of $\beta$ emission, first proposed by E. Fermi in
1934 and later improved by E.J. Konopinski, H.J. Lipkin, G. Uhlenbeck and H. Yukawa, is a very general one which
may be also applied to other type of interactions. Taking into account previous theoretical studies, as already
said mainly due to W. Heisenberg, E.U. Condon and E.P. Wigner, this theory assumes proton and neutron as two
distinct quantum states of a unique particle, the quantum of the nucleonic field, called \it nucleon, \rm which
can go either into one, or into the other, of these two states just through a $\beta$ decay, emitting one
positive/negative electron and one neutrino\footnote{In this historical account, it is no possible to consider
the problem of \it helicity \rm of the neutrino as well as the related questions inherent the existence or not
of the antineutrino, the alternative theory of E. Majorana, and so on. See (Fermi 1963, Chap. 1) and (Yang 1969,
Chap. 1).}. To be precise, we have a $\beta^-$ emission in the transformation of a neutron ($n$) into a proton
($p$) according to a decay process of the type $n\rightarrow p+e^-+\nu$, with the emission of one (negative)
electron ($e^-$) and one neutrino ($\nu$); we have a $\beta^+$ emission in the transformation of a proton into a
neutron according to a decay process of the type $p\rightarrow n+e^++\nu$, with the emission of one positive
electron ($e^+$) and of one neutrino. Nevertheless, as it will be proved later by L.M. Lederman and co-workers
as well as by other workers, there exists another type of neutrino different from the one produced by the above
proton and neutron decays (denoted by $\nu_e$), namely the neutrino produced by $\mu$ and $\pi$ meson decays
(denoted by $\nu_{\mu}$). On this last point, we shall return later.

For a certain time, it was supposed that the nuclear forces could be explained through a nucleonic field
associated to the pair electron-neutrino (see (Polara 1949, Chap. IV, § 7), (Segrè 1976, Chap. X) and (Weinberg
1999, Chap. 1, § 1.2)) and whose quantum was the neutrino. It was hypothesized that proton and neutron were
linked together by means of an exchange of one neutrino, like in case of the ionized molecule $H_2^+$ where the
force between the atom $H$ and the ion $H^+$ became attractive at distances of the magnitude of $10^{-15}$ just
thanks to a periodic exchange of the unique available electron\footnote{Indeed, the notion of \it exchange force
\rm comes from the quantum theory of chemical bond (see (Slater 1980)). It will be later extended first to
nuclear physics thanks to the work of E.P. Wigner (see (Eisenbud and Wigner 1960, Chaps. 5 and 11)), then to the
particle physics context.}. Nevertheless, in this way it wasn't possible to account for nucleus stability
questions, so that this hypothesis (which had also been considered by E. Fermi in his 1934 theory) had to be
rejected. All that, however, will be one of the starting points of the subsequent pioneering 1935 Yukawa's work
(see (Yukawa 1935)). Following (Castelfranchi 1959, Chaps. XX and XXI, §§ 231, 262), the positive electrons
$e^+$ are emitted only in artificial radioactive decays and were, almost at the same time, discovered in
1932-33, by many researches, independently of each other, amongst whom C.D. Anderson, R.A. Millikan and P.M.S.
Blackett with G. Occhialini in experiences with cosmic rays, as well as by I. Curie with F. Joliot, by L.
Meitner, C.Y. Chao, H.H. Hupfeld, J.R. Richardson and J. Chadwick in experiments on induced radioactivity. It
was later observed, above all by Blackett and Occhialini, that these positive electrons just were the
antielectrons expected by the Dirac's electron theory. At the same time in which it was carried out the above
experiences on $\beta$ decay, a prominent role began to have the study of a new type of very high energy
radiation, most highly penetrating, that is to say, the \it cosmic radiation. \rm Between the 1940s and 1950s,
the unique available high energy sources were the cosmic rays, until the coming of particle accelerators in the
1950s which allowed more controllable energetic sources. In that period, there was a research competition
between experiences made on cosmic rays and those through particle accelerators.

Following (Polara 1949, Chap. V), (Castelfranchi 1959, Chap. XXIII), (Rossi 1964), (Tolanski 1966, Chap. 18),
(Born 1976, Chap. 2, § 15), (Brown and Hoddeson 1983), (Schlaepfer 2003) and (Carlson and De Angelis 2010), the
cosmic radiation was discovered, in the early years of 20th Century, by J. Elster with H. Geitel in German and
by C.T.R. Wilson in England, from the observation of a weak residual electrification in perfectly isolated
electroscopes. Some year later, E. Rutherford with H. Lester and H.L. Cooke, together to J.C. McLennan and E.F.
Burton, showed that 5 cm of lead reduced this mysterious radiation by 30\% while an additional 5 tonnes of
unrefined lead failed to reduce the radiation further. Such a phenomenon was immediately attributed to a not
well identified external strong penetrating radiation, maybe coming from the Earth. Thanks to a new electroscope
made by T. Wulf in 1907, it was possible to observe that this external radiation did not decrease with the
altitude but, in some cases, even increased, so that it could not come from the Earth as it was later confirmed
first by A. Gockel in 1909-10, then both by V. Hess with W. Kolh\"{o}rster in 1911-14 and by D. Pacini in 1911,
the former with a series of experiments made with flight balloons equipped with electroscopes, and the
latter\footnote{Very few are the textbooks which quote the Italian physicist Domenico Pacini (1878-1934) as one
of the pioneers of cosmic-ray research; amongst them (Castelfranchi 1959, Chap. XXIII) - where, inter alia, it
is also possible to find interesting historical notes throughout the text - and (Gliozzi 2005, Chap. 16, Section
16.12). For this historical case, and for a modern general historical revisitation of the cosmic ray story, see
(Carlson and De Angelis 2010).} by means of deep sea immersions of electroscopes. The World War I interrupted
the researches on this strange type of penetrating radiation, then retaken later in 1920s and 1930s with the
experiences of A. Millikan, E. Regener, G. Pfotzer, I.S. Bowen, H.V. Neher, H. Tizard, A. Piccard, M. Cosyns, W.
Bothe, J. Clay, A. Corlin, D. Hoffmann, D.V. Skobeltzyn, E. Steinke, G.H. Cameron, P.S. Gill, G.L. Locher, E.J.
Williams, C.F. Von Weizs\"{a}cker, L. Nordheim, J.B. Street, H. Kulenkampff, E.C. Stevenson and others,
continued until the 1940s and 1950s pioneering experimental works of T.H. Johnson, L.W. Alvarez, A.H. Compton,
C.D. Anderson, D.A. Glaser, S. Neddermeyer, G.E. Roberts, R.E. Marshak, H.A. Bethe, B. Rossi, F. Rasetti, G.
Bernardini, S. De Benedetti, C. St\o rmer, G. Lema\^{i}tre, M.S. Vallarta, V. Bush, G. Clark, P. Bassi, M.
Schein, H.L. Bradt, B. Peters, P.M.S. Blackett, G. Occhialini, C.F. Powell, C.M.G. Lattes, M. Conversi, E.
Pancini, O. Piccioni, H. Muirhead, J.F. Carlson, J.R. Oppenheimer, P. Auger, P. Ehrenfest, L. Leprince-Ringuet,
S.I. Tomonaga, G. Araki, G.D. Rochester, C.C. Butler and others (see (Brown and Hoddeson 1983, Part III) and
(Rossi 1964)). As regards the related experimental techniques employed, the first group of researches were
conducted by means of ionization chambers and, above all, Wilson chambers, these last first used by D.
Skobeltzyn in 1929, in the version improved by P.M.S. Blackett and under the action of strong magnetic fields.
In the second group of researches, instead, besides Wilson chambers, sequential Geiger-M\"{u}ller counters were
also used as well as photographic emulsion chambers in the version improved by C.F. Powell and G. Occhialini on
the basis of the previous 1937 works made by M. Blau and H. Wambacher on nuclear emulsions.

From the experimental data provided by all these notable works, in particular from the various absorption curves
related to this cosmic radiation and related geomagnetic effects, it was possible to identify two secondary
components, departing from a primary one, which have different nature according to the results of the \it
azimuthal \rm and \it latitude \rm effects, both then characterized, but in a different manner, by the so-called
\it east-west asymmetry \rm phenomenon which is closely connected with an asymmetry related to cosmic ray
intensity distributions, in turn related to the geometry of the so-called \it St\o rmer cones \rm which give the
allowed trajectories of cosmic rays under the action of the geomagnetic field. For this, it was identified both
a \it hard \rm component, much penetrating, and a \it soft \rm component, little penetrating. The terrestrial
magnetic field and the atmosphere, constitute two protective layers against the reaching of cosmic rays on the
Earth's surface: once that high energy primary cosmic rays hit terrestrial magnetosphere (involving too the Van
Allen belts - see (Rossi 1964, Chap. XIII)) and the upper atmosphere, they interact with the encountered atoms
(above all, the nitrogen and oxygen ones), the resulting collisions producing fragment's \it stars \rm (that is
to say, multiple traces outgoing from a same collision point) and atmospheric \it showers \rm of many particles,
at that time most unknown, according to \it multiple production processes \rm theorized by W. Heisenberg and G.
Wataghin. Theoretical attempts to explain the related phenomenology led to the so-called \it cascade theory of
cosmic showers \rm which was worked out, in 1930s and 1940s, mainly by J.F. Carlson with J.R. Oppenheimer in the
United States and by M. Blau, H. Wambacher, L. J\'{a}nossy, H.A. Bethe, W. Heitler, J. Hamilton, H. Peng and
H.J. Bhabha in Europe, but also with notable contributions by L.D. Landau, I.E. Tamm, V.L. Ginzburg, S.Z.
Belenky, H.S. Snyder, R. Serber, W.H. Furry and S.K. Chakrabarty. Thanks to this theory, it was possible to
ascertain that the soft component of cosmic radiation was mainly made by high energy electrons and photons,
whereas the determination of the particle composition of the hard component was more difficult to achieve; and,
at this point, the cosmic radiation and $\beta$ decay research pathways meet.

Following (Polara 1949, Chaps. V and VI), (Rossi 1964), (Muirhead 1965, Chap. 1), (Yang 1969, Chap. 2), (Segrè
1976, Chap. XII), (Born 1976, Chap. 2), (Zichichi 1981), (Brown et al., 1989), (Segrè 1999, Part III), (Zichichi
2000, II.1-3$^{a}$-II.1-4$^b$) and (Gliozzi 2005, Chap. 16, Section 16.12), whilst the (local) cosmic radiation
soft component was ascertained to be mainly made by high energy electrons and photons, many perplexities yet
held as regards the hard component whose constituents seemed do not belong to the set of elementary particles
then known. Indeed, the latter appeared to possess either positive or negative electric charge, while the
analysis of experimental data strongly indicated the existence of a particle having mass intermediate between
that of the proton and of the electron, and probably in the region of 100-200 electron masses ($m_e$). This
suspicion was verified by S.H. Neddermeyer with C.D. Anderson and by E.C. Stevenson with J.C. Street, in the
years between 1936 and 1938, who photographed quite instable particles with masses estimated (above all by R.B.
Brode and co-workers) to be about 200-240 $m_e$, stopping in a cloud chamber. These particles were generically
called \it mesotrons \rm by C.D. Anderson or \it mesons \rm by W. Heisenberg, because of their mass value; it
ended then to prevail the second name\footnote{Following (Gamow 1966, Chap. VIII), the name mesotron was
discarded because, under advise of Heisenberg father, a professor of classical language, the right etymology of
the term was inclined towards the term meson.}. As said above, notwithstanding many difficulties subsisted in
the research field devoted to cosmic radiation, the tenacity of researchers led to the conclusion that this type
of radiation (namely, the hard one) was formed by new particles both positively and negatively charged with mass
intermediate between the electron and proton ones. Nevertheless, these authors did not know the 1935 Yukawa work
and what there was predicted\footnote{The discovery of the meson, as well as that of the positron, has been
preceded by theoretical forecasts respectively due to H. Yukawa in the first case, and to P.A.M. Dirac in the
second one.}, mainly due to the fact that it was published in a journal not widely known outside Japan (see
(Kragh 2002, Chap. 13)). In such a paper, starting from the Heisenberg work on nuclear structure and from the
Fermi theory on $\beta$ decay, Yukawa supposed that proton and neutron could interact through a quantized field
(the mesonic one) of which, in analogy with the electromagnetic case, he computed too the main physical
properties of the related quantum.

Yukawa made a further suggestion about the properties of his hypothetical particle: indeed, in order to
simultaneously account for nuclear $\beta$ decay and for the fact that the meson had not been observed (at that
time), he suggested that it decayed spontaneously into one electron and one neutrino in a time which was
estimated to be about $10^{-7}$ sec. In 1938, with some first experiences made by H. Kuhlenkampff, the question
related to meson decay was one of the most debated of the period between the 1930s and the 1940s, which had, as
main protagonists, W. Heisenberg, P.M.S. Blackett, H. Euler, A.H. Compton, B. Rossi, D.B. Hall, N. Hilberry,
J.B. Hoag, W.M. Nielsen, H.V. Neher, M.A. Pomerantz, G. Bernardini, G. Cocconi, O. Piccioni, M. Conversi and
others. An apparent verification of this property was obtained by E.J. Williams and G.E. Roberts in 1940,
observing a $\beta$ decay of a particle of mass about 250 $m_e$ into a cloud chamber. In this period, attempts
to identify the generic meson observed in the cosmic radiation by C.D. Anderson and co-workers, with the
Yukawa's particle were done, notwithstanding that will reveal out to be false\footnote{The occurred mistaken
particle's identifications will be mainly due to the experimental difficulties to identify the related spin
values which are the only ones that allow to discern between particles having equal mass and charge values (see
(Villi et al. 1971, Introduction)).}. Furthermore, an apparent experimental evidence for such an identification
was provided by the measurements of the meson lifetimes by F. Rasetti in 1941 and by many others, amongst whom
B. Rossi, N.G. Nereson, K.I. Greiser, R. Chaminade, A. Freon and R. Maze. At the same time, the comparison of
the Yukawa nucleonic theory with the cosmic radiation one (mainly due to J.R. Oppenheimer and J.F. Carlson) in
the light of the obtained experimental data, above all those made by M. Conversi, E. Pancini and O. Piccioni in
Italy in the years 1943-1945 and by R. Chaminade, A. Freon and R. Maze in France in 1945, led R.E. Marshak and
H.A. Bethe to suggest in 1947 the possible existence of two different types of mesons, also on the wake of what
previously envisaged by E. Fermi, E. Teller and V.F. Weisskopf. Nevertheless, due to the World War II
circumstances, the Japanese physicists worked in an almost full isolation and most of their researches of that
time were recognized only later. Indeed, S.I. Tomonaga, Y. Tanikawa, S. Sakata and T. Inoue, already in 1943 had
proposed the hypothesis of the possible existence of two different types of meson. The main conclusion of the
above mentioned experiences was that the negative and positive mesons differently interacted with matter: in
fact, measuring the related capture rates $\lambda_c$, the positive ones decayed as they were more or less free,
whereas the negative ones were attracted by the nuclei, reacting in a strong manner with heavy nuclei and in a
weak manner with the light ones, and this wasn't what predicted by S.I. Tomonaga and G. Araki in 1940 on the
basis of Yukawa's theory.

The clarification of such a question came from the technological developments which have even been historically
connected with the scientific progress of ideas. Indeed, starting from the previous experimental techniques due
to S. Kinoshita and C. Waller, it was set up new nuclear emulsion detectors in 1940s by the Bristol group made
by C.F. Powell, G. Occhialini, C.M.G. Lattes and H. Muirhead, thanks to which it was possible to effectively
identify two types of mesons. This conclusion was further confirmed, in 1948, by other experiments run both by
the above Bristol group with also Y. Goldschmidt-Clermont, D.T. King and D.M. Ritson, and, at Berkeley, by E.
Gardner and C.M.G. Lattes with a particle accelerator. These two types of mesons, detected by the above
fundamental experiences, led to identify two first classes of mesons: in one, it was included those mesons at
first called \it primary mesons, \rm then \it meson $\mu$ \rm or \it muon\rm; in the other, it was included
lighter mesons at first called \it secondary mesons, \rm then \it meson $\pi$, \rm or \it pion. \rm In the
former falls the meson foreseen by C.D. Anderson with S.D. Neddermeyer and detected by one of the celebrated
Conversi-Pancini-Piccioni experiences, while in the latter it should fall the Yukawa's one. Thus, the $\pi$
meson provided the glue for nuclear forces and undergone to the following main decay-chain-reaction
$\pi\rightarrow\mu\rightarrow e$ where the first decay scheme $\pi\rightarrow\mu+\nu_{\mu}$ was first studied,
in 1948, by U. Camerini, H. Muirhead, C.F. Powell and D.M. Ritson as well as by J.R. Richardson. Then, the
various experiences performed on negative and positive counterparts of cosmic rays led to the conclusion
according to which both these two types of meson may be either positively or negatively charged, denoted by
$\mu^{\pm}$ and $\pi^{\pm}$, the Yukawa's one being of the type $\pi^{-}$. The $\pi$ meson significatively and
strongly interacts with atomic nuclei, contrarily to the $\mu$ meson which is mainly subjected to weak
interactions; the former has mass about 270-300 $m_e$ while the latter has mass about 200-210 $m_e$. The $\pi$
mesons decay in $\mu$ mesons and these, in turn, decay in electrons, by means of reactions of the type
$$\pi^+\rightarrow\mu^++\nu_{\mu},\ \pi^-\rightarrow e^-+\bar{\nu}_{\mu},\ \mu^+\rightarrow e^++\nu_{\mu}+
\bar{\nu}_{\mu},\ \mu^-\rightarrow e^-+\nu_{\mu}+\bar{\nu}_{\mu}$$where $\nu_{\mu}$ denotes the neutrino and
$\bar{\nu}_{\mu}$ the related antineutrino. As said above, the neutrino $\nu_e$ was determined to be different
from the neutrino $\nu_{\mu}$ by experiences made by G. Danby, J.M. Gaillard, K. Goulianos, L.M. Lederman, N.
Minstry, M. Schwartz and J. Steinberger in 1962 at Brookhaven. Nevertheless, the possible existence of two
different types of neutrino was first theoretically proposed by G. Puppi in 1948 in studying the universality of
Fermi weak interactions (\it Puppi triangle\rm) on the basis of decay processes involving $\mu$ and $\pi$
mesons: indeed, following (Hughes and Wu 1977, Vol. I, Chapter I), the most important contribution resulting
from the study of the muon so far, was probably the revelation of the close relationships between muon decay,
muon capture and nuclear beta decay, just known as the three sides of the Puppi triangle. Moreover, further
experiences made by F. Reines and C.L. Cowan Jr. in 1959, by M.G. Inghram and J.H. Reynolds in 1950, by C.S. Wu
in 1960 and by G. Bernardini\footnote{See (Zichichi 2008) for brief recalls on the work of Gilberto Bernardini.}
and co-workers in 1964, showed that $\nu_e\neq\bar{\nu}_e, \nu_{\mu}\neq\bar{\nu}_{\mu}, \nu_e\neq\nu_{\mu}$.

Afterwards, to explain the experimental evidence for the charge independence of nuclear forces as well as to
account for the soft component in cosmic radiation, independently of each other, N. Kemmer in 1938, H. Tamaki in
1942 and H.W. Lewis, J.R. Oppenheimer with S.A. Wouthuysen in 1947, pointed out that in addition to Yukawa's
charged meson, a neutral meson had to exist, whose experimental evidence was obtained in the years 1950-1951 by
A.G. Carlson, J.E. Hooper and D.T. King, by R. Bjorkland, W.E. Crandall, B.J. Moyer and H.F. York and by W.K.H.
Panofsky, R.L. Aamodt, H.F. York and J. Hadley. Such a meson, denoted by $\pi^0$, decays according to a law of
the type $\pi^0\rightarrow\gamma+\gamma$, being $\gamma$ a photon. The $\mu$ and $\pi$ mesons will be
generically called too \it L mesons. \rm At this point, it was generally felt that the neutral pion discovery
marked the end of particle searches, whereas the decay $\pi^0\rightarrow\gamma+\gamma$ marked instead the
opening of new horizons in subnuclear and theoretical physics: for instance, on the basis of this decay process,
J. Schwinger formulated the so-called \it partial conservation of the axial current \rm (PCAC) hypothesis in
quantum electrodynamics and quantum chromodynamics, opening new fruitful chapters in current algebra theory.
Indeed, studying in-depth the nuclear interactions of the particles of cosmic rays, it was possible to discover
other elementary particles. As said above, the $\mu$ mesons weakly interact with matter whereas the $\pi$ mesons
are nucleary active particles together other ones which were discovered at high altitudes in many mountain
laboratories located in different world areas, amongst which those at \it Aiguille du Midi \rm in the French
Pyrenees (Chamonix), at \it Testa Grigia \rm and \it Plateau Rosa \rm in the Italian Alps, at \it Chacaltaya \rm
in the Bolivian Andes, at \it Mount Evans \rm in Colorado, at \it Jungfraujoch \rm in the Bernese Alps, and so
on. Thanks to nuclear emulsion techniques set up by the Bristol group headed by C.F. Powell and by further
experiences made by R.H. Brown, U. Camerini, P.H. Fowler, H. Muirhead, C.F. Powell and D.M. Ritson, in the years
1947-1950 new particles having mass intermediate between the $\pi$ meson mass and the proton one, were detected.
They observed the decay of a charged particle into three charged mesons, one of these appearing to be a
$\pi$-particle. The parent particle was called a \it $\tau$ meson \rm and its mass was estimated to be about
1000 $m_e$, the first heavy meson. So, these researchers identified two classes of new heavy unstable particles,
that of \it heavy mesons \rm (lighter than the protons and heavier than the $\pi$ mesons) and that of \it
hyperons \rm (heavier than the protons), which may be electrically charged or neutral and never isolated. The
heavy mesons were also generically called \it K mesons \rm or \it kaons, \rm while the hyperons were also
generically called \it Y mesons, \rm so that a new hierarchy of mesons had to present. It was also customary to
indicate the nature and number of the decay products by subscripts; thus, for example, the $\tau$ meson was also
called a $K_{\pi 3}$ due to one of its decay schemes $\tau^+\rightarrow\pi^++\pi^++\pi^-$. In the 1950s, besides
the usual experimental research on cosmic radiation, with the advent of particle accelerators new experiences
begun too, in such a manner that new particles were discovered and considerable further researches were
accomplished to classify these new particles according to masses, lifetimes and decay schemes. All this
represented one of the most important period of the physics of 1950s.

The first heavy mesons and hyperons were observed by L. J\'{a}nossy in the years 1943-1946 at Dublin and by G.D.
Rochester and C.C. Butler in 1947 at Manchester. At first, such new particles were variously called \it k
particles \rm or $V$ \it particles, \rm due to the V-shaped tracks leaved by the non-neutral decay particles
observed into cloud chambers. Amongst these, there were those particles which will be later called
$K^0,\Lambda^0$, $\tau$ and $\theta$ \it particles, \rm these last two could be either neutral or electrically
charged. As recalled above, the Bristol group, headed by C.F. Powell, detected in 1949 the first positively
charged heavy meson, at first called $\tau^+$ \it meson, \rm then \it $K^+$ meson, \rm that undergo different
decay processes, amongst which $K^+\rightarrow\pi^++\pi^0$, whose experimental evidences were obtained, in the
years 1951-1954, by C. O'Ceallaigh, by the Paris group of B.P. Gregory, A. Laggarigue, L. Leprince-Ringuet, F.
Muller and Ch. Peyrou, by J. Crussard, M.F. Kaplon, J. Klarmann and J.H. Noon and by A.L. Hodson, J. Ballam,
W.H. Arnold, D.R. Harris, R.R. Rau, G.T. Reynolds and S.B. Treiman. Further researches made in cloud chambers
with magnetic fields will show that both $K^+$ and $K^-$ mesons exist. Then, the neutral heavy meson, at first
called $\theta^0$ \it particle, \rm then \it $K^0$ particle, \rm was first observed by C. O'Ceallaigh in 1950.
At the same time, R. Armenteros, K.H. Barker, C.C. Butler and A. Cachon as well as L.M. Lederman K. Lande, E.T.
Booth, J. Impeduglia and W. Chinowsky in 1956, were able to show that at least two types of neutral particles
existed, one is the $\Lambda^0$ hyperon decaying according to the scheme $\Lambda^0\rightarrow p+\pi^-$, and the
other probably decayed as follows $\theta^0\rightarrow\pi^++\pi^-$. At that time, the Bristol research group
discovered two heavy mesons, called \it $\tau$ \rm and \it $\theta$ mesons, \rm which initially seemed to be the
same particle because they had same mass and mean lifetime, but underwent distinct decay processes and had
different parity, so that, in the years 1953-1956, d'après R.H. Dalitz, it spoke of a \it $\theta-\tau$ puzzle.
\rm The analysis of further experimental data led T.D. Lee and C.N. Yang to assume in 1956 a \it parity
violation \rm of weak interactions, thanks to which it was possible to establish that $\tau$ and $\theta$ mesons
are the same particle, thereafter called \it K particle. \rm The discovery of the breaking of the symmetry
operators \it parity \rm (\it P\rm) and \it charge \rm (\it C\rm) received first experimental evidence by C.S.
Wu, E. Ambler, R.W. Hayward and D.D. Hoppes in 1957. Subsequent 1957 works made by R. Garwin, L. Lederman and M.
Weinrich and by J.J. Friedman and V.L. Telegdi, showed further evidence for a non-conservation of parity and
charge in the decay of kaons and hyperons, attaining a deeper theoretical knowledge on the \it C, P \rm and \it
T \rm invariance properties. Subsequently, many decay modes for kaons were found even if, at first, it was not
realised that they represented alternative decay modes of the same particle. Then, other types of hyperons were
also found in cosmic radiation: amongst these, C.M. York, R.B. Leighton and E.K. Bjornerund, in 1952, were able
to experimentally ascertain a new type of hyperon, called \it $\Sigma^+$ particle, \rm which decays according to
a reaction of the type $\Sigma^+\rightarrow n+\pi^0$. A further confirmation of the existence of
$\Sigma$-hyperons was obtained by A. Bonetti, R. Levi-Setti, M. Panetti and G. Tomasini in 1953, who also
identified the alternative decay mode $\Sigma^+\rightarrow n+\pi^+$, while the negative counterpart $\Sigma^-$
was observed by W.B. Fowler, R.P. Shutt, A.M. Thorndike and W.L. Whittemore at Brookhaven in 1954. Another
hyperon of mass $\sim 2600 m_e$, called $\Xi^-$ \it meson, \rm was detected by E.W. Cowan in 1954, which decays
according to the scheme $\Xi^-\rightarrow\Lambda^0+\pi^-$. The neutral $\Xi^0$ and $\Sigma^0$ hyperons were then
experimentally revealed by L.W. Alvarez, P. Eberhard, M.L. Good, W. Graziano, H.K. Ticho and S.G. Wojcicki in
1959, only after having been theoretically predicted as follows. This was the situation around middle of 1950s,
where the emphasis shifted from work using cosmic radiations to work on large accelerators. The attention was
also focused on the classification of the various particles so far discovered according to their masses,
lifetimes and decay schemes.

It was observed that the various kaons and hyperons discovered in the 1950s (above all the $\Lambda^0$), had a
strange behavior respect to their decay and production processes, so that they were given the collective
appellation of \it strange particles. \rm Namely, following (Yang 1961, Chapter III) and (Muirhead 1965, Chapter
1, Section 1.4), it was experimentally found that these strange particles have production times of about
$10^{-23}$ sec and decay times of about $10^{-10}$ sec, so that the forces involved in their production
processes were stronger than those present in the decay ones; furthermore, this strange fact did not occur when
such particles were isolated, in which only weak interactions taken place. To explain this incompatibility
between experimental data and theoretical framework, A. Pais proposed in 1952 the hypothesis of \it associated
production \rm according to which the decay and production processes are not inverses of each other, but rather
they differ for the presence of another (associated) particle, or rather, at least two strange particles should
be involved in the production process in order that a strong interaction could occur, whilst a weak interaction
occurs if only one strange particle is presents, as in the decay process. The Pais hypothesis received
experimental confirmation with the 1953-55 works of W.B. Fowler, R.P. Shutt, A.M. Thorndike and W.L. Whittemore.
At this point, it is necessary to reconsider the above mentioned notion of isospin and related conservation law.
Following (Muihread 1965, Chapter 1, Sections 1.3 and 1.4), the concept of conservation of isotopic spin
(isospin) is associated with the experimental evidence for the principle of charge independence of nuclear
forces according to which, at identical energies, the forces between any of the pairs of nucleons \it n-n, n-p
\rm and \it p-p, \rm depend only on the total angular momentum and parity of the pair and not upon their charge
state. So, B. Cassen and E.U. Condon, in 1936, showed that the principle of charge independence could be
elegantly expressed by the isospin concept. The isospin of a system is formally similar to angular momentum but
is linked to the charge states of the system. If a group of nuclei or particles exist in \it n \rm charge
multiplets, then the \it isospin number T \rm for this group is given by $2T+1=n$. The charge state of a
particle or nucleus in the multiplet is related to the third (along the \it z \rm axis) component of an \it
isospin operator \rm via the relations $Z={Q}/{e}=(T_3+\frac{1}{2}A)$ for nucleons and nuclei, and $Q/e=T_3$ for
pions, where $Z$ is the atomic number, $Q$ the total charge (\it hypercharge\rm) and $A$ the atomic weight of
the system. For instance, if $\chi_p$ and $\chi_n$ denote the isospin functions for the proton and the neutron
respectively, then $T_3$ has eigenvalues $1/2$ and $-1/2$ respectively, that is to say $T_3\chi_p=(1/2)\chi_p$
and $T_3\chi_n=-(1/2)\chi_n$. The isospin quantum numbers were assigned to the strange particles produced,
independently by M. Gell-Mann and by T. Nakano and K. Nishijima, in the years between 1952 and 1956. In regards
to a cluster of elementary particles, they observed invariance properties when the charge center of the
multiplet of strange particles was displaced respect to the center of the multiplet of non-strange particles.
Furthermore, they considered a scheme assuming that the conservation laws for $T$ and $T_3$ were conserved or
broken in dependence on the given interaction: to be precise, $T$ and $T_3$ were conserved in strong
interactions, $T$ was broken whilst $T_3$ was conserved for electromagnetic interactions and, finally, $T$ and
$T_3$ were broken in weak interactions. The satisfactory nature of Gell-Mann, Nakano and Nishijima scheme lays
in the fact that it predicted the existence of two new elementary particles which were later experimentally
found.

Later, again following (Muirhead 1965, Chapter 1, Section 1.4), it was pointed out by Gell-Mann and Nishijima,
independently of each other in the years 1955-56, that a more elegant classification of the strongly interacting
particles than that based on isospin alone, could be made if a new parameter $S$, called the \it strangeness
number, \rm was introduced and defined by the relation $Q/e=T_3+(B+S)/2$ where $B$ is the \it baryon number\rm;
baryon is a generic name for nucleons and hyperons. Such a relation shows that the associated production
phenomena and the isospin symmetry are related together. As it has been said above, the success of the
Gell-Mann, Nakano and Nishijima scheme lies in the fact that it predicted, on the basis of isospin and
strangeness conservation laws, various charge multiplets; in particular, new particles were predicted like the
$\Sigma^0$, $\Xi^0$, $K^+$, $K^-$ and related antiparticles. All these theoretical assumptions received
experimental confirmation by the works of Y. Nambu, K. Nishilima, Y. Yamaguchi and W.B. Fowler in the years
1953-1955, as well as many of the above predicted particles (besides the $\Xi^0$ and $\Sigma^0$ above
remembered) were detected by W.B. Fowler, R.P. Shutt, A.M. Thorndike, W.L. Whittemore and W.D. Walker, in the
years 1955-1959. However, the elegant semi-empirical classification scheme of Gell-Mann, Nakano and Nishijima,
envisaged the existence of many other new particles which will be later discovered, whose static and dynamic
properties will be of fundamental importance for the subsequent 1960s theoretical work of M. Gell-Mann and Y.
Ne'eman (\it eihgtfold way\rm). In conclusion, following\footnote{Such a leading textbook has been one of main
references herein followed.} (Muirhead 1965, Chapter 1, Section 1.5), the discoveries of new particles have
occurred sometimes as a result of theoretical insights and sometimes by accident, the most strange particle
falling into the latter category: from what has been said above, at this stage, we have that the weak
interactions are associated with electrons, muons and neutrinos $(e, \mu, \nu)$, collectively called \it leptons
\rm (which are not subject to strong interactions) and with certain decay processes for mesons and hyperons; the
nucleons and hyperons $(n, p, \Lambda, \Sigma, \Xi)$ are then collectively called \it baryons. \rm Following
(Roman 1960, Introduction), the particles recalled so far may be also classified into two classes according to
their spin values, identifying a first class in which fall particles with integer spin $(\gamma, \pi, K)$, said
to be \it bosons, \rm and a second one in which fall particles with half-integer spin $(\nu, e, \mu, p, n,
\Lambda, \Sigma, \Xi)$ said to be \it fermions. \rm The fermions again fall into two rather distinct groups,
namely leptons and baryons. Apart from the photons, the bosons fall too into two groups: the lighter
$\pi$-mesons (or pions) and the heavier $K$-mesons (or kaons). Thus, our classification scheme is tentatively
Photon-Leptons-Pions-Kaons-Baryons, and represents one of the main achievement of the physics in the decade
1950-1960.

\section*{2. Some early works on cosmic rays}\addcontentsline{toc}{section}
{2. Some early works on cosmic rays}

In this section, according to what has been said in the Preface, in delineating some early works on cosmic rays
we consider the first research papers of A. Zichichi and co-workers, dating back to 1950s, which concerned
physics of $K$ mesons, with particular attention to the experimental context. To be precise, as a research
fellow at CERN of Geneva, he joined the heavy meson decay research group which, in turn, belonged to the wider
investigation activity on cosmic rays, according to the research program policy of that period of this worldwide
renowned research institution. The first papers of Zichichi on cosmic rays mainly concern with the various
properties of strange particles.

As said in the previous section 1, in that period there were still many unsolved questions concerning the
so-called \it strange \rm or $V$ particles. In 1., it is just discussed, together the related experimental
arrangements, certain peculiar features showed by only two out of twenty events observed into a cloud chamber
located at Testa Grigia laboratories (3,500 m.a.s.l.) in which about 150 $V^0$-particles related to decay of a
single charged unstable fragment (like an unstable hydrogen isotope, or an excited deuteron or triton) have been
found. From a deep phenomenology analysis of the experimental data and taking into account the already existent
related literature, it emerged that, very likely, one of these two observed peculiar events, say the Event I,
could be due to a certain $\Lambda^0$-decay of an unstable hydrogen isotope rather than a nuclear interaction
(which perhaps has would been quite unusual), whereas, instead, the second one, say the Event II, led to the
conclusion according to which, under certain further hypotheses, it could be a $\tau$-decay, even supposing that
it is not a nuclear interaction.

The second work 2. is a contribution to the experimental evidence for the existence of the neutral
$\tau^0$-meson which had been predicted but not established. To be precise, it is reported and discussed the
result of an experiment made at Jungfraujoch Research Station and consisting of four tracks originating at a
point in the cloud chamber gas which may be interpreted either as the radiative decay of a $\theta^0$-meson or
as the decay of a $\tau^0$-meson. Two out of these are tracks of positive and negative electrons electrons,
while the other two are tracks of fast particles resembling a typical $V$-event which is most readily explained
as the decay of a neutral $\tau$-meson, rather than a $\theta^0$-meson, with the subsequent decay of the
secondary $\pi^0$-meson into an electron pair and a $\gamma$-ray, which may be respectively written as
$\tau^0\rightarrow\pi^++\pi^-+\pi^0$ and $\pi^0\rightarrow e^++e^-+\gamma$. After having discussed on the
various possibilities, this four-pronged event is deemed to be geometrically associated with a small nuclear
interaction with, in turn, can be interpreted, on the basis of the experimental data, as a charge exchange
reaction of the type $K^++n\rightarrow\tau^0+p$, albeit it is also not excluded a possible double production
according to the scheme $\pi^++n\rightarrow\tau^0+\Sigma^+$ in which, in turn, the $\Sigma^+$-particle decays
into one proton and one yet undetected $\pi^0$-meson.

Taking into account the previous work 2. discussed above, the work 3. mainly shows some results coming from
certain emulsion experiments which provide the first two remarkable examples of $K$-meson pairs of the type
$(K^0, \bar{K}^0)$ and $(K^+,\bar{K}^0)$, produced in elementary neutron-proton interactions whose production
reactions respectively are $n+p\rightarrow K^0+\bar{K}^0+n+p$ and $n+p\rightarrow K^++\bar{K}^0+n+n$; these
allowed to extend the knowledge on the phenomenology of heavy mesons, to further confirmation of some of the
various hypotheses suggested by M. Gell-Mann and A. Pais in the years 1952-54 about associated productions, in
particular, the prediction according to which the $K^0$-mesons should exist in two states as particle and
antiparticle with $S=+1$ and $S=-1$. To be precise, from a systematic study of the associated production of
heavy mesons and hyperons in a cosmic-ray cloud chamber, in 3. examples of very simple nuclear interactions
giving rise to pairs of $K$-mesons have been found. The importance of these observations is that they provide
experimental evidence to support the theoretical prediction that $K^0$-mesons should exist in two states with
opposite strangeness $S=+1$ and $S=-1$, that is to say, these events are evidence that $K^0$-mesons with both
positive and negative strangeness exist. In relation to the well-known $\theta-\tau$ puzzle, the authors also
argued on the possible identification or not of the produced $K^0$-mesons with the $\theta^0$-particles, but the
experimental measurements weren't of great usefulness for this.

The work 4. is a brief research note in which it is determined, on the basis of previous works made by J.A.
Newth and M.S. Bartlett, the mean lifetime estimates of the two decaying particles $\Lambda_0$ and $\theta_0$,
isolated amongst 115 $V^0$-events observed in a multi-plate cloud chamber triggered for penetrating showers, and
respectively interested to the following main decay reactions $\Lambda^0\rightarrow p+\pi^-+37$ MeV and
$\theta^0\rightarrow\pi^++\pi^-+214$ MeV. Moreover, neglecting the existence of other types of unstable neutral
particles with a two-body decay, it has been possible to classify the above 115 $V^0$-particles.

The paper 5. is a research report, presented at CERN Scientific Policy Committee on 21 October 1957, after the
CERN research activity on cosmic rays taken the decision to stop the so-called Geneva experiment on $K$-meson
decays, whose related motivations were exposed in the subsequent CERN report No. CERN/SPC/52 (B). In 5., the
director-general of the Jungfraujoch Research Station in detail proposed a new experimental apparatus
specifically designed to prosecute the research activity on high energy interactions with a mountain experiment
based on the nuclear interaction of protons at energies in the neighborhood of 100 GeV, through a large magnet
cloud chamber. The novel feature of this apparatus was a magnetic spectrometer which measured the momenta of the
primary particles. One of the main aims of this experiment was also that to fathom new directions on particle
physics as, for instance, the search for new unstable strange particles having very short lifetimes.

As it has been said in section 1, in 1950s gradually started to run the first particle accelerators of
synchrotron type which will be intended to replace the cosmic-ray researches. But this conclusion did not yet
apply to the study of the production processes of strange particles. Now, the results described in the work 6.
come from an experiment designed to study the production of strange particles in materials of low and high
atomic weight, precisely carbon and copper consecutively used, through the interaction of energetic secondaries,
sprung out by nuclear interactions in passing through the targets (of carbon and copper), which produce 79
neutral $V$-particles. The division of all the $V^0$-events into $\Lambda^0$- and $\theta^0$- decays was made in
order to determine their lifetime estimates. If one denotes with $N(\Lambda^0)$ and $N(\theta^0)$ the numbers of
$\Lambda^0$- and $\theta^0$-mesons so produced, then a significant difference between the values of the ratio
$N(\Lambda^0):N(\theta^0)$ for their production in carbon and copper has been found; this asymmetry's fact
occurred in the decay of the $\Lambda^0$ particles respect to the short-lived $\theta^0$ ones, was also
explained by H. Blumenfeld, E.T. Booth, L.M. Ledermann and W. Chinowsky, who conducted similar experiences in
1956 with carbon and lead, reaching to almost equal results, through the associated production of pairs of
$K$-mesons through which it is possible to increase the number of $\Lambda^0$ particles so slowly produced, with
also non-conservation of strangeness. Thus, the results achieved in 6. as well as by Blumenfeld and co-workers,
may be taken as further evidence for the great importance of the pair production of $K$-mesons in cosmic-ray
experiments.

The decay asymmetry detected in the previous work 6. will be deeper studied in the next work 7. where many other
properties of $\Lambda^0$- and $\theta^0$- particles, like for example spin, mean lifetime, behavior with
respect to inversion operators and anisotropy effects on geometrical distributions, have carried out on 107
$\Lambda^0$ and $\theta^0$ particles produced in iron plates of a multiple cloud chamber exposed to cosmic
radiation at an altitude of 3,500 m.a.s.l. Likewise, the work 8. reports the first results of an experimental
study of the nuclear interaction of cosmic rays (mainly of the type proton-proton) with a magnet cloud chamber
based at an altitude of 3,500 m.a.s.l. and operating at energy of about 100 GeV, showing that such a type of
nuclear interaction study is feasible.

\section*{3. Historical introduction: II}\addcontentsline{toc}{section}{3. Historical introduction: II}

In this section, we recall the main events and facts of that historical path which goes from the introduction of
the spin to the notion of anomalous magnetic moment, with particular attention to the leptonic case. The
necessarily limited historical framework so outlined in this section, covers a temporal period which roughly
goes up from early 1920s to 1960s.

\subsection*{3.1 On Landé separation factors}\addcontentsline{toc}{subsection}{3.1 On Landé separation factors}

Following (Muirhead 1965, Chapter 2) and (Tomonaga 1997), when a fundamental interaction is taken into account
then the experimental determination of the basic particle data, like masses, lifetimes, spins and magnetic
moments, is necessarily required. The most accurately known properties of the particles are those which can be
associated with their magnetic moments. Magnetic properties of elementary particles have been and yet are of
paramount importance both to theoretical and experimental high energy physics. One of the main intrinsic
properties of the elementary particles is the spin, which can be inferred from the conservation laws for angular
momentum. Following (Landau 1982, Chapter VIII), in both classical and quantum mechanics, the laws of
conservation of angular momentum are a consequence of the isotropy of space respect to a closed system, so that
it depends on the transformation properties under rotation of the coordinate system. Therefore, all quantum
systems, like atomic nuclei or composite systems of elementary particles, besides the orbital angular momentum,
show to have as well an intrinsic angular momentum, called \it spin, \rm which is unconnected with its motion in
space and to which it is also associated a magnetic moment whose strengths are not quantized and may assume any
value. The spin disappears in the classical limit $\hbar\rightarrow 0$ so that it has no classical counterpart.
The spin must be meant as fully distinct from the angular momentum due to the motion of the particle in space,
that is to say, the orbital angular momentum. The particle concerned may be either elementary or composite but
behaving in some respect as an elementary particle (e.g. an atomic nucleus). The spin of a particle (measured,
like the orbital angular momentum, in units of $\hbar$) will be denoted by $\vec{s}$. Following (Rich and Wesley
1972), (Bertolotti 2005, Chapter 8), (Miller et al. 2007) and (Roberts and Marciano 2010, Chapter 1), the
physical idea that an electron has an intrinsic angular momentum was first put forward independently of each
other by A.H. Compton in 1921 to explain ferromagnetism\footnote{Furthermore, Compton acknowledges A.L. Parson
for having first proposed the electron as a spinning ring of charge. Compton modified this idea considering a
much smaller distribution of charge mainly concentrated near the center of the electron. The Compton's paper is
almost unknown (see (Compton 1921)) albeit it is quoted by the 1926 Uhlenbeck and Goudsmit paper. Following
(Roberts and Marciano 2010, Chapter 3, Section 3.2.1), also R. Kronig proposed, in 1925, the spin as an internal
angular momentum responsible for the electron forth's quantum number (see (Bertolotti 2005, Chapter 8).} and by
G. Uhlenbeck and S. Goudsmit in 1925 to explain spectroscopic observations in relation to the anomalous Zeeman
effect, while spin was introduced into quantum mechanics by W. Pauli in 1927 as an additional term to the \it
Pauli equation \rm which is obtained by the non-relativistic representation of the Dirac equation to small
velocities (see (Jegerlehner 2008, Part I, Chapter 3, Section 3.2)) to account for the quantum mechanical
treatment of the spin-orbit coupling of the anomalous Zeeman effect (see also (Haken \& Wolf 2005, Chapter 14,
Section 3)). An equation similar to the Pauli's one, was also introduced by C.G. Darwin in 1927 (see (Roberts
and Marciano 2010, Chapter 3, Section 3.2.1)).

Following (Jegerlehner 2008, Part I, Chapter 1), (Melnikov and Vainshtein 2006, Chapter 1) and (Shankar 1994,
Chapter 14), leptons have interesting static (classical) electromagnetic and weak properties like the magnetic
and electric dipole moments. Classically, dipole moments may arise either from electrical charges or currents.
In this regards, an important example which may turns out to be useful to our purposes is the circulating
current, due to an orbiting particle with electric charge $Q$ and mass $m$, which exhibits the following orbital
magnetic dipole moment
\begin{equation}\vec{\mu}_L=\frac{Q}{2c}\vec{r}\wedge\vec{v}=\frac{Q}{2mc}\vec{L}=\Gamma\vec{L}\end{equation}
where $\Gamma=Q/2mc$ is the \it classical gyromagnetic ratio\footnote{Usually, the gyromagnetic ratio is denoted
by lower case $\gamma$, but here we prefer to use the upper case $\Gamma$ to distinguish it by the well-known
\it Lorentz factor $\gamma=1/\sqrt{1-\beta^2}$ \rm with $\beta=v^2/c^2$.} \rm and $\vec{L}=
m\vec{r}\wedge\vec{v}=\vec{r}\wedge\vec{p}$ is the orbital angular momentum whose corresponding quantum
observable is the operator $-i\hbar\vec{r}\wedge\nabla=\hbar\vec{l}$, so that we have the following orbital
magnetic dipole moment operator (see (Jegerlehner 2008, Part I, Chapter 3) and (Shankar 1994, Chapter 14))
\begin{equation}\vec{\mu}_l=g_l\frac{Q\hbar}{2mc}\vec{l}\end{equation}where $g_l$ is a constant introduced by the
usual quantization transcription rules. For $Q=e$, the quantity $\mu_0=e\hbar/2mc$ is normally used as a unit
for the magnetic moments and is called the \it Bohr magneton. \rm The electric charge $Q$ is usually measured in
units of $e$, so that $Q=-1$ for leptons and $Q=+1$ for antileptons; therefore, we also may rewrite (2) in the
following form
\begin{equation}\vec{\mu}_l=g_l\frac{Qe\hbar}{2mc}\vec{l}=g_lQ\mu_0\vec{l}.\end{equation}
Both electric and magnetic properties have their origin in the electrical charges and their currents, apart from
the existence or not of magnetic charges. Following (Jegerlehner 2008, Part I, Chapter 1) and (Muirhead 1965,
Chapter 9, Section 9.2(d)), whatever the origin of magnetic and electric moments are, they contribute to the
electromagnetic interaction Hamiltonian (interaction energy) of the particle with magnetic and electric fields
which, in the non-relativistic limit, is given by
\begin{equation}\mathcal{H}_{em}=-(\vec{\mu}_m\cdot\vec{B}+\vec{d}_e\cdot\vec{E})\end{equation}
where $\vec{\mu}_m$ and $\vec{d}_e$ are respectively the magnetic and electric dipole moments (see (Jegerlehner
2008, Part I, Chapter 1)).

If one replaces the orbital angular momentum $\vec{L}$ with the spin $\vec{s}$, then we might search for an
analogous (classical) magnetic dipole moment, say $\vec{\mu}_s$, associated with it and, therefore, given by
$(Q/2mc)\vec{s}$. Nevertheless, following (Born 1969, Chapter 6, Section 38) and (Muirhead 1965, Chapter 2,
Section 2.5)), to fully account for the anomalous Zeeman effect, we should consider this last expression
multiplied by a certain scalar factor, say $g_s$ (often simply denoted by $g$), so that
\begin{equation}\vec{\mu}_s=g_s\frac{Q}{2mc}\vec{s}\end{equation}which is said to be the \it spin magnetic
moment. \rm Now, introducing, as a corresponding quantum observable, the \it spin operator \rm defined by
$\vec{S}=\hbar\vec{s}=\hbar\vec{\sigma}/2$, where $\vec{\sigma}$ is the \it Pauli spin operator, \rm it is
possible to consider both the \it spin magnetic moment operator \rm and the \it electric dipole moment operator
\rm (see (Jegerlehner 2008, Part I, Chapter 1)), respectively defined as follows
\begin{equation}\vec{\mu}_s\doteq g_sQ\mu_0\frac{\vec{\sigma}}{2},\ \ \ \ \ \ \ \ \ \
\vec{d}_e\doteq\eta Q\mu_0\frac{\vec{\sigma}}{2},\end{equation}where $\eta$ is a constant, the electric
counterpart of $g_s$. Following (Caldirola et al. 1982, Chapter XI, Section 3), the attribution of a $s=1/2$
spin value to the electron, led to the formulation of the so-called \it vectorial model \rm of the atom. In such
a model, amongst other things, the electron orbital angular moment $\vec{L}$ composes with the spin $\vec{s}$
through well-defined spin-orbit coupling rules (like the Russell-Saunders ones) to give the (classical) \it
total angular moment \rm defined to be $\vec{j}\doteq\vec{L}+\vec{s}$, while the (classical) \it total magnetic
moment \rm is defined to be $\vec{\mu}_{total}\doteq\vec{\mu}_L+\vec{\mu}_s$, so that, taking into account (3)
and (6), the corresponding quantum observable counterpart, in this vectorial model, is
\begin{equation}\vec{\mu}_{total}\doteq\vec{\mu}_l+\vec{\mu}_s=g_lQ\mu_0\vec{l}+g_sQ\mu_0\frac{\vec{\sigma}}{2}=
Q\mu_0(g_l\vec{l}+g_s\vec{S})\end{equation}which is said to be the \it total magnetic moment \rm of the given
elementary particle with charge $Q$ and mass $m$; since $g_s\neq 1$, it follows that it is not, in general,
parallel to the total angular moment operator $\vec{J}\doteq\vec{l}+\vec{S}$, so that it undergoes to precession
phenomena when magnetic fields act.

The existence of the various above constants $g_l,g_s$ and $\eta$ is mainly due to the fact that, in the
vectorial model of anomalous Zeeman effect, the direction of total angular moment $\vec{j}$ does not coincide
with the direction of total magnetic moment, so that these scalar factors just take into account the related
non-zero angles which are called \it Landé separation factors \rm because first introduced by A. Landé
(1888-1976) in the early 1920s (see (Born 1969, Chapter 6, Section 38)). To be precise, only the parallel
component of $\vec{\mu}_{tot}$ to $\vec{j}$, say $\vec{\mu}_{tot}^{\|}$, is efficacious, so that we should have
\begin{equation}\vec{\mu}_{tot}^{\|}=g_j\frac{Q\hbar}{2mc}\vec{j}\end{equation}where the scalar factor $g_j$
(or simply $g$) takes into account the difference between the vectorial model of anomalous Zeeman effect and the
theory of the normal one. To may computes this factor, we start from the relation
\begin{equation}{\mu}_{tot}^{\|}=\mu_l\cos(\widehat{\vec{l},\vec{j}})+\mu_s\cos(\widehat{\vec{s},\vec{j}})\end{equation}
with\begin{equation}\mu_l=g_l\frac{Q\hbar}{2mc}l,\ \ \ \ \ \ \ \ \ \
\mu_s=g_s\frac{Q\hbar}{2mc}s\end{equation}where $g_l$ and $g_s$ are known to be respectively the \it orbital \rm
and \it spin factors, \rm which, in turn, represent the ratios respectively between the orbital and spin
magnetic and mechanic moments. Replacing (10) into (9), we have
\begin{equation}g_j=g_l\frac{l}{j}\cos(\widehat{\vec{l},\vec{j}})+
g_s\frac{s}{j}\cos(\widehat{\vec{s},\vec{j}})\end{equation}from which (see (Born 1969, Chapter 6, Section 38))
it is possible to reach to the following relation
\begin{equation}g_j=g_l\frac{(j^2+l^2-s^2)}{2j^2}+g_s\frac{(j^2+s^2-l^2)}{2j^2}\end{equation}Experimental
evidences dating back to 1920s and mainly related to the anomalous Zeeman effect, seemed suggesting that $g_l=1$
and $g_s=2$ for the electron, that is, the atomic vectorial model explains the fine structure features of alkali
metals and the anomalous Zeeman effect if one supposes to be $g_s\neq 1$, that is to say, a spin intrinsic
gyromagnetic ratio anomalous respect to the orbital one ($g_l=1$), so speaking of a \it spin anomaly. \rm
Following (Bohm 1993, Chapter IX, Section 3), the deviations from the $g_s=2$ value for the electron comes from
the radiative corrections of quantum electrodynamics and is of the same order as, and of analogous origin to,
the Lamb shift. The value $g_s=2$ was first established as far back as 1915 by a celebrated experiment of A.
Einstein and W.J. de Haas which led to the formulation of the so-called \it Einstein-de Hass effect \rm and that
was also incorporated in the spin hypothesis put forward in the 1920s (see (\u{S}polskij 1986, Volume II,
Chapter VII, Section 70)). Following (Jegerlehner 2008, Part I, Chapter 1), the anomalous magnetic moment is an
observable which may be studied through experimental analysis of the motion of leptons. The story started in
1925 when Uhlenbeck and Goudsmit put forward the hypothesis that an electron had an intrinsic angular momentum
of $\hbar/2$ and that associated with this there were a magnetic dipole moment equal to $e\hbar/2mc$, i.e. the
Bohr magneton $\mu_0$. According to E. Back and A. Landé, the question which naturally arose was whether the
magnetic moment of the electron $(\mu_m)_e$ is precisely equal to $\mu_0$, or else $g_s=1$ in $(10)_2$, to which
them tried to answer through a detailed study of numerous experimental investigations on the Zeeman effect made
in 1925, reaching to the conclusion that the Uhlenbeck and Goudsmit hypothesis was consistent although they did
not really determine the value of $g_s$. In 1927, Pauli formulated the quantum mechanical treatment of the
electron spin in which $g_s$ remained a free parameter, whilst Dirac presented his revolutionary relativistic
theory of electron in 1928, which, instead, unexpectedly predicted $g_s=2$ and $g_l=1$ for a free electron. The
first experimental evidences for the Dirac's theoretical foresights for electrons came from L.E. Kinster and
W.V. Houston in 1934, albeit with large experimental errors at that time. Following (Kusch 1956), it took many
more years of experimental attempts to descry that the electron magnetic moment could exceed 2 by about 0.12,
the first clear indication of the existence of a certain anomalous contribution to the magnetic moment given by
\begin{equation}a_i\doteq\frac{(g_s)_i-2}{2},\ \ \ \ \ \ \ \ \ \ i=e,\mu,\tau.\end{equation} With the new
results on renormalization of QED by J. Schwinger, S.I. Tomonaga and R.P. Feynman of 1940s, the notion of \it
anomalous magnetic moment \rm (AMM) falls into the wider class of QED radiative corrections.

\subsection*{3.2 On Field Theory aspects of AMM}\addcontentsline{toc}{subsection}{3.2 On Field Theory aspects of AMM}

Following (Jegerlehner 2008, Part I, Chapter 3), for the measurement of the anomalous magnetic moment of a
lepton, it is necessary to consider the motion of a relativistic point-particle $i$ (or Dirac
particle\footnote{That is to say, a particle without internal structure.}) of charge $Q_ie$ and mass $m_i$ in an
external electromagnetic field $A_{\mu}^{ext}(x)$. The equations of motion of a charged Dirac particle in an
external field are given by the \it Dirac equation
\begin{equation}\big(i\hbar\gamma^{\mu}\partial_{\mu}+Q_i\frac{e}{c}\gamma^{\mu}(A_{\mu}+A_{\mu}^{ext}(x))-m_ic\big)
\psi_i(x)=0,\end{equation}\rm and by the second order wave equation\begin{equation}\big(\Box
g^{\mu\nu}-(1-\xi^{-1}\big)\partial^{\mu}\partial^{\nu})A_{\nu}(x)=-Q_ie\bar{\psi}_i(x)\gamma^{\mu}\psi_i(x).
\end{equation}The first step is now to find a solution to the relativistic one-particle problem given by the
Dirac equation (14) in the presence of an external field, neglecting the radiation field in first approximation.
In such a case, the equation (14) reduces to
\begin{equation}i\hbar\frac{\partial\psi_i}{\partial t}=\big(-c {\vec{\alpha}}(i\hbar{\vec{\nabla}}-
Q_i\frac{e}{c}{\vec{A}})-Q_ie\Phi+\beta m_ic^2\big)\psi_i\end{equation}where
\begin{equation}\beta = \gamma^0 = \left(
\begin{array}{cc}
1 & 0 \\
0 & -1 \\
\end{array}
\right),\ \ \ \vec{\alpha} = \gamma^0\vec{\gamma}= \left(
\begin{array}{cc}
0 & {\vec{\sigma}} \\
\vec{\sigma} & 0 \\
\end{array}
\right)\end{equation} are the Dirac matrices, $A^{\mu\ ext}=(\Phi,\vec{A})$ is the electromagnetic
four-potential with scalar and vector potential respectively given by $\Phi$ and $\vec{A}$ (of the external
electromagnetic field) and $i=e,\mu,\tau$. For the interpretation of the solution to the last Dirac equation
(16), the non-relativistic limit plays an important role because many relativistic QFT problems may be most
easily understood and solved in terms of the non-relativistic problem as a starting point. To this end, it is
helpful and more transparent to work in natural units, the general rules of transcription being the following:
$p^{\mu}\rightarrow p^{\mu}, d\mu(p)\rightarrow \hbar^{-3}d\mu(p), m\rightarrow mc, e\rightarrow e/(\hbar c),
\exp(ipx)\rightarrow\exp(ipx/\hbar)$ and, for spinors, $^t(u,v)\rightarrow {^t(u/\sqrt{c},v/\sqrt{c})}$;
furthermore, we shall consider a generic lepton $e^-, \mu^-, \tau^-$ with charge $Q_i$, dropping the index $i$.
Moreover, to get, from the Dirac spinor $\psi$, the two-component Pauli spinors $\varphi$ and $\chi$ in the
non-relativistic limit, one has to perform an appropriate unitary transformation, the so-called \it
Foldy-Wouthuysen transformation\footnote{It is a unitary transformation introduced around the late 1940s by L.L.
Foldy and S.A. Wouthuysen to study the non-relativistic limits of Dirac equation as well as to overcome certain
conceptual and theoretical problems arising from the relativistic interpretations of position and momentum
operators. Following (Foldy and Wouthuysen 1950), in the non-relativistic limit, where the momentum of the
particle is small compared to $m$, it is well known that a Dirac particle (that is, one with spin 1/2) can be
described by a two-component wave function in the Pauli theory. The usual method of demonstrating that the Dirac
theory goes into the Pauli theory in this limit makes use of the fact that two of the four Dirac-function
components become small when the momentum is small. One then writes out the equations satisfied by the four
components and solves, approximately, two of the equations for the small components. By substituting these
solutions in the remaining two equations, one obtains a pair of equations for the large components which are
essentially the Pauli spin equations. See (Bjorken and Drell 1964, Chapter 4).}, \rm upon the Dirac equation
(16) rewritten as follows
\begin{equation}i\hbar\frac{\partial\psi}{\partial t}=\vec{H}\psi, \ \ \ \ \ \vec{H}=c\vec{\alpha}
\big(\vec{p}-\frac{Q}{c}\vec{A}\big)+\beta mc^2+Q\Phi,\end{equation}with $\vec{\alpha}$ and $\beta$ given by
(17) (see (Bjorken and Drell 1964, Chapter 1, Section 4, Formula (1.26)).

Then, following (Bjorken and Drell 1964, Chapter 1, Section 4) and (Jegerlehner 2008, Part I, Chapter 3), in
order to obtain the non-relativistic representation for small velocities, we should split off the phase of the
Dirac field $\psi$, which is due to the rest energy of the lepton
\begin{equation}\psi=\tilde{\psi}\exp\big(-i\frac{mc^2}{\hbar}t\big),\ \ \ \ \ \tilde{\psi}=\left(
\begin{array}{c}
\tilde{\varphi} \\
\tilde{\chi}
\end{array}
\right)\end{equation}so that the Dirac equation takes the form
\begin{equation}i\hbar\frac{\partial\tilde{\psi}}{\partial t}=(\vec{H}-mc^2)\tilde{\psi}\end{equation}and
describes the following coupled system of equations
\begin{equation}\big(i\hbar\frac{\partial}{\partial
t}-Q\Phi\big)\tilde{\varphi}=c\vec{\sigma}\big(\vec{p}-\frac{Q}{c}\vec{A}\big)\tilde{\chi},\end{equation}
\begin{equation}\big(i\hbar\frac{\partial}{\partial t}-Q\Phi+2mc^2\big)\tilde{\chi}=c\vec{\sigma}\big(\vec{p}
-\frac{Q}{c}\vec{A}\big)\tilde{\varphi}\end{equation}which, respectively, provide the Pauli description in the
non-relativistic limit and the one of the negative-energy states. As $c\rightarrow\infty$, it is possible to
prove that
\begin{equation}\tilde{\chi}\cong\frac{1}{2mc}\vec{\sigma}\big(\vec{p}-\frac{Q}{c}\vec{A}\big)\tilde{\varphi}
+O(1/c^2),\end{equation}by which we have\begin{equation}\big(i\hbar\frac{\partial}{\partial
t}-Q\Phi\big)\tilde{\varphi}\cong\frac{1}{2m}\big(\vec{\sigma}(\vec{p}-\frac{Q}{c}\vec{A})\big)^2\tilde{\varphi}
\end{equation}and since $\vec{p}$ does not commute with $\vec{A}$, we may use the relation
\begin{equation}(\vec{\sigma}\vec{a})(\vec{\sigma}\vec{b})=\vec{a}\vec{b}+i\vec{\sigma}(\vec{a}\wedge\vec{b})
\end{equation}to obtain\begin{equation}\big(\vec{\sigma}(\vec{p}-\frac{Q}{c}\vec{A})\big)^2=
\big(\vec{p}-\frac{Q}{c}\vec{A}\big)^2-\frac{Q\hbar}{c}\vec{\sigma}\cdot\vec{B}\end{equation}where $\vec{B}=rot\
\vec{A}$, so finally reaching to the following 1927 \it Pauli equation
\begin{equation}i\hbar\frac{\partial\tilde{\varphi}}{\partial t}=\tilde{H}\tilde{\varphi}=
\Big(\frac{1}{2m}\big(\vec{p}-\frac{Q}{c}\vec{A}\big)^2+Q\Phi-\frac{Q\hbar}{2mc}\vec{\sigma}\cdot\vec{B}\Big)
\end{equation}\rm which, up to the spin term, is nothing but the non-relativistic Schr\"{o}dinger equation. Following too
(Muirhead 1965, Chapter 3, Section 3.3(f)), the last term of (27) has the form of an additional potential
energy. Now, by (4), since the potential energy of a magnet of moment $\vec{\mu}_m$, in a field of strength $B$,
is $-\vec{\mu}_m\cdot\vec{B}$, equation (27) shows that a Dirac particle with electric charge $Q$ should possess
a magnetic moment equal to $(Q\hbar/2mc)\vec{\sigma}=2Q\mu_0\vec{\sigma}/2$ that, compared with $(6)_1$, would
imply $g_s=2$. This is what Dirac theory historically provided for an electron. Later, Pauli showed as the Dirac
equation could be little modified to account for leptons of arbitrary magnetic moment by adding a suitable term.

Indeed, in\footnote{See also (Pauli 1973, Chapter 6, Section 29).} (Pauli 1941, Section 5)), the author
concludes his report with some simple applications of the theories discussed in (Pauli 1941, Part II, Sections
1, 2(d) and 3(a)), concerning the interaction of particles of spin 0, 1, and 1/2 with the electromagnetic field.
In the last two cases we denote the value $e\hbar/2mc$ of the magnetic moment as the normal one, where $m$ is
the rest mass of the particle. The assumption of a more general value $g_s(e\hbar/2mc)$ for the magnetic moment
demands the introduction of additional terms, proportional to $g_s-1$, into the Lagrangian or Hamiltonian. Pauli
concludes his report with some simple applications of the theories discussed in (Pauli 1941, Part II, Sections
1, 2(d) and 3(a)) concerning the interaction of particles of spin 0, 1, and 1/2 with the electromagnetic field.
In the last two cases, Pauli denotes the value $e\hbar/2mc$ of the magnetic moment as the normal one, where $m$
is the rest mass of the particle. The assumption of a more general value $g(e\hbar/2mc)$ for the magnetic moment
demands the introduction of additional terms, proportional to $g-1$, in the Lagrangian or Hamiltonian. To be
precise, following (Dirac 1958, Chapter 11, Section 70), (Corinaldesi and Strocchi 1963, Chapter VII, Section
4), (Muirhead 1965, Chapter 3, Section 3.3(f)) and (Levich et al. 1973, Chapter 8, Section 63 and Chapter 13,
Section 118), Pauli modified the basic Dirac equation, written in scalar form as follows
\begin{equation}i\hbar\gamma_{\mu}\frac{\partial}{\partial
x_{\mu}}\psi+mc^2\psi-i\hbar\frac{Q}{c}\gamma_{\mu}A_{\mu}\psi=0,\end{equation}to get the following Lorentz
invariant \it Dirac-Pauli equation \rm
\begin{eqnarray}& & i\hbar\gamma_{\mu}\frac{\partial}{\partial
x_{\mu}}\psi+mc^2\psi-i\hbar\frac{Q}{c}\gamma_{\mu}A_{\mu}\psi-i\hbar
a_{\mu}\gamma_{\mu}\gamma_{\nu}(A_{\mu,\nu} -A_{\nu,\mu})\nonumber\\ & & =
i\hbar\gamma_{\mu}\frac{\partial}{\partial x_{\mu}}\psi+mc^2\psi-i\hbar\frac{Q}{c}\gamma_{\mu}A_{\mu}\psi-i\hbar
a_{\mu}\sigma_{\mu\nu}q_{\nu}A_{\mu}=0\end{eqnarray}replacing the gauge invariant interaction term
$-i\hbar\sigma_{\mu\nu}q_{\nu}A_{\mu}$ with the following phenomenological term (see also (Sakurai 1967, Chapter
3, Section 3-5) $-i\hbar a_{\mu}\sigma_{\mu\nu}q_{\nu}A_{\mu}$ called an \it anomalous moment interaction \rm
(or \it Pauli moment\rm), where $a_{\mu}$ represents the anomalous part of the magnetic moment of the particle,
$q$ is the momentum transfer and $\hat{\sigma}=-(i/2)[\vec{\gamma},\vec{\gamma}]$ is the spin $1/2$ momentum
tensor. In the non-relativistic limit, this last expression reduces to the following equation (compare with
(27))
\begin{equation}i\hbar\frac{\partial\psi}{\partial t}=
\Big(\frac{1}{2m}\big(\vec{p}-\frac{Q}{c}\vec{A}\big)^2+Q\psi-\big(a_{\mu}+\frac{Q\hbar}{2mc}\big)
\vec{\sigma}\cdot\vec{B}\Big)\end{equation}so justifying the use of the term 'anomalous' to denote a deviation
from the classical results. Thus, the transition from the non-relativistic approximation of the Dirac equation
goes over into the Pauli equation; furthermore, from this reduction there results not only the existence of the
spin of particles but also the existence of the intrinsic magnetic moment of particle and its anomalous part.
Namely, we should have $g_s=2(1+a_{\mu})$, where its higher order part $a_{\mu}=(g_s-2)/2\geq 0$ just measures
the deviation's degree respect to the value $g_s=2$ (Dirac moment) as predicted by the 1928 Dirac theory for
electron\footnote{Following (Roberts and Marciano 2010) and (Miller et al. 2007, Section 1), the
non-relativistic reduction of the Dirac equation for an electron in a weak magnetic field $\vec{B}$, is as
follows $i\hbar(\partial\psi/\partial t)=[(p^2/2m)-(e/2m)(\vec{L}+2\vec{S})\cdot\vec{B}]\psi$, by which it
follows that $g_s=2$.} as well as by H.A. Kramers in 1934 (see (Farley and Semertzidis 2004, Section 1))
developing Lorentz covariant equations for spin motion in a moving system. Later, this Pauli \it ansatz \rm was
formally improved and generalized by L.L. Foldy and S.A. Wouthuysen in the forties to obtain a generalized Pauli
equation which will be the theoretical underpinning of further experiments. Indeed, at the first order in $1/c$,
the lepton behaves as a particle which has, other than a charge, also a magnetic moment given by
$\mu_m=(Q\hbar/2mc)\vec{\sigma}=(Q/mc)\vec{S}$, as said above. Following (Corinaldesi and Strocchi 1963, Chapter
VII, Section 5), (Bjorken and Drell 1964, Chapter 4, Section 3) and (Jegerlehner 2008, Part I, Chapter 3), from
an expansion in $1/c$ of the Dirac Hamiltonian given by $(18)_2$, we have the following effective third order
Hamiltonian obtained applying a third canonical Foldy-Wouthuysen transformation to $(18)_2$
\begin{eqnarray}\vec{H}'''_{FW}&=&\beta\Big(mc^2+\frac{\big(\vec{p}-(Q/c)\vec{A}\big)^2}{2m}-\frac{\vec{p}^4}
{8m^3c^2}\Big)+Q\Phi-\beta\frac{Q\hbar}{2mc}\vec{\sigma}\cdot\vec{B}+\nonumber \\& &-\frac{Q\hbar^2}{8m^2c^2}div
\vec{E}-\frac{Q\hbar}{4m^2c^2}\vec{\sigma}\cdot\big[(\vec{E}\wedge\vec{p}+\frac{i}{2}rot
\vec{E})\big]+O(1/c^3)\end{eqnarray}where each term of it, has a direct physical meaning: see (Bjorken and Drell
1964, Chapter 4, Section 3) for more details. In particular, the last term takes into account the spin-orbit
coupling interaction energy and will play a fundamental role in setting up the experimental apparatus of many
$g-2$ later experiments. The last Hamiltonian, to the third order, gives rise to the following \it generalized
Pauli equation $i\hbar(\partial\tilde{\varphi}/\partial t)=\vec{H}'''_{FW}\tilde{\varphi}$, \rm which is a
generalized version, including high relativistic terms via the application of a Foldy-Wouthuysen transformation,
of the first form proposed by Pauli in 1941 (see (Pauli 1941)) and that leads to the second approximation
Schr\"{o}dinger-Pauli equation as a non-relativistic limit of the Dirac equation (see (Corinaldesi and Strocchi
1963, Chapter VIII, Section 1)).

Our particular interest is the motion of a lepton in an external field under consideration of the full
relativistic quantum behavior which is ruled by the QED equations of motions (14) and (15) that, in turn, under
the action of an external field, reduce to (16). For slowly varying field, the motion is essentially determined
by the generalized Pauli equation which besides also serves as a basis for understanding the role of the
magnetic moment of a lepton at the classical level. The anomalous magnetic moment roughly estimates the
deviations from the exact value $g_s=2$, because of certain relativistic quantum fluctuations in the
electromagnetic field (initially called \it Zitterbewegung\rm) around the leptons and mainly due, besides weak
and strong interaction effects, to QED higher order effects as a consequence of the interaction of the lepton
with the external (electromagnetic) field and which are usually eliminated through the so-called \it radiative
corrections. \rm At present, we are interested to QED contributions only. Following (Muirhead 1965, Chapter 11,
Section 11.4), (Jegerlehner 2008, Part I, Chapter 3) and (Melnikov and Vainshtein 2006, Chapter 2), the QED
Lagrangian of interaction of leptons and photons is (see also (Muirhead 1965, Chapter 8, Section 8.3(a)))
\begin{equation}\mathcal{L}^{QED}_{int}=-\frac{1}{4}F_{\mu\nu}F^{\mu\nu}+\bar{\psi}(i{\gamma_{\mu}\partial_{\mu}}-m)
\psi-QJ^{\mu}A_{\mu}\end{equation}where $\psi$ is the lepton field, $A^{\mu}=(\Phi,\vec{A})$ is the vector
potential of the electromagnetic field, $F^{\mu\nu}=\partial^{\mu}A^{\nu}-\partial^{\nu}A^{\mu}$ is the
field-strength tensor of the electromagnetic field, $J^{\mu}(x)=\bar{\psi}(x)\gamma^{\mu}\psi(x)$ is the
electric current and $Q$ is the lepton charge. Let us consider an incoming lepton $l(p_1^{\mu},r_1)$, with
4-momentum $p_1^{\mu}$, rest mass $m$, charge $Q$ and $r_1$ as third component of spin, which scatters off the
external electromagnetic potential $A_{\mu}$ towards a lepton $l(p_2^{\mu},r_2)$ of 4-momentum $p_2^{\mu}$ and
third component of spin $r_2$. To the first order in the external field and in the classical limit of
$q^2=p_2^2-p_1^2\rightarrow 0$, the interaction is described by the following scattering
amplitude\begin{equation}\mathcal{M}(x;p)=\langle
l(p_2^{\mu},r_2)|J^{\mu}(x)|l(p_1^{\mu},r_1)\rangle\end{equation}where $\vec{q}=\vec{p}_2-\vec{p}_1$ is the
momentum transfer. In practice, it will be more convenient to work, through Fourier transforms, with invariant
momentum transfers rather than spatial functions. So, in momentum space, due to space-time translation
invariance for which $J^{\mu}(x)=\exp(iPx)J^{\mu}(0)\exp(-iPx)$, and to the fact that the lepton states are
eigenstates of 4-momentum, that is to say $\exp(-iPx)|l(p_i,r_i)\rangle=\exp(-ip_ix)|l(p_i;r_i)\rangle, i=1,2$,
we find the following Fourier transform of the scattering matrix
\begin{eqnarray}\tilde{\mathcal{M}}(q;p)&=&\int\exp(iqx)\langle l(p_2,r_2)|J^{\mu}(x)|l(p_1,r_1)\rangle
d^4x=\nonumber\\&=&\int\exp[i(p_2-p_1-q)x]\langle l(p_2,r_2)|J^{\mu}(0)|l(p_1,r_1)\rangle d^4x=\nonumber\\&=&
(2\pi)^4\delta^{(4)}(q-p_2+p_1)\langle l(p_2,r_2)|J^{\mu}(0)|l(p_1,r_1)\rangle\end{eqnarray}which is
proportional to the Dirac $\delta$-function of 4-momentum conservation. Therefore, the $T$-matrix element is
given by\begin{equation}\langle l(p_2,r_2)|J^{\mu}(0)|l(p_1,r_1)\rangle.\end{equation}Via the current
conservation law $\partial_{\mu}J^{\mu}(\vec{x})=0$ and the parity conservation in QED, the most general
parametrization of the $T$-matrix element has the following QED relativistically covariant decomposition
\begin{equation}\langle l(p_2)|J^{\mu}(0)|l(p_1)\rangle
=\bar{u}(p_2)\Gamma^{\mu}(p_2,p_1)u(p_1)\end{equation}where $\Gamma^{\mu}$, called lepton-photon \it vertex
function, \rm is any expression (or group of expression) which has the transformation properties of a 4-vector
and is also a $4\times4$ matrix in the spin space of the lepton. Following (Muirhead 1965, Chapter 11, Section
11.4(c)) and (Roberts and Marciano 2010, Chapter 2, Section 2.2; Chapter 3, Section 3.2.2), we shall have the
following Lorentz structure for the scattering amplitude
\begin{equation}\bar{u}(p_2)\Gamma^{\mu}(p_2,p_1)u(p_1)=-iQ\bar{u}(p_2)
\Big(F_D(q^2)\gamma^{\mu}+F_P(q^2)\frac{i\sigma^{\mu\nu}q_{\nu}}{2m}\Big)u(p_1)\end{equation}where $u(p)$
denotes the Dirac spinors, while
$\sigma^{\mu\nu}=(i/2)(\gamma^{\mu}\gamma^{\nu}-\gamma^{\nu}\gamma^{\mu})=(i/2)[\gamma^{\mu},\gamma^{\nu}]$ are
the components of the Dirac spin operator $\hat{\sigma}=-(i/2)\vec{\gamma}\wedge\vec{\gamma}$ or else the spin
$1/2$ angular momentum tensor. $F_D(q^2)$ (or $F_E(q^2)$) is the \it Dirac \rm(or \it electric charge\rm) \it
form factor, \rm while $F_P(q^2)$ (or $F_M(q^2)$) is the \it Pauli \rm (or \it magnetic\rm) \it form factor, \rm
which roughly are connected respectively with the distribution of charge over the lepton and with the anomalous
magnetic moment to the interaction lepton-electromagnetic field. We now need to know the relationships between
these form factors and the anomalous part of the lepton magnetic moment.

In the non-relativistic quantum mechanics, a lepton interacting with an electromagnetic field is described by
the Hamiltonian
\begin{equation}H=\frac{(\vec{p}-Q\vec{A})^2}{2m}-\vec{\mu}_s\cdot\vec{B}+Q\Phi, \ \ \ \ \ \ \vec{B}=rot\vec{A}
\end{equation}which is nothing that $\tilde{H}$ of (27). To find the relations between the lepton magnetic moment
$\mu_s$ and the Dirac and Pauli form factors, we consider the scattering of the lepton off the external vector
potential $A_{\mu}$ in the non-relativistic approximation, using the Hamiltonian (38) and comparing the results
with (33). Following (Melnikov and Vainshtein 2006, Chapter 2), the non-relativistic scattering amplitude in the
first order Born approximation is given by
\begin{equation}\Omega=-\frac{m}{2\pi}\int{\bar{\psi}}(\vec{p}_2)V\psi(\vec{p}_1)d^3\vec{r}\end{equation}where
$\psi(\vec{p}_{1})=\tilde{\varphi}\exp(i\vec{p}_1\cdot\vec{r})$ and
$\psi(\vec{p}_2)=\tilde{\chi}\exp(i\vec{p}_2\cdot\vec{r})$ are the wave functions of the lepton described by the
two components of Pauli spinors (see (19)) $\tilde{\varphi}$ and $\tilde{\chi}$, and
\begin{equation}V=-\frac{Q}{2m}(\vec{p}\cdot\vec{A}+\vec{A}\cdot\vec{p})-\mu_s\vec{\sigma}\cdot\vec{B}+Q\Phi.
\end{equation}By a Fourier transform, we have\begin{equation}\Omega=-\frac{m}{2\pi}\tilde{\chi}
\Big(-\frac{Q}{2m}\vec{A}_q\cdot(\vec{p}_2+\vec{p}_1)+Q\Phi_q-i\mu_s\vec{\sigma}\cdot(\vec{q}\wedge\vec{A}_q)\Big)
\tilde{\varphi}\end{equation}where $\Phi_q$ and $\vec{A}_q$ stands for the Fourier transforms of the electric
potential $\Phi$ and of the vector potential $\vec{A}$. Therefore, we will derive (41) starting from the
relativistic expression for the scattering amplitude (33) and taking then the non-relativistic limit. If the
Dirac spinors are normalized to $2m$, the relation between the two oscillating amplitudes in the
non-relativistic limit, is given by\begin{equation}-i\lim_{|\vec{p}|\ll
m}\mathcal{M}(x;p)=4\pi\Omega.\end{equation}To derive the non-relativistic limit of the scattering amplitude
$\mathcal{M}$, we use the explicit representation of the Dirac matrices, given by
\begin{equation}\gamma^0 = \left(
\begin{array}{cc}
I & 0 \\
0 & -I \\
\end{array}
\right),\ \ \ \ \ \gamma^i= \left(
\begin{array}{cc}
0 & {{\sigma}_i} \\
-{\sigma}_i & 0 \\
\end{array}
\right)\ \ \ i=1,2,3,\end{equation}and the Dirac spinors $u(p)$. Using these expressions in $\mathcal{M}$ and
working at first order in $|\vec{p}_i|/m \ i=1,2$, we obtain
\begin{eqnarray}\mathcal{M}&=&-2iem\tilde{\chi}\Big[F_D(0)\Big(\Phi_q-\frac{\vec{A}_q\cdot(\vec{p}_1+\vec{p}_2)}{2m}\Big)
+\nonumber\\ & &
-i\frac{F_D(0)+F_P(0)}{2m}\vec{\sigma}\cdot(\vec{q}\wedge\vec{A}_q)\Big]\tilde{\varphi}.\end{eqnarray}Using
(41), (42) and (44), we find\begin{equation}F_D(0)=1,\ \ \ \ \ \ \ \ \ \
\mu_s=\frac{Q}{2m}\big(F_D(0)+F_P(0)\big)\end{equation}which compared with (5) and (6), give
\begin{equation}g_s=2(1+F_P(0))\end{equation}so that, if the Pauli form factor $F_P(q^2)$ does not vanish for
$q=0$, then $g_s$ is different from 2, the value predicted by Dirac theory of electron. It is conventional to
call this difference the muon anomalous magnetic moment and write it as
\begin{equation}a_{\mu}=F_P(0)=\frac{g_s-2}{2}\end{equation}so that, in the static (classical) limit we have too
\begin{equation}F_D(0)=1,\ \ \ \ \ \ \ \ \ \ F_P(0)=a_{\mu}\end{equation}where the first relation is the
so-called \it charge renormalization condition \rm (in units of $Q$), while the second relation is the finite
prediction for $a_{\mu}$ in terms of the pauli form factor. In QED, $a_{\mu}$ may be computed in the
perturbative expansion in the fine structure constant\footnote{Following (Muirhead 1965, Chapter 1, Section
1.3(b)), the interaction of the elementary particles with each other can be separated into three main classes,
each with its own coupling strength. To be precise, the common parameter appearing in the electromagnetic
processes is the \it fine structure constant $\alpha=e^2/4\pi\hbar c$\rm; the strength of strong interactions is
characterized by the dimensionless coupling term $g^2/4\pi\hbar c$, while the weak interactions are ruled by the
Fermi coupling constant $G_F$.} $\alpha=Q^2/4\pi$ as follows
\begin{equation}a_{\mu}^{QED}=\sum_{i=1}^{\infty}a_{\mu}^{(i)}=\sum_{i=1}^{\infty}c_i
\Big(\frac{\alpha}{\pi}\Big)^i.\end{equation}The first term in the series is $O(\alpha)$ since, when radiative
corrections are neglected, the Pauli form factor vanishes. This is easily seen from the QED Lagrangian
$\mathcal{L}_{int}^{QED}$ given by (32), which implies that, through leading order in $\alpha$, the interaction
between the external electromagnetic field and the lepton, is given by
$-iQ\bar{u}(p_2)\gamma^{\mu}u(p_1)A_{\mu}$. A consequence of the current conservation, is the fact that the
Dirac form factor satisfies the condition $F_D(0)=1$ to all orders in the perturbation expansion. The
renormalization constants influence the Pauli form factor only indirectly, through the mass, the charge and the
fermion wave function renormalization, because there is no corresponding tree-level operator in QED Lagrangian.
Therefore, the anomalous magnetic moment is the unique prediction of QED; moreover, the $O(\alpha)$ contribution
to $a_{\mu}$ has to be finite without any renormalization. The QED radiative corrections provide the largest
contribution to the lepton anomalous magnetic moment. The one-loop result was computed by J. Schwinger in 1948
(see (Schwinger 1948)), who found the following lowest-order radiative (or one-loop) correction to the electron
anomaly (see (Rich and Wesley 1972) and (Roberts and Marciano 2010, Chapter 3, Section 3.2.2.1))
\begin{equation}a_{e}^{(2)}=F_P(0)=\alpha/2\pi\cong 0.00116.\end{equation}In 1949, F.J. Dyson showed that
Schwinger's theory could be extended to allow calculation of higher-order corrections to the properties of
quantum systems. Since Dyson showed too that the one-loop QED contribution to the anomalous magnetic moment did
not depend on the mass of the fermion, the Schwinger's result turned out to be valid for all leptons, so that we
have $a_i^{(2)}=F_P(0)=\alpha/2\pi, \ i=e,\mu,\tau$. Currently, QED calculations have been extended to the
four-loop order and even some estimates of the five-loop contribution exist. It is interesting however to remark
that Schwinger's calculation was performed before the renormalizability of QED were understood in details;
historically, this provided a first interesting example of a fundamental physics result derived from a theory
that was considered to be quite ambiguous at that time. Therefore, the anomalous magnetic moment of a lepton is
a dimensionless quantity which may be computed order by order as a perturbative expansion in the fine structure
constant $\alpha$ in QED and beyond this, in the Standard Model (SM) of elementary particles or extensions of
it. As an effective interaction term, the anomalous magnetic moment is mainly induced by the interaction of the
lepton with photons or other particles, so that it has a pure QED origin. It corresponds to a dimension 5
operator (see (51)) and since any renormalizable theory is constrained to exhibit terms of dimension 4 or less
only, it follows that such a term must be absent for any fermion in any renormalizable theory at tree (or
zero-loop) level. It is the absence of such a \it Pauli term \rm that leads to the prediction $g_s=2+O(\alpha)$.
Therefore, at that time, it was necessary looking for other theoretical tools and techniques to experimentally
approach the determination of the anomalous magnetic moment of leptons. Following (Jegerlehner 2008, Part I,
Chapter 3), in higher orders the form factors for the muon in general acquires an imaginary part. Indeed, if one
considers the following effective dipole moment Lagrangian with complex coupling
\begin{equation}\mathcal{L}_{eff}^{DM}=-\frac{1}{2}\Big[\bar{\psi}\sigma^{\mu\nu}\Big(D_{\mu}\frac{1+\gamma_5}{2}
+\bar{D}_{\mu}\frac{1-\gamma_5}{2}\Big)\psi\Big]F_{\mu\nu}\end{equation}with $\psi$ the muon field, we have
\begin{equation}\Re D_{\mu}=a_{\mu}\frac{Q}{2m_{\mu}},\ \ \ \ \ \Im D_{\mu}=d_{\mu}=\frac{\eta}{2}
\frac{Q}{2m_{\mu}},\end{equation}so that the imaginary part of $F_P(0)$ corresponds to an electric dipole moment
(EDM) which is non-vanishing only if we have $T$ violation. The equation (51) provides as well the connection
between the magnetic and electric dipole moments through the dipole operator $D$. As we will see later, the
incoming new ideas on symmetry in QFT will turn out to be of extreme usefulness to approach and to analyze the
problem of determination of the anomalous magnetic moment of the leptons, the equation (51) being just one of
these important results.

\subsection*{3.3 Experimental determinations of the lepton AMM: a brief historical sketch}
\addcontentsline{toc}{subsection}{3.3 Experimental determinations of the lepton AMM: a brief historical sketch}

\subsubsection*{3.3.1 On the early 1940s experiences}\addcontentsline{toc}
{subsubsection}{3.3.1 On the early 1940s experiences}

Following (Kusch 1956), (Rich and Wesley 1972), (Farley and Picasso 1979), (Hughes 2003) and (Jegerlehner 2008,
Part I, Chapter 1), in the same period in which appeared the famous 1948 Schwinger seminal research note, thanks
to the new molecular-beams magnetic resonance spectroscopy methods mainly worked out by the research group
leaded by I.I. Rabi in the late of 1930s, P. Kusch and H.M. Foley detected, in 1947, a small anomalous
$g_L$-value for the electron within a 4\% accuracy (see also (Weisskopf 1949)), analyzing the $^{2}P_{3/2}$ and
$^{2}P_{1/2}$ state transition of Gallium: to be precise they found the values $g_s=2.00229\pm 0.00008$ and
$g_l=0.99886\pm 0.00004$; later, J.E. Nafe, E.B. Nelson and Rabi himself were able, in May 1947, to detect a
discrepancy between theoretical and predicted values of about 0.26\% by the measurements of the hyperfine
structure level splitting of hydrogen and deuterium in the ground state on the accepted Dirac $g$-factor of 2,
which was quickly confirmed in the same year by D.E. Nagle, R.S. Julian and J.R. Zacharias (see also (Schweber
1961, Chapter 15, Section d)). In this regards, in September 1947, G. Breit (1947a,b) suggested that such
discordances between theoretical expectations and experimental evidences could be overcome if one had supposed
$g\neq 2$. Independently by Breit, also J.M. Luttinger (1948) (as well as T.A. Welton and Z. Koba - see (Rich
and Wesley 1972) and references therein - between 1948 and 1949) stated that some experiments of then, seemed to
require a modification in the $g$-factor of the electron. In this regards, Schwinger suggested that the coupling
between the electron and the radiation field could be the responsible of this, calculating the effect on the
basis of a general subtraction formalism for the infinities of quantum electrodynamics. Luttinger, instead,
shown that the possible change in the electron magnetic moment could be derived very simply without any
reference to an elaborate subtraction formalism. Soon after, P. Kusch, E.B. Nelson and H.M. Foley presented, in
1948, another precision measurement of the magnetic moment of the electron, just before Schwinger's theoretical
result whose 1948 paper besides quotes them, which together the discovery of the fine structure of hydrogen
spectrum (\it Lamb shift\rm) by W.E. Lamb Jr. and R.C. Retherford in 1947, as well as the corresponding
calculations by H.A. Bethe, N.M. Kroll, V. Weisskopf, J.B. French and W.E. Lamb Jr. in the same period, were the
main triumphs of testing the new level of QED theoretical understanding with precision experiments. All that was
therefore a stimulus for the development of modern QED. These successes had a strong impact in establishing the
QFT as a general formal framework for the theory of elementary particles and for our understanding of
fundamental interactions. The late 1940s were characterized by a close intertwinement between theory and
experiment which greatly stimulated the rise of the new QED. On the theoretical side, a prominent role was
gradually undertaken by the new non-Abelian gauge theory proposed by C.N. Yang and R.L. Mills in 1954 as well as
by the various relativistic local QFT symmetries amongst which the discrete ones of charge conjugation $(C)$,
parity $(P)$ and time-reversal $(T)$ reflection which are related amongst them by the well-known $CPT$ theorem,
according to which the product of the these three discrete transformations, taken in any order, is a symmetry of
any relativistic QFT (see (Streater and Wightman 1964)). Actually, in contrast to the single transformations
$C$, $P$ and $T$, which are symmetries of the electromagnetic and strong interactions only (d'après T.D. Lee and
C.N. Yang celebrated work), $CPT$ is a universal symmetry and it is this symmetry which warrants that particles
and antiparticles have identical masses as well as equal lifetimes; but also the dipole moments are very
interesting quantities for the study of the discrete symmetries mentioned above.

\subsubsection*{3.3.2 Some previous theoretical issues}\addcontentsline{toc}
{subsubsection}{3.3.2 Some previous theoretical issues}

The celebrated 1956 paper of T.D. Lee and C.N. Yang (see (Lee and Yang 1956)) on parity violation, has been an
invaluable source of theoretical insights. The paper discusses the question of the possible failure of parity
conservation in weak interactions taking into account what experimental evidences existed then as well as
possible proposal of experiments for testing this hypothesis. Amongst these last, they discuss, since the
beginning, on some experiments concerning polarized proton beams which would have led to an electric dipole
moment if the parity violation were occurred. The related important consequences were too discussed, like the
proton and neutron EDM, taking into consideration the previous early 1950s experiences made by E.M. Purcell,
N.F. Ramsey and J.H. Smith for the proton who made an experimental measurement of the electric dipole moment of
the neutron by a neutron-beam magnetic resonance method, finding a value less than $10^{-20}$ $e$-cm ca. in
agreement with parity conservation for strong and electromagnetic interactions. Nevertheless, Lee and Yang
argued that yet lacked valid experimental confirmations of parity conservation for weak interactions suggesting,
to this end, to consider the measure of the angular distribution of the electrons coming from $\beta$ decays of
oriented nuclei like those of $Co^{60}$, thing that will be immediately done, with success, by C.S. Wu and
co-workers, furnishing a first experimental evidence for a lack of parity conservation in $\beta$ decays.
Subsequently, Lee and Yang also argue on the question of parity conservation in meson and hyperon decays, as
well as in those strange particle decays having the following features: 1) the strange particle involved has a
non-vanishing spin and (2) it decays into two particles at least one of which has a non-vanishing spin or rather
it decays into three or more particles. Thus, what conjectured by Lee and Yang could be also applied to the
decay processes a) $\pi\rightarrow\mu+\nu$ and b) $\mu\rightarrow e+2\nu$. So, in the sequential decay
$\pi\rightarrow\mu\rightarrow e$, starting from a $\pi$ meson at rest, one might study the distribution of the
angle $\theta$ between the $\mu$-meson momentum and the electron momentum, the latter being in the
center-of-mass system of the $\mu$ meson. The decay b) is then a pure leptonic one, so no hadronic phenomenon is
involved, this making easier the related calculations (see (Okun 1986, Chapter 3)). Lee and Yang then argue
that, if parity is conserved in neither a) nor b), then the distribution will not in general be identical for
$\theta$ and $\pi-\theta$ directions. To understand this, one may consider first the orientation of the muon
spin. If a) violates parity conservation, then the muon would be in general polarized along its direction of
motion. In the subsequent decay b), the angular distribution problem with respect to $\theta$ is therefore
closely similar to the angular distribution problem of $\beta$ rays from oriented nuclei, as discussed before,
so that, in this way, it will be also possible to detect possible parity violations in this type of decays.
These last remarks on $\pi\mu e$ sequence will be immediately put in practice in the celebrated 1956 experiences
pursued by R.L. Garwin, L.M. Lederman with M. Weinrich and by J.L. Friedman with V.L. Telegdi, which will
further confirm Lee and Yang hypothesis of parity violation in weak interactions. Following (Sakurai 1964,
Chapter 7, Section 2) and (Schwartz 1972, Chapter 4, Section 11), polarized muons slow down and stop before they
decay, but depending on the material (graphite, aluminium, etc.) the muon spin direction is still preserved, so
we have a source of polarized muons. Negative muons are emitted with their angular momenta pointing along their
directions of motion, whereas positive muons are emitted with their angular momenta pointing opposite to their
directions of motion. Furthermore, if these positive muons were stopped in matter and allowed to decay, then the
direction of this angular momentum (or spin) at the moment of decay could be determined by the distribution in
directions of the emitted decay electron which follow the former. If parity is not conserved in muon decay
either, then there will be a forward-backward asymmetry in the positron distribution with respect to the
original $\mu^+$ direction. The just above mentioned experiences showed more positrons emitted backward with
respect to the $\mu^+$ direction, showing that parity is not conserved in both $\pi$ and $\mu$ decays.

As it has said above, Lee and Yang already argued on electric dipole moments in relation to parity conservation
law for fundamental interactions, in some respects enlarging the discussion to the general framework of discrete
symmetry transformations. To understand about the properties of the dipole moments under the action of such
transformations, in particular the behavior under parity and time-reversal, we have to look at the interaction
Hamiltonian (4) and, above all, at the equations (6) which both depend on the axial vector $\vec{\sigma}$, so
that also $\vec{\mu}_m$ and $\vec{d}_e$ will be also axial vectors. On the other hand, the electric field
$\vec{E}$ and the magnetic one $\vec{B}$ transform respectively as a (polar) vector and as an axial vector.
Then, an axial vector changes sign under $T$ but not under $P$, while a (polar) vector changes sign under $P$
but not under $T$. Furthermore, since electromagnetic and strong interactions are the two dominant contributions
to the dipole moments, and since both preserve $P$ and $T$, it follows that the corresponding contributions to
(4) must conserve these symmetries as well. Indeed, following (Muirhead 1965, Chapter 9, Section 9.2(d)), we
have
\begin{eqnarray}P\vec{\sigma}P^{-1}=\sigma, \ \ \ \ \ T\vec{\sigma}T^{-1}=-\vec{\sigma},\ \ \ \ \
P\vec{H}P^{-1}=\vec{H},\\ T\vec{H}T^{-1}=-\vec{H},\ \ \ \ \ P\vec{E}P^{-1}=-\vec{E},\ \ \ \ \
T\vec{E}T^{-1}=\vec{E},\nonumber\end{eqnarray}whence it follows that
\begin{eqnarray}P(\vec{\sigma}\cdot\vec{H})P^{-1}=\vec{\sigma}\cdot\vec{H},\ \ \ \ \ \ \ \ \ \
T(\vec{\sigma}\cdot\vec{H})T^{-1}=\vec{\sigma}\cdot\vec{H},\\
P(\vec{\sigma}\cdot\vec{E})P^{-1}=-\vec{\sigma}\cdot\vec{E},\ \ \ \ \ \ \ \ \ \
T(\vec{\sigma}\cdot\vec{E})T^{-1}=-\vec{\sigma}\cdot\vec{E}\nonumber.\end{eqnarray}Therefore, as L.D. Landau and
Ya.B. Zel'dovich pointed out (see (Landau 1957) and (Zel'dovich 1961)), due to these symmetry rules on $P$ and
$T$, the magnetic term $-\vec{\mu}_m\cdot\vec{B}$ is allowed, while an electric dipole term
$-\vec{d}_e\cdot\vec{E}$ is forbidden so that we should have $\eta=0$ in $(6)_2$. Now, $T$ invariance (that, by
$CPT$ theorem, is equivalent to $CP$ invariance) is also violated by weak interactions, which however are very
small for light leptons. Nevertheless, for non-negligible second order weak interactions (as for heavier leptons
- see (Chanowitz et al. 1978) and (Tsai 1981)), an approximate $T$ invariance will require the suppression of
electric dipole moments, i.e. $d_e\rightarrow 0$. Thus, electric dipole interaction cannot occur unless both $P$
and $T$ invariance breaks down in electrodynamics. Following (Roberts and Marciano 2010, Chapter 1, Section
1.3), P.A.M. Dirac discovered, in 1928, an electric dipole moment term in the relativistic equations involved in
his electron theory. Like the magnetic dipole moment, the electric dipole moment had to be aligned with spin, so
that we have an expression of the type $\vec{d}=\eta(Q\hbar/2mc)\vec{s}$ (see $(6)_2$) where, as already said,
$\eta$ is a dimensionless constant which is the analogous to $g_s$. Whilst the magnetic dipole moment is a
natural property of charged particles with spin, electric dipole moment are forbidden both by parity and time
reversal symmetries as said above. Nevertheless, from a historical viewpoint, the search for an EDM dates back
to suggestions due to E.M. Purcell and N.F. Ramsey since 1950 who however pointed out that the usual parity
arguments for the non-existence of electric dipole moments for nuclei and elementary particles, albeit appealing
from the standpoint of symmetry, weren't necessarily valid. They questioned about these arguments based on
parity and tried, in 1957, to experimentally measure the EDM of the neutron through a neutron-beam magnetic
resonance method, finding a value for $d$ of about $(-0.1\pm 2.4)\cdot 10^{-20}$ $e$-cm. This result was
published only after the discovery of parity violation although their arguments were provided in advance of the
celebrated 1956 T.D. Lee and C.N. Yang paper on parity violation for weak interactions. Once parity violation
received experimental evidence, other than L.D. Landau, soon after also N.F. Ramsey, in 1958, pointed out that
an EDM would violate both $P$ and $T$ symmetries.

\subsubsection*{3.3.3 Further experimental determinations of the lepton AMM}\addcontentsline{toc}
{subsubsection}{3.3.3 Further experimental determinations of the lepton AMM}

\it A) Some introductory theoretical topics\\\\ i) On resonance spectroscopy methods. \rm Amongst special
devices and techniques of experimental physics, a fundamental role is played by magnetic resonance spectroscopic
techniques through which Zeeman level transitions are induced by magnetic dipole radiations by means of the
application of an external static magnetic field $\vec{B}$. The spontaneous transitions with $\Delta l=\pm 1$
(electric dipole) are more probable than those with $\Delta l=0$ and $\Delta m=\pm 1$ (magnetic dipole).
Nevertheless, the presence of a resonant electromagnetic field increases the latter. With the action of this
perturbing field the probability of induced transitions is proportional to the square of the intensity of the
electromagnetic field, so that magnetic dipole transitions may be easily induced through suitable
radio-frequency (RF) values provided by a RF oscillator with an imposed constant magnetic field which has the
main role to select the desired RF frequencies to be put in resonance with the precession ones. As an extension
of the original method of the famous Stern-Gerlach experiment, the above mentioned technique was first proposed
by I.I. Rabi, together his research group at Chicago around the late 1930s, who made important experiments on
atomic beams that, amongst other things, led to the precise determination of the atomic hyperfine structure; in
particular, the Lamb shift between hydrogen $2S_{1/2}$ and $2P_{1/2}$ gave an accurate measurement of the
electron anomalous magnetic moment. Independently by Rabi's research group works, also L.W. Alvarez and F. Bloch
set up, in 1940, a similar technique. The nuclear magnetic moments have been measured through \it nuclear
magnetic resonance \rm (NMR) techniques that, thanks to relaxation mechanisms which release thermal energy in
such a manner to warrant a weak thermal contact between nuclear spins and liquid or solid systems to which they
belong, allow to determine fundamental physical properties of the latter. The \it electron paramagnetic
resonance \rm (EPR) or \it electron spin resonance \rm (ESR) refers to induced transitions between Zeeman levels
of almost free electrons in liquids and solids. It has been first observed by E.K. Zavoiskij in 1945 and usually
runs into the microwaves frequencies and it has been applied to determine anomalous magnetic moment values. Both
in NMR and EPR, in which an external inhomogeneous magnetic field $\vec{B}_0$ is acting, the transitions between
Zeeman levels are induced by an additional homogeneous alternating weak magnetic field $\vec{B}_1$ (for
instance, acting upon a $x$-$y$ plane), oscillating transversally to $\vec{B}_0$ (for instance, directed along
the $z$ axis) with an angular frequency $\omega_1$ which may be, or not, in phase with Larmor precession
frequency; for instance, if $\vec{B}_1$ acts along the $x$ axis, then an induced e.m.f. will be detectable along
the $y$ axis. Thanks to the 1949 N.F. Ramsey works, it is also possible to apply a second alternating static
magnetic field $\vec{B}_2$, even perpendicularly to $\vec{B}_0$ (\it double resonance \rm techniques), and so on
(\it multiple resonance \rm techniques); the possible reciprocal geometrical dispositions of the various
involved magnetic fields $\vec{B}_0,\vec{B}_1,\vec{B}_2$ and so on, give rise to different resonance
experimental methods also in dependence on the adopted relaxation methods and related detected times: amongst
them, the \it Bloch decay \rm and the \it spin echoes. \rm In single resonance techniques, the perturbing
alternating field $\vec{B}_1$ must be in resonance with the separation between two adjacent Zeeman levels (i.e.
with $\Delta m=\pm 1$). The resulting statistical coherence will imply a macroscopic value (roughly $N\mu_{ct}$)
quite high to may be detected by a coil, with the symmetry axis belonging in the equatorial plane and, for
instance, oriented along the $y$ axis, also thanks to electronic devices which will amplify the initial value.

Following (Dekker 1958, Chapter 20), (Kittel 1966, Chapter 16), (Kastler 1976, Part III, Chapter V),
(Cohen-Tannoudij et al. 1977, Volume I, Complement $F_{IV}$), (Bauer et al. 1978, Chapters 12 and 13), (Pedulli
et al. 1996, Chapters 7, 8 and 9), (Humphreis 1999, Chapter 14), (Bertolotti 2005, Chapter 9) and (Haken and
Wolf 2005, Chapter 12), for particles having a non-zero spin, the application of the field $\vec{B}_0$ only,
implies a torque acting upon the cyclotron (or orbital) magnetic moment $\vec{\mu}_L$ so giving rise to two
non-zero components, namely a longitudinal component $\vec{\mu}_{cl}$ (directed along $\vec{B}_0$) and a
transversal one $\vec{\mu}_{ct}$ (belonging to the plane having $\vec{B}_0$ as normal vector). This torque will
imply too a Larmor precession, with angular frequency given by $\omega_0=g(eB_0/2mc)$ (for elementary particles
with rest mass $m$), that causes a rotation of $\vec{\mu}_{ct}$ in the equatorial plane around the $z$ axis.
Nevertheless, in general there is no statistical coherence amongst these transversal components, also due to the
thermal excitation. But, as showed by F. Bloch, W.W. Hansen and M. Packard as well as by E.M. Purcell, H.C.
Torrey, N. Bloembergen and R.V. Pound in the years 1945-46, the application of a perturbing (alternating)
magnetic field $\vec{B}_1$, transversally arranged respect to $\vec{B}_0$ and usually induced by the passage,
along a transmissive spire, of a direct current (DC) into a variable RF oscillator, gives rise to a coherent and
ordered precession of the transversal components of magnetic moment when the frequency of the perturbing field,
say $\omega_1$, is equal to $\omega_0$ (\it magnetic resonance \rm condition or \it resonance equation\rm);
this, in turn, will imply either spin-orbit decouplings as well as resonating Zeeman magnetic level transitions,
in agreement with the well-known Bohr's correspondence principle according to which the concept of quantum level
transition should correspond, in the classical electrodynamics, to the periodic variation either of an atomic
electric or magnetic moment (in our case, the rotation of $\vec{\mu}_{ct}$ in the equatorial plane). The weak
perturbing magnetic field $\vec{B}_1$ is usually applied, above all in NMR techniques, in such a manner that its
values verify $B_1\ll B_0$ which nevertheless imply long storage times; often, as in the original (Chicago) I.I.
Rabi research group experiences, a second opposed (to $\vec{B}_0$) inhomogeneous magnetic field is also applied
next to the RF oscillator group, to refocalize the particle beam until the receiver device. In such a manner, a
very weak rotating magnetic field is able to reverse the spin direction of the beam particles, whilst
$\vec{\mu}_L$ precesses (\it Rabi's precession\rm), in the rotating frame, about a well-precise 'effective'
magnetic field $\vec{B}_{eff}$, given by the superposition of the various applied magnetic fields, according to
particular equations of motion called \it Bloch's equations. \rm In dependence on the RF oscillator chosen as an
energy source, we have either \it continuous wave \rm (CW) or \it pulsed wave \rm (PW) resonance techniques: the
intensity of the resulting signal is measured in function of the magnetic field or frequency values for the
former and in function of the time for the latter. As we shall see later, the resonance spectroscopy methods
have played a fundamental role in determining magnetic ed electric properties of atomic and nuclear systems
(see, for instance, (Bloch 1946)): for instance, through a suitable formulation of a resonance condition, it
will be possible to experimentally determine the anomalous magnetic moment of elementary constituents as
electrons, neutrons, protons and muons.\\\\ \it ii) On spin precession motion. \rm Following (Schwartz 1972,
Chapter 4, Section 10), (Rich and Wesley 1972, Section 3.1.1), (Cohen-Tannoudij et al. 1977, Volume I,
Complement $F_{IV}$), (Ohanian 1988, Chapter 11, Section 11.1), (Kinoshita 1990, Chapter 11, Sections 1-4),
(Picasso 1996, Section 2), (Farley and Semertzidis 2004, Section 3) and (Barone 2004, Chapter 6, Section 6.10),
a general \it precession problem \rm is identified by a kinematical equation of the form
$d\vec{\Phi}/dt=\vec{\Omega}(t)\wedge\vec{\Phi}$, where $\vec{\Phi}$ is the vectorial quantity that precesses
around the given vector $\vec{\Omega}$; for instance, $\vec{\Phi}$ may be a magnetic moment, an angular momentum
or the spin, which precesses around the direction given by the force lines of the perturbing field
$\vec{\Omega}$ (as, for example, a magnetic field), with angular velocity $\Omega(t)$. The related experienced
torque $\vec{\tau}$, is given by $\vec{\Omega}(t)\wedge\vec{\Phi}$. In case of an elementary spinning particle
having charge $Q$ and mass $m$, in a (uniform) magnetic field $\vec{B}$, we may put $\vec{\Phi}=\vec{\mu}_s$,
where $\vec{\mu}_s$ is the spin magnetic moment given by $g_sQ\mu_0\vec{\sigma}/2$ the $(6)_1$. In this case,
$\vec{\Omega}=k\vec{\mu}_s=(gQ/2mc)\vec{\mu}_s$, so that we have, in the particle rest frame, the following
Larmor precession equation $d\vec{\mu}_s/dt=k\vec{\mu}_s\wedge\vec{B}$ (see (Cohen-Tannoudij et al. 1977, Volume
I, Complement $F_{IV}$), (Bloch 1946, Equation (11)) and (Bargman et al. 1959, Equation (3))) related to the
precession of $\vec{\mu}_s(t)$ around $\vec{B}$; $\vec{\sigma}$ is said to be the \it polarization vector. \rm
The relativistic generalization of the last precession equation will lead to the so-called \it
Bargman-Michel-Telegdi equation \rm (see (Bargman et al. 1959)). Following (Gottfried 1966, Chapter VI, Section
49), for beams of elementary particles, said $\vec{\sigma}$ the Pauli operator whose components are the Pauli
matrices, the \it beam polarization \rm is defined to be $\langle\vec{\sigma}\rangle$ and shall often be written
as $\vec{P}$; it is zero for an incoherent and equal mixture of $|1/2\rangle$ and $|-1/2\rangle$, whereas
$|\vec{P}|=1$ for pure spin states.\\\\\it B) The first experimental determinations of the electron AMM\\\\\rm
Following (Kusch 1956), (Rich and Wesley 1972), (Crane 1976), (Farley and Picasso 1979), (Combley et al. 1981),
(Kinoshita 1990, Chapters 8 and 11) and (Jegerlehner 2008, Part I, Chapter 1) and as it has already said above,
P. Kusch and H.M. Foley, in November 1947, measured $a_e$ for the electron with a precision of about 5\%,
obtaining the value $a_e=0.00119(5)=0.00119\pm 0.00005$ at one standard deviation. The establishment of the
reality of the anomalous magnetic moment of the electron and the precision determination of its magnitude, was
part of an intensive programme of postwar research with atomic and molecular beams which seen actively involved
P. Kusch at Columbia, together to I.I. Rabi research group. All that was crowned by success with the assignment
of Nobel Prize for Physics in 1955, shared with W.E. Lamb, whose related Nobel lecture is reprinted in (Kusch
1956). Other attempts to estimate the anomalous magnetic moment either of the electron and of the proton were
carried out by J.H. Gardner and E.M Purcell in 1949 and 1951, by R. Karplus and N.M. Kroll in 1950, by S.H.
Koenig, A.G. Prodell with P. Kusch in 1952, by R. Beringer with M.A. Heald and by J.B. Wittke and R.H. Dicke in
1954, by P.A. Franken and S. Liebes Jr. in 1956 as well as by W.A. Hardy and E.M. Purcell in 1958, in any case
reaching to an accuracy of about 1\% for the various anomalous moment values. The Gardner and Purcell
experiments (see (Gardner and Purcell 1949) and (Gardner 1951)) introduced, for the first time, a new
experimental method to determine $a_e$, based on a comparison of the cyclotron frequency of free electrons with
the nuclear magnetic resonance (NMR) frequency of protons, so opening the way to the application of resonance
techniques to measure the lepton anomalous moments on the wake of the pioneering Rabi's molecular beam resonance
method for measuring nuclear magnetic moments (see (Rabi et al. 1938, 1939)) recalled above. To be precise, an
experimental determination of the ratio of the precession frequency of the proton, $\omega_p=\mu_pH_0$, to the
cyclotron frequency, $\omega_e=eH_0/mc$, of a free electron in the same magnetic field, was carried out. The
result, $\omega_p/\omega_e$, is the magnitude of the proton magnetic moment, $\mu_p$, in Bohr magnetons $\mu_0$.
Finally, by the comparison between $\mu_p/\mu_0$ and $\mu_e/\mu_p$, it was possible to determine $\mu_e/\mu_0$.
Possible sources of systematic error were carefully investigated and in view of the results of this
investigation and the high internal consistency of the data, it was felt that the true ratio, uncorrected for
diamagnetism, lie within the range $\omega_e/\omega_p=657.475\pm0.008$. If the diamagnetic correction to the
field at the proton for the hydrogen molecule was applied, the proton moment in Bohr magnetons became
$\mu_p=(1.52101\pm0.00002)×10^{-3}(e\hbar/2mc)$. In (Koenig et al. 1952), the ratio of the electron spin $g_e$
value and the proton $g_p$ value was measured with high precision. It was found that
$g_e/g_p=658.2288\pm0.0006$, where $g_p$ is the $g$ value of the proton measured in a spherical sample of
mineral oil. This result, when combined with the previous measurement by Gardner and Purcell of the ratio of the
electron orbital $g_e$ value and the proton $g_p$ value, yielded for the experimental value of the magnetic
moment of the electron $µ_s=(1.001146\pm0.000012)\mu_0$. The result was in excellent agreement with the
theoretical value calculated by Karplus and Kroll, namely $\mu_s=(1.0011454)\mu_0$. However, all these methods
were related to electrons bound in atoms, this implying, amongst other things, a lower accuracy level due to the
corrections necessary to account for atomic binding effects. Thus, anomalous moment experimental determinations
on free electrons were more suitable.

Following (Rich and Wesley 1972), (Kinoshita 1990, Chapter 8), in the years 1953-54, H.R. Crane, W.H. Louisell
and R.W. Pidd at Michigan, for the first time, determined $a_e$ for free electrons from measurements of $g-2$
(not $g$ itself) by means of the precession of the electron spin in a uniform magnetic field, obtaining the
result $g=2.00\pm 0.01$, that is to say, $g$ must be within 10\% of 2.00. They introduced, on the basis of the
previous basic work made by N.F. Mott in 1930s, a new pioneering technique which will be later called the \it
$(g-2)$ precession method, \rm so opening the way to the precession methods for determining lepton g factors.
Following (Louisell et al. 1954), (Hughes and Schultz 1967, Chapter 3), (Rich and Wesley 1972), (Combley and
Picasso 1974) and (Crane 1976), we briefly recall the main stages which led to the experimental methods for
measuring the magnetic moment of the free electron according to this $(g-2)$ precession method. A first attempt
was based, after a N.H. Bohr argument\footnote{Arguing upon the unobservability of the magnetic moment of a
single electron on the basis of the well-known Heisenberg indetermination principle. Therefore, we must consider
a statistical approach in such a manner that the average behavior of the spins of a large ensemble of particles
can be treated, to a large extent, as a classical collection of spinning bar magnets.}, on a statistical fashion
of the well-known 1924 Stern-Gerlach experiment on the atomic magnetic moments, applied to free electrons and
consisting in sending a large number of electrons through a magnetic field and by attempting to use the detailed
line shape to reveal the effects of the magnetic moment. Nevertheless, such a method appeared particularly
unpromising in connection to a precise solution to the electron moment problem. A second attempt, instead, was
based on the previous 1929 N.F. Mott double-scattering method for studying the polarization of particles beams.
The Louisell, Pidd and Crane principle of the method employed a Mott double-scattering method roughly consisting
in producing polarized electrons by shooting high-energy electrons upon a gold foil; hence, the part of the
electron bunch which is scattered at right angles, is then partially polarized and trapped in a constant
magnetic field where spin precession takes place for some time. The bunch is afterwards released from the trap
and allowed to strike a second gold foil, which allows to analyze the relative polarization. To be precise, this
method depend on the fact that a beam of electrons is partially polarized along a direction normal to the plane
defined by the incident beam and the emerging scattering direction. Furthermore, a second scattering process
exhibited an azimuthal asymmetry in scattering intensity, if measured in the same plane, mainly due to
polarization perpendicular to the plane of the incident and scattered beams. Mott defined the amplitude of this
asymmetry as $\delta$ and provided some its estimates. To explain this effect, both on the basis of the above
Bohr' argument and in taking into account the Stern-Gerlach results, Mott put forward the hypothesis that
electron spins had to be thought of as precessing around the direction of a magnetic field rather than as
aligned parallel or anti-parallel to this, like in the Stern-Gerlach experiment\footnote{Following (Miller et
al. 2007) and (Roberts and Marciano 2010, Chapter 1), the study of atomic and subatomic magnetic moments began
in 1921 first with a paper by O. Stern then with the famous 1924 O. Stern and W. Gerlach experiment in which a
beam of silver atoms was done pass through a gradient magnetic field to separate the different magnetic quantum
states. From this separation, the magnetic moment of the silver atom was determined to be one Bohr magneton
$\mu_0$ within 10\%. This experiment was carried out to test the Bohr-Sommerfeld quantum theory. In 1927, T.E.
Phipps and J.B. Taylor repeated the experiment with a hydrogen beam and they also observed two bands from whose
splitting they concluded that, like silver, the magnetic moment of the hydrogen atom was too one $\mu_0$.
Subsequently, in 1933, R.O. Frisch and O. Stern determined the anomalous magnetic moment of the proton, while in
1940, L.W. Alvarez and F. Bloch determined the anomalous magnetic moment of the neutron, and both turned out to
be quite different from the value 2, because of their internal structure.}. Therefore, the asymmetry observed
along the second scattering should be due to this precession because, if the spin were aligned parallel and
anti-parallel to the direction of a magnetic field parallel to the beam incident on the scatterer of the
experimental apparatus, then it would be enough to apply a weak magnetic field to remove such an asymmetry
effect. In this sense, the spin had to be meant as a physical observable rather than a mathematical device
(d'après Pauli). Furthermore, since this 1954 Louisell-Pidd-Crane method essentially requires a simultaneous
measurement of the electron position and of a single spin component, it follows that the uncertainty principle
is not violated. Crane says that Mott's way out of his dilemma was, perhaps, the first break toward thinking of
electrons as precessing magnets. Nevertheless, this far seeing Mott's hint didn't took by nobody at that time
until the 1953-54 pioneering works of Louisell, Pidd and Crane. They extended this Mott double-scattering method
inserting, between the first and second scatterers, a constant magnetic field, parallel to the path to the path
between the scatterers, in the form of a magnetic mirror trap which permitted the electrons to undergo several
hundred $(g-2)$ precessions between scatterings. This causes the electron to precess and rotates the
polarization plane of maximum asymmetry after the second scattering no longer coincides with the plane of the
first scattering. By measuring the angle of rotation and knowing the magnetic field, the electron energy and the
distance, the gyromagnetic ratio for the electron may be found. A fact which had a dominating influence was that
the orbital, or cyclotron, angular frequency of the electron in the magnetic field differs from the angular
frequency of precession of the spin direction although in higher-order correction terms, these respectively
being given by $\omega_o=eB/(2mc)$ and $\omega_s=g(eB/(2mc))$ with $g=2(1+ \alpha/2\pi+...)$ (d'après
Schwinger). This fact turns out to be useful to determine $g$ whose value may be therefore determined from a
direct comparison of the rotation of the plane of polarization and the cyclotron rotation. Moreover, all
observed asymmetries in the beam, whether they are associated with the spin or not, rotate around together, so
that it was needed for discriminating amongst them. Certain sources of asymmetry have nothing to do with the
polarization effect notwithstanding they follow the polarization asymmetry itself as it rotates around. However,
Louisell, Pidd and Crane were able to determine and isolate the non-spin asymmetry, mainly due to scattering
nonlinearities, from spin asymmetry that was experimentally detected with very small measurement errors. Due to
the action of the Lorentz force, if $\phi_c$ (or $\phi_o$) is the cyclotron (or orbital) rotation angle between
scatterers, $\phi_d$ is the sum of deflection angles at entry and exit to the solenoid field, and $\phi_s$ is
the angle through which the spin asymmetry was rotated relative to the direction of the beam before entry into
the solenoid field, then an estimate to $g$ is given by $2(\phi_s-\phi_d)/\phi_c$, whose experimentally detected
values were reported in Table I of (Louisell et al. 1954), computed at different values of $B$. Nevertheless,
Louisell, Pidd and Crane concluded that the precision of which their method is capable (they obtained an
accuracy of 1\%) was not enough to reveal the correction to the $g$ factor at about one part in a thousand, so
that their result wasn't sufficiently precise to be useful in comparison with the theoretical prediction.
Meanwhile, or in parallel, the results so found have been ascertained to be coherent with Dirac theory of
electron by H. Mendlowitz with K.M. Case, who also calculated the possible effects of a uniform magnetic field
on a Mott double-scattering experiment showing that they can be used to measure $a_e$ as in the
Louisell-Pidd-Crane experience. Coherence with Dirac theory also came from a previous 1951 work of H.A. Tolhoek
and S.R. De Groot which concerned another parallel research area on hyperfine structures oriented towards
precision measurements on $g$ of the free electron; the latter proposed, in 1951, a scheme in which a magnetic
field and a RF field were interposed between the first and second Mott scatterers, and in which destruction of
the asymmetry indicated resonance. A notable research group based at the University of Columbia and directed by
I.I. Rabi since 1940s, followed another line of attack to measure the gyromagnetic ratio for the free electron,
based upon the magnetic resonance method, proposing new experiments in two somewhat different forms respect to
the previous research line based on Mott scattering method. In both these forms, polarized electrons are trapped
in stable orbits into a magnetic field. A radio-frequency (RF) perturbing field is then applied and the
frequency which destroys the polarization is determined. From the frequency which destroys the polarization and
the strength of the magnetic field, the value of the gyromagnetic ratio is obtained. Since 1956, H.G. Dehmelt
group at Washington demonstrated that spin-exchange collisions between oriented sodium atoms and free, thermal
energy electrons could be used to measure $a_e$ via a direct RF resonance technique, so contributing to the
first determinations of the free electron anomalous magnetic moment.

The two above mentioned methods mainly differ in the way in which the electrons are polarized, giving priority
to trapping, and in the way in which the presence or absence of polarization is determined after the application
of the magnetic or RF perturbing field and the subsequent escaping from the trapping phase carried out by the
latter. The essence of the method consists essentially in finding the frequency of the feeble beat between the
rotation of the spin direction (in the trap or well) and the orbital, or cyclotron, rotation when the particles
are trapped in a well-determined magnetic well. Afterwards, a careful determination of electron energies as well
as a precise control of fields and potentials are also demanded. Forerunners of resonance methods, other than
the above mentioned one, may be also retraced in some previous experiences made by R.H. Dicke and F. Bloch in
the early 1940s. In any case, following (Louisell et al. 1954), in both methods in which resonance is involved,
the strong coupling to the cyclotron motion due to the fact that the required perturbing frequency is almost
identical to the cyclotron one with consequent transfer of energy from the perturbing field to the cyclotron
motion, might introduce serious difficulties in order to achieve the right accuracy with the increasing of the
cyclotron revolutions. Furthermore, it is very difficult to control the particle while it is into the trap
inside which it oscillates (along the $Z$ direction, parallel to the perturbing field). Nevertheless, Louisell,
Pidd and Crane state that the magnetic resonance methods, together their experimental extension to the Mott
double-scattering method, seem to be the only ones\footnote{Besides some other experimental attempts to get
polarized beams of electrons, by F.E. Myers and R.T. Cox as well as by E. Fues and H. Hellman, at the end of
1930s.} able to give really quantitative results of sufficient accuracy to reveal the correction to the electron
moment. Some problems occur when we consider electrons and positrons which both require to be previously
polarized: for the former, the above mentioned Mott scattering method is used, while for the latter, a suitable
radioactive source is used for their initial polarization whereas the final one is found through a clever scheme
first proposed by V.L. Telegdi (see .... (Grodzins 1959) and references therein). As regards muons, instead,
this last problem does not subsist since them born already polarized and reveal their final polarization through
the direction of the related decay products. Following (Crane 1976) and (Hughes and Schultz 1967, Chapter 3,
Section 3.5.3.1), in 1958, P.S. Farago proposed a method\footnote{Besides also quoted by (Bargmann et al. 1959,
Case (E))).} for comparing the orbital and the spin precession of electrons moving in a magnetic field, which
will turn out to be useful to directly measure radiative corrections to the free-electron magnetic moment.
Indeed, the Farago's principle of the method consisted in considering initially polarized electrons, emitted by
a $\beta$ active source and moving perpendicular to a strong uniform magnetic field $\vec{B}$, hence using a
Mott scattering for analysis. A uniform weak vector field $\vec{E}$ is also applied perpendicularly to $\vec{B}$
in such a manner that the beam walks enough to miss the back of the source of the first turn. The beam continues
walking towards right for a distance almost equal to the orbital diameter. After the order of about some
hundreds of revolutions, it then encounters a Mott scattering foil at which the final direction of polarization
perpendicular is determined from the intensity asymmetry in the direction perpendicular to the orbit plane. If
the final polarization direction is measured as a function of the transit time between source and target
(consisting of about 250 orbital revolutions or turns), then a sine curve is obtained whose frequency is equal
to the difference between the spin precession frequency and the orbital frequency of the circulating electrons.
To the extent that $E/B\ll 1$ (electron trochoidal motion), this difference frequency is proportional to
$(\mu_e/\mu_0-1)=g/2-1=a_e$, so that the Farago's method measures directly the radiative correction to the free
electron magnetic moment $\mu_e$, hence $a_e$ (see (Farago 1958)). The Farago's method was later improved and
experimentally realized by his research group at the University of Edinburgh (see (Farago et al. 1963)); it
constituted, at that time, the first method that allowed a continuous measurement rather than by pulses.
Nevertheless, the Farago's method couldn't compete in accuracy with experiments in which the particles are
trapped and allowed to make a far larger number of revolutions. In any case, its principle of the method, in
some respects, has preempted certain basic methods underpinning some later storage techniques (amongst which the
one based on polynomial magnetic fields). Other determinations of $a_e$ were later realized, in the early 1960s,
by D.T. Wilkinson, D.F. Nelson, A.A. Schupp, R.W. Pidd and H.R. Crane (\it Michigan group\rm) even improving
their principle of the method of 1954 and mainly based upon the remark that, if polarized electrons were caused
to move with their velocities perpendicular to a uniform magnetic field, then, at a fixed azimuth on the
cyclotron orbits, one would observe the polarization precessing at a rate equal to the difference between the
spin precession rate ($\omega_s$) and the orbital cyclotron rate ($\omega_c$), just this difference precession
rate (anomalous or spin-cyclotron-beat frequency $\omega_a=\omega_s-\omega_c$) being directly proportional to
$a_e$. This method will be generically called the (\it Michigan\rm) \it principle of $(g-2)$ spin motion, \rm or
simply \it spin precession method \rm (or also \it free-precession method\rm), and will lead to the next basic
equation (59).

Following (Rich and Wesley 1972) and (Crane 1976), meanwhile the spin precession methods were further pursued as
a result of the pioneering works made by the above Michigan group, other techniques were employed to approach
$g-2$, above all for electrons. As it has already said above, H.A. Tolhoek and S.R. De Groot proposed, since
1951, a scheme in which a magnetic field, coupled with a RF field, would be interposed between the first and the
second Mott scatterers, even if themselves were aware that such an apparatus wasn't able to provide enough
cycles of the spin precession to give a well defined frequency, mainly because of the absence of a trap. In
1953, F. Bloch proposed a novel resonance-type experiment to measure $a_e$ using electrons occupying the lowest
Landau level in a magnetic field. In the years 1956-58, H.G. Dehmelt performed an experiment in which free
thermal electrons in argon buffer gas, at the mean temperature of 400$^{o}K$, become polarized in detectable
numbers by undergoing exchange collisions with oriented sodium atoms during which the atom orientation is
transferred to the electrons. Such collisions establish interrelated equilibrium values among the atom and the
electron polarizations which depend on the balance between the polarizing agency acting upon the atoms (optical
pumping) and the disorienting relaxation effects acting both on atoms and electrons. When the electrons were
furthermore artificially disoriented by gyromagnetic spin resonance, an additional reduction of the atom
polarization ensued, which was detected by an optical monitoring technique (with an optical pumping cell rather
than a quadrupole trap), so allowing to the determination of the free-electron spin $g$ factor and opening the
way to experimentally use the so-called \it Penning trap \rm consisting of a uniform axial magnetic field
$\vec{B}=B_{0z}\hat{z}$ and a superimposed electric quadrupole field generated by a pair of hyperbolic
electrodes surrounding the storage region. The magnetic field confines the electrons radially, while the
electric field confines them axially. The essential novel feature of the this Dehmelt's techniques consisted,
following an idea of V.L. Telegdi and co-workers (see (Ford et al. 1972)), in the fact that a RF induced pulse
(or beat) frequency, rather than a spin precession frequency, was the main responsible to rotate the
polarization. The principle of the method is quite similar to the known \it spin echoes \rm of E.L. Hahn (1950)
in which an intense RF power in the form of pulses is applied to an ensemble of spins in a large static magnetic
field. The frequency of the pulsed RF power is applied through a RF current circulating in a wire stretched
along the center axis of the trapping chamber, producing lines of force that are circles concentric with the
orbits. If the RF is held on for the right length of time, then the polarization is turned from the plane
perpendicular to the applied magnetic field towards the direction parallel to it. Afterwards, it comes back
again if the RF pulse is held on twice as long, just like spin echoes.

Following (Gr\"{a}ff 1971), (Rich and Wesley 1972) and (Holzscheiter 1995), the precision measurements of lepton
$g$-factor anomalies can be classified as being either \it precession experiments \rm and \it resonance
experiments \rm in dependence on the technique employed, in both of which the main involved problem being that
concerning the trapping of polarized charged particles. The main dynamical features of the problem are as
follows: the momentum $\vec{p}$ of the particle, which is exactly perpendicular to $\vec{B}$, revolves with the
cyclotron (or orbital) angular frequency $\omega_c=QB/mc$, the spin precesses about $\vec{B}$ with Larmor
angular frequency $\omega_s=(1+a_l)\omega_c$ with $a_l=(g-2)/2$, while the difference between these angular
frequencies is the one at which the spin rotates about the momentum, that is to say
$\omega_{a_l}=\omega_s-\omega_c=a_lQB/mc=\theta/T$ where $\theta$ is the angle between spin and momentum and $T$
the time. Consequently, to get the lepton anomaly $a_l$, it is thus necessary to measure the quantities
$\omega_{a_l}$ and $B$, assuming $Q/mc$ to be known. Thus, we have $a_l=\omega_{a_l}/\omega_c$ (see also
(Kinoshita 1990, Chapter 11, Section 4.1, Equation (4.8)). If the particle velocity has a small angle relative
to the orbital plane $x$-$y$ of motion particle, then the particle will follow a spiral path, along the axial
direction given by the $z$-axis, with \it pitch angle \rm $\psi$, spiralling in the main (not necessarily
constant) magnetic field $B_z$; the $(g-2)$ frequency is consequently altered. In any real storage system, the
pitch angle is corrected by suitable vertical focusing forces which prevent the particles to be lost.
Furthermore, the pitch angle changes periodically between positive and negative values, so that the correction
to the $(g-2)$ frequency become more complex. All the $(g-2)$ experiments for electrons and muons are in
principle subject to a pitch correction and, as we will see later, this problem will be successfully overcome,
for the first time, with the introduction of the so-called \it polynomial magnetic fields. \rm An arbitrary
experiment which attempts to measure the anomalous magnetic moment of a free lepton necessarily encounters the
following problems: \it a) \rm trapping of the particle; \it b) \rm measurement of the trapping field either by
nuclear magnetic resonance (NMR) or by measuring $\omega_s$ or $\omega_c$; \it c) \rm polarization of the spin
of the particle; \it d) \rm determination of the anomaly frequency \rm either \it i) \rm by detection of the
spin polarization vector relative to the momentum vector of the particle as a function of the time in a magnetic
field, calling this type of experiment a \it geometrical experiment\footnote{Roughly corresponding to the above
precession experiment type.}\rm, or, alternatively, \it ii) \rm by induction and detection of the relevant RF
transition $\omega_s$ and $\omega_c$ or, if possible, $\omega_s$ or $\omega_c$ and the difference angular
frequency $\omega_a$ directly, calling this type of experiment a \it RF spectroscopic
experiment\footnote{Roughly corresponding to the above resonance experiment type.}. \rm To trap particles, it
has been used: 1) the \it magnetic bottle method \rm consisting in imposing a homogeneous magnetic field with a
superimposed relatively weak inhomogeneous magnetic field as first used by the above mentioned Michigan group;
2) a RF quadrupole trap starting from the first studies on electric quadrupole mass separator made by F. v.
Bush, W. Paul, H.P. Reinhard with U. v. Zahn and by E. Fisher, in the 1950s, for separating isotopes. To detect
the ions, a resonance detection technique is used, taking advantage of the fact that for given parameters of the
trap each charge-to-mass ratio exhibits a certain unique ''eigenfrequency''. In addition to the radio-frequency
quadruple field, a RF dipole field at the frequency $\omega_{res}$ is applied as well to the end caps. If
through proper choice of the parameters $a$ and $q$, respectively representing the amplitudes of the RF
component and the direct current (DC) component of the quadruple field, the ions are brought to resonance with
this dipole field, then the amplitude of the ion motion is increased, absorbing energy from the drive field, and
can be detected. The important fact is that different ions will have different frequencies for a given set of
$a$ and $q$, or, that at a fixed frequency, one can bring all different ion species to resonance subsequently by
slowly varying the DC potential at a constant RF amplitude. This made the quadruple trap an ideal tool for
precision mass spectrometry or residual gas analysis, areas in which RF traps have gained high respect over the
last decades. At first glance, the RF drive field seems to be a disturbance to the system, and in effect it is.
Due to the continuously applied drive force stored particles are heated permanently, leading to 2nd order
doppler broadening of spectral lines. This effect can be counteracted by cooling mechanisms, either collisions
with residual gas molecules, or far more powerful and selective than this, by laser cooling. Nevertheless, due
to this ''micromotion'', the Paul's research group trap has always been a second choice respect to the so-called
\it Penning trap \rm if one desired an ultrahigh precision work. Based on this last new device, dating back to
the late 1930s F.M. Penning works, D.H. Dehmelt group at Seattle (Washington), P.S. Farago group at Edinburgh
and G. Gr\"{a}ff group at Bonn/Mainz have performed various electron $g-2$ experiments. As concerns, instead,
the polarization problem, in experiments of geometrical type, polarized muons are produced by the forward decay
of pions, polarized electrons by Mott double-scattering and polarized positrons by beta decays, while, as
regards experiments of RF spectroscopic type, electrons are polarized by means of spin exchange with a polarized
atomic beam as well as electrons of low energy are created in pulses in a high magnetic field. Finally, as
regards the determination of the lepton anomaly, in the geometrical experiments the angle $\theta$ between the
spin vector and momentum of the particle is measured at a fixed orbital point as a function of time. The
polarization of electrons is detected by Mott double-scattering, the polarization of positrons by exploiting the
spin dependence by ortho- and para-positronium formation, whilst the muon polarization is measured using the
fact that in the rest frame, the decay electrons are preferentially emitted along the spin direction. As the
momentum of a particle in a magnetic bottle is no longer perpendicular to the magnetic field, the
Bargmann-Michel-Telegdi (BMT) formula for $\omega_a$ (see (Bargmann et al. 1959, Equation (9))) has to be used.
Instead, in the RF spectroscopic measurements, the transition at frequency $\omega_a$ has to be induced and
observed. Nevertheless, this level transition corresponds to a combination of a magnetic and electric dipole
transition with $\Delta n=\pm 1$ and $\Delta m_s=\pm 1$ at the same time\footnote{For instance, a quantum state
transition from $|n,m_s=-1/2\rangle$ to $|n-1,m_s=+1/2\rangle$ is forbidden being a second order (two-photon)
transition because it involves a simultaneous change of the spin quantum number ($m_s$) and of the orbital (or
cyclotron) quantum number ($n$). But, with a proper choice of the electromagnetic configuration by means of the
application of a suitable perturbing field, this transition can be driven.}; such a transition if forbidden to
first order, but it can be enforced by an inhomogeneous magnetic RF field which, in turn, necessarily must be
accompanied by a homogeneous magnetic RF field. This last field, nevertheless, may produce line shifts and line
asymmetries. Furthermore, the transition at frequency $\omega_a$ involves a jump from one cyclotron orbit to
another with a spin flip at the same time; likewise for the induction of the Larmor frequency. The main
limitations of RF spectroscopic experiments lie just in this transition prohibition and in the presence of
unwanted homogeneous magnetic RF fields; another limitation is also provided by the limited energy of the
trapped particles. In conclusion, the principle of the method of almost all $g-2$ experiments roughly consists
in measuring the interaction between the magnetic moment of the particle and a homogeneous magnetic field
superimposed by an inhomogeneous magnetic or electric trapping field. The latter, nevertheless reduces the
accuracy of the experiments which may be improved decreasing the relative inhomogeneity even if, for technical
reasons, this is not possible in the $g-2$ experiments of the muons through further substantial increase of the
homogeneous magnetic field.

Therefore, to sum up (following (Rich and Wesley 1972)), the precession experiments include measurements of the
electron, positron and muon anomalies, the distinguishing feature of these experiments (as those made at
Michigan for electrons and at CERN for muons) being a direct observation of the spin precession motion of
polarized leptons in region of static magnetic field. The resonance technique instead has mainly been used to
measure lepton anomaly (prior to electrons), its characteristic feature being the presence of an oscillating
electromagnetic field used to induce transitions between the energy eigenstates of a lepton interacting with a
static magnetic field by applying a microwave field at the spin precession frequency $\omega_c$ and subsequently
a RF field at the spin-cyclotron difference frequency $\omega_a$.\\\\ \it c) Towards the first experimental
determinations of muon AMM\\\\\rm In the same period in which the above mentioned electron AMM determinations
were achieved, many further experimental evidences were also accumulated in confirming that the muon behaved as
a heavy electron of spin $1/2$, so that the former were taken as models to set up possible experiences for the
latter. But, before to outline these, what were the theoretical motivations underlying the researches towards
muon? In 1956, V.B. Berestetskii, O.N. Krokhin and A.X. Klebnikov, in providing, through processes involving
photons and leptons, a sensitive test of the limit for the (R.P. Feynman) UV cut-off (or QED-breaking)
$\Lambda_l$, which represents a measure for the distance at which QED breaks down, pointed out that the
measurement of the muon anomalous magnetic moment could accomplish this in a more sensitive manner than that of
the electron. Indeed, if one supposes that the muon is not completely point-like in its behavior, but has a form
factor\footnote{The dependence on $q^2$ of the form factors, experimentally enables us to get information about
charge radial distributions and magnetic moments of charged leptons (see (Povh et al. 1995, Part I, Chapter 6,
Section 6.1)). For instance, for a generic Dirac particle, we have $F(q^2)=1$.}
$F_{\mu}(q^2)=\Lambda_{\mu}^2/(q^2+\Lambda_{\mu}^2)$, then it can be show that an expression for the sensitivity
of $a_{\mu}$ is given by\begin{equation}\frac{\delta
a_{\mu}}{a_{\mu}}=-\frac{4m_{\mu}^2}{3\Lambda_{\mu}^2}\end{equation}which may be generalized for leptons as
follows\begin{equation} \frac{\delta a_l}{a_l}\sim\frac{m_l^2}{\Lambda_l^2},\ \ \ \ \ \
l=e,\mu,\tau.\end{equation}Berestetskii, Krokhin and Klebnikov emphasized that the high muon mass could imply a
significant correction to $a_{\mu}$ even when $\Lambda_{\mu}$ is large. Therefore, due to its high mass, the
muon allows to explore very small distances (of the order of $10^{-15}$ cm) because of the simple fact that
$q^2\sim m$ and the higher it is the momentum $q^2$, the higher it is the energy involved and, therefore, the
shorter it is the involved distance scale due to uncertainty principle. Furthermore, mainly because of the
vastly different behavior of the three charged leptons mainly due to the very different masses $m_l$ implying
completely different lifetimes $\tau_e\simeq\infty$ and $\tau_l=1/\Gamma_l\varpropto 1/(G_F^2m_l^5)\
l=\mu,\tau$, as well as vastly different decay patterns, it was clear that the anomalous magnetic moment of the
muon would be a much better probe for possible deviations from QED. In 1957, J. Schwinger thought that the muon
could have an extra interaction which distinguished it from the electron and gave it its higher mass. This could
be a coupling with a new massive field or some specially mediated coupling to the nucleon. Whatever the source
be, the new field would have had its own quantum fluctuations, and therefore gives rise to an extra contribution
to the anomalous moment of the muon. Thus, the principle of $(g-2)$ spin motion was also recognized as a very
sensitive test of the existence of such fields and potentially a crucial signpost to the so-called $\mu-e$
puzzle (see later). But, at that time, there wasn't any possibility to descry some useful principle of the
method for pursuing this\footnote{For instance, the parity violation of weak interactions was not yet known at
that time.}, so that nobody had an idea how to measure $a_{\mu}$. Albeit the $(g-2)$ spin motion principle will
turn out to be, a priori, very similar to those later developed to measure $a_{\mu}$, nevertheless it was
immediately realized that handling the muons in a similar way was impossible, and this raised the difficult task
of how to may polarize such short lived particles like muons, in comparison with the long lifetimes of electrons
which allowed to measure $a_e$ directly by atomic spectroscopy in magnetic fields. As we shall see later, this
was pursued, for the first time, by the pioneering works of the first CERN research groups on $g-2$ since the
late 1950s, above all thanks to new magnetic storage techniques set up just to this end. Nevertheless, behind
this last pioneering research work, there was a great and considerable previous work of which a brief outline we
are however historically obliged to remember.

The principle of the method of the Michigan group experiments has been applied to determine the muon $g$-factor
in some experiments performed, since the middle 1950s, by a notable research group of the Columbia University
headed by L.M. Lederman in the wake of the previous work of his maestro I.I. Rabi (see (Lederman 1992)). In
1958, T. Coffin, R.L. Garwin, S. Penman, L.M. Lederman and A.M. Sachs (see (Coffin et al. 1958)) made a RF
spectroscopic experiment with stopped muons in which the magnetic moment of the positive $\mu$ meson was
measured in several target materials by means of a solid-state nuclear magnetic resonance technique with
perturbing RF pulses. Muons were brought to rest with their spins parallel to a magnetic field. A
radio-frequency (RF) pulse was applied to produce a spin reorientation which was detected by counting the decay
electrons emerging after the pulse in a fixed direction. The experimental results were expressed in terms of a
$g$-factor which for a spin 1/2 particle is the ratio of the actual moment to $e\hbar/2m\mu c$. The most
accurate result obtained in a $CHBr_3$ target, was $g=2(1.0026\pm0.0009)$ compared to the theoretical prediction
of $g=2(1.0012)$, while less accurate measurements yielded $g=2.005\pm0.005$ in a copper target and
$g=2.00\pm0.01$ in a lead target. After the well-known above mentioned 1956 proposal of parity violation in weak
transitions by T.D. Lee and C.N. Yang, it was immediately realized that muons produced in weak decays of the
pion $\pi^+\rightarrow\mu^++\nu_{\mu}$ (see Section 1) could be longitudinally polarized, while the decay
positron of the muon $\mu^+\rightarrow e^++2\nu_{\mu}$\footnote{Only after 1960, it was ascertained that
$\nu_{\mu}\neq\bar{\nu}_{\mu}$, whereupon we might more correctly write $\mu^+\rightarrow
e^++\nu_{\mu}+\bar{\nu}_{\mu}$ (see Section 1).} could indicate the muon spin direction. This was confirmed by
R.L. Garwin, L.M. Lederman and M. Weinrich (see (Garwin et al. 1957)), as well as by J.I. Friedman and V.L.
Telegdi (see (Friedman and Telegdi 1957)), in the same year of\footnote{For technical reasons, the paper of
Friedman and Telegdi was delayed to the Physical Review Letters issue next to the one in which was published the
paper of Garwin, Lederman and Weinrich, notwithstanding both papers were received almost contemporaneously, the
former on January 17, 1957 and the latter on January 15, 1957. Nevertheless, following (Cahn and Goldhaber 2009,
Chapter 6), the Friedman and Telegdi emulsion experiment at Chicago was started before others but has employed
more time to be completed because of the laborious scanning procedure.} 1957. The first researchers, who
achieved an accuracy of 5\%, started from certain suggestions, made in the remarkable works of T.D. Lee, R.
Oehme and C.N. Yang, according to which their hypotheses on violation of $C$, $P$ and $T$ symmetries had to be
sought in the study of the successive reactions $1)\ \pi^+\rightarrow\mu^++\nu_{\mu}$ and $2)\ \mu^+\rightarrow
e^++\nu_{\mu}+\bar{\nu}_{\mu}$. To be precise, they pointed out that the parity violation would have implied a
polarization of the spin of the muon emitted from stopped pions in the first decay reaction along the direction
of the motion; furthermore, the angular distribution of electrons in the second decay reaction could serve as an
analyzer for the muon polarization. Moreover, in a private communication, Lee and Yang also suggested to Garwin,
Lederman and Weinrich that the longitudinal polarization of the muons could offer a natural way of determining
their magnetic moment, partial confirmations of the validity of this idea having already been provided by the
preliminary results of the celebrated C.S. Wu and co-workers experiments on $Co^{60}$ nuclei. By stopping, in a
carbon target puts inside a magnetic shield, the polarized $\mu^+$ beam formed by forward decay in flight of
$\pi^+$ mesons inside the cyclotron, Garwin and co-workers established the following facts: \it i) \rm a large
asymmetry was found for electrons in 2), establishing that the $\mu^+$ beam was strongly polarized; \it ii) \rm
the angular distribution of the electrons was given by $1+a\cos\theta$ where $\theta$ was measured from the
velocity vector of the incident muons, founding $\theta=100^o$ $a=-1/3$ with an estimated error of 10\%; \it
iii) \rm in both reactions, parity was violated; \it iv) \rm by a theorem of Lee, Oheme and Yang (see (Lee et
al. 1957)), the observed asymmetry proves that invariance under charge conjugation is not conserved; \it v) \rm
the $g$ value for free $\mu^+$ particles was found to be $+2.00\pm 0.10$; and \it vi) \rm the measured $g$ value
and the angular distribution in 2), led to the very strong probability that the $\mu^+$ spin was 1/2. The
magnetizing current, induced by applying a uniform small vertical field in the magnetic shielded enclosure about
the target, produced as a main effect the precession of muon spins, so that a road based on muon spin precession
principle to seriously think about the experimental investigation of $a_{\mu}$, was finally descried. Amongst
other things, the work of Garwin, Lederman and Weinrich opened the way to the so-called \it muon spin resonance
\rm ($\mu$SR), a widespread tool in solid state physics and chemical physics. In 1957, their result was improved
to an accuracy of about 4\% by J.M. Cassels, T.W. O'Keele, M. Rigby, A.M. Wetherall and J.R. Wormald.

Likewise, following the celebrated suggestion of Lee and Yang on non-conservation of parity in weak
interactions, Friedman and Telegdi (1957) investigated the correlation between the initial direction of motion
of the muon and the direction of emission of the positron in the main decay chain
$\pi^+\rightarrow\mu^+\rightarrow e^+$ produced in nuclear emulsions just to detect a possible parity
non-conservation in the latter decay interactions. Following Lee and Yang arguments, violation of parity
conservation may be inferred essentially by the measurement of the probability distribution of some pseudoscalar
quantity, like the projection of a polar vector along an axial vector. For instance, Lee and Yang themselves
suggested several experiments in which a spin direction is available as a suitable axial vector; in particular,
they pointed out that the initial direction of motion of the muon in the decay process
$\pi^+\rightarrow\mu^++\nu_{\mu}$ can serve for this purpose, as the muon will be produced with its spin axis
along its initial line of motion if the Hamiltonian responsible for this process does not have the customary
invariance properties. If parity is further not conserved in the decay process $\mu^+\rightarrow
e^++2\nu_{\mu}$, then a forward-backward asymmetry in the distribution of angles, say $W(\theta)$, between this
initial direction of motion and the moment of the decay electron, is predicted. To this end, positive pions from
the University of Chicago synchrocyclotron were brought to rest in emulsion carefully shielded from magnetic
fields, as well as over 1300 complete decay events were measured. A correlation $W(\theta)=1+a\cos\theta$ was
found, with $a=-0.174±0.038$, clearly indicating a backward-forward asymmetry, that is to say a violation of
parity conservation in both decay processes. Following an argument of T.D. Lee, R. Oehme\footnote{Reinhard Oehme
(1928-2010) was an influential theoretical physicist who gave notable contributions mainly in mathematical and
theoretical physics. Amongst these, Oehme was the first to realize that every time the $CPT$ symmetry must be
obeyed, then if $P$ was violated, $C$ and/or $T$ had to be violated as well. He proved that if the various
experiments suggested by Lee and Yang showed a $P$ violation, then $C$ had to be violated too. In this regards,
Oehme sent a letter to Yang and Lee explaining this insight, and they immediately suggested that all three
together would have written a paper (Lee et al. 1957)). See above all (Yang 2005) where this historical event,
often misunderstood, has definitively been clarified.} and C.N. Yang, this asymmetry would have implied a
non-invariance of either decay reactions with respect to both space inversion $P$ and charge conjugation $C$,
taken separately. Furthermore, Friedman and Telegdi given a detailed discussion of a depolarization process
specific to $\mu^+$ mesons, i.e. the possible formation of muonium $(\mu^+e^-)$. The results of this and similar
experiments were also compared with those obtained with muons originating from $p^+$ decays in flight and the
implications of such a comparison were discussed too. Therefore, the Friedman and Telegdi work, for the first
time, pointed out, also thanks to a private communication with R. Oehme, that $P$ and $C$ were violated
simultaneously, or rather, to be precise, $P$ was normally violated while $CP$ was to very good approximation
conserved, in the decay processes analyzed by them.

Following (Farley and Picasso 1979) and (Jegerlehner 2008, Part I, Chapter 1), it should be mentioned that until
the end of 1950s, the nature of the muon was quite a mystery. In that period, the possible deviations from the
Dirac moment $g=2$ were ascribed to the interaction of leptonic particle with its own electromagnetic field. Any
other field coupled to the particle would produce a similar effect and, in this regards, the calculations have
been made for scalar, pseudoscalar, vector and axial-vector fields, using an assumed small coupling constant $f$
to a certain boson of mass $M$. For example, for the case of a vector field, the above mentioned work of
Berestetskii, Krokhin and Klebnikov as well as the 1958 work of W.S. Cowland, provided the estimate $\delta
a_{\mu}^{Vec}=(1/3\pi)(f^2/M^2)m^2_{\mu}$ so that a precise measurement of $a_{\mu}$ could therefore reveal the
presence of a new field, but, before this, it had to be discovered all the known fields, comprising the weak and
strong interactions, and hereupon taken into account. Following (Picasso 1996) and references quoted therein,
the theoretical value for $a_{\mu}$ can be expressed as follows
$a_{\mu}^{(th)}=a_{\mu}^{QED(th)}+a_{\mu}^{QCD(th)}+a_{\mu}^{Weak(th)}$. In the 1950s, the only contribution
which could be measured with a certain precision was the QED one, while both the strong and weak interaction
contributions will be determined only later\footnote{The first ones who pointed out on the importance of
hadronic vacuum-polarization contributions to $a_{\mu}$ were C. Bouchiat and L. Michel in 1961 as well as L.
Durand in 1962 (see (Roberts and Marciano 2010, Chapter 3, Section 3.2.2.2)).}. In any case, the QED
contribution turns out to be the dominant one for $a_e$ while as of today, good estimates have been achieved for
weak interaction contributions to $a_{\mu}$ but not for the hadronic ones. While today it is well-known that
there exist three lepton-quark families with identical basic properties except for differences in their masses,
decay times and patterns, at that time it was very hard to believe that the muon is just a heavier version of
the electron, so giving rise to the so-called \it $\mu-e$ puzzle, \rm paraphrasing the previous well-known \it
$\theta-\tau$ puzzle \rm which was brilliantly solved by the celebrated work of T.D. Lee and C.N. Yang on the
parity violation for weak interactions. For instance, it was expected that the muon exhibited some unknown kind
of interaction, not shared by electron and that would have due to explain the much higher mass. All this
motivated and stimulated the experimental research to explore $a_{\mu}$. As it has already been said above, the
big interest in the muon anomalous magnetic moment was motivated by the above mentioned Berestetskii, Krokhin
and Klebnikov argument in relation to the main fact according to which the anomalous magnetic moment of leptons
mediates spin-flip transitions whose amplitudes are proportional to the masses of particles, so that they are
particularly appreciable for heavier ones via a generalization of (55) given by\begin{equation}\frac{\delta
a_l}{a_l}\varpropto\frac{m_l^2}{M_l^2}\ \ \ \ \ (M_l\gg m_l)\end{equation}where $M_l$ is a parameter which may
be either an energy scale or an ultraviolet cut-off where QED ceases to be valid (QED-breaking) or as well the
mass of a hypothetical heavy state or of a new heavier particle. The relation (57) also allows us to ascertain
whether an elementary particle has an internal structure: indeed, if the lepton $l$ is made by hypothetical
components of mass $M_l$, then the anomaly $a_l$ would be modified by a quantity $\delta a_l$ given by the
relation $\delta a_l=O(m_l^2/M_l^2)$ so that the measurements of $a_l$ might provide a lower limit for $M_l$
which, at the current state of research, has a magnitude of about 1 TeV, which imply strong limitations to the
possible hypotheses on the internal structure of a lepton (see (Picasso 1985)). On the other hand, the relation
(57) also implies that the heavier the new state or scale, the harder it is to see. Therefore, from (57), it
follows that the sensitivity to high-energy physics grows quadratically with the mass of the lepton, which means
that the interesting effects are magnified in $a_{\mu}$ compared to $a_e$ by a factor of about
$(m_{\mu}/m_e)^2\sim 4\cdot 10^4$, and this is just what has made and still makes $a_{\mu}$ the elected
monitoring fundamental parameter for the new physics also because of the fact that the measurements of
$a_{\tau}$ go out of the present experimental possibilities due to the very short lifetime of $\tau$.

As also reported in (Garwin et al. 1957), if $g=2$ then the direction of muon polarization would remain fixed
relatively to the direction of motion throughout the trajectory, while if $g\neq 2$ then a phase angle $\delta$
opens up between these two directions. Following (Muirhead 1965, Chapter 2, Section 2.5(a,e)), (Farley and
Picasso 1979) and (Picasso 1996), to estimate $\delta$, let us assume that we have longitudinally polarized
charged leptons slowly moving in a magnetic field and we know their direction of polarization. If they are
allowed to pass into a system with a magnetic field of strength $B$, they experience a torque given by
$\vec{\tau}=\vec{\mu}_s\wedge\vec{B}$ which, in turn, implies the execution of helical orbits about the
direction of $\vec{B}$ which lead to a \it Larmor precession \rm about the direction of $\vec{B}$ with the
following angular velocity (in natural units) calculated in the particle rest frame
\begin{equation}\omega_s=g\frac{Q}{2mc}B=\Gamma B\end{equation}where $\Gamma=g(Q/2mc)$ is the \it
gyromagnetic ratio. \rm If the charged particle is also in motion, then it will execute spiral orbits about
$\vec{B}$ which possess the characteristic \it cyclotron frequency \rm $\nu_c$ given by $\omega_c=2\pi
\nu_c=(Q/mc)B$. In one defines the laboratory rotation frequency of the spin relative to the momentum vector as
$\omega_{a_i}\doteq\omega_s-\omega_c$, then the phase angle $\delta$, after a time $t$, is given by
\begin{equation}\delta=\omega_{a_i}t=(\omega_s-\omega_c)t=\frac{g-2}{2}\frac{Q}{mc}Bt=a_i\frac{Q}{mc}Bt\end{equation}
where $g=2(1+a_i)\ i=e,\mu,\tau$. Hence, if $g=2$, then $\omega_s=\omega_c$ and the charged leptons will always
remain longitudinally polarized. But if $g>2$ as predicted, then the spin starts to precess and turns faster
than the momentum vector. Therefore, it is immediately realized that a measurement of the phase angle $\delta$
after a time $t$, may estimate the magnitude of the deviation of the $g$-value from 2. Equation (59) will be the
basic formal tool for the so-called $(g-2)$ experiments and that will be carried out later: if the charged
lepton is kept turning in a known magnetic field $\vec{B}$ and the angle between the spin and the direction of
motion is measured as a function of time $t$, then $a_i$ may be estimated. The value of $Q/mc$ is obtained from
the precession frequency of the charged leptons at rest, via equation (58). Furthermore, the fundamental
equation (59) has been derived only in the limit of low velocities but it has been proved to be exactly true as
well at any speed as, for example, made in (Bargmann et al. 1959) using a covariant classical formulation of
spin-motion. It has also been proved that the $(g-2)$ precession is not slowed down by time dilation even for
high-velocity muons.

Following (Farley and Picasso 1979) and (Brown and Hoddeson 1983, Part III, Chapter 8), after the celebrated
experience made by Garwin, Lederman and Weinrich in 1957, the possibility of a $(g-2)$ experiment for muon was
finally envisaged. In 1959, as recalled by (Jegerlehner 2008, Part I, Chapter 1), the Columbia research group
made by L.M. Lederman, R.L. Garwin, D.P. Hutchinson, S. Penman and G. Shapiro, performed a measurement of
$a_{\mu}$ with a precision of about 5\%, even using a precession technique applied to a polarized muon beam
whose directions are determined by means of their asymmetric decay modes. In the same years, many other research
groups at Berkeley, Chicago, Liverpool and Dubna started as well to study the problem. If the muon had a
structure giving a form factor less than one for photon interactions, then the value of $a_{\mu}$ should be less
than predicted. Nevertheless, compared with the measurement on the electron, the muon $(g-2)$ experiment was
much more difficult because of the low intensity, diffusive nature and high momentum of available muon sources.
All this, together the possibility to get a reasonable number of precession cycles, entailed, amongst other
things, the need to have large volumes of magnetic field. One solution, adopted by A.A. Schupp, R.W. Pidd and
H.R. Crane in 1961, was to scale up the original Michigan $(g-2)$ method for electrons whose spin directions was
established with the aid of a double scattering experiment in which the first and second scatterings were
performed respectively before and after the passage of the electrons through a solenoid. However, out of the
many attempts to approach such a problem (see also (Garwin 2003)), the first valuable results were achieved by
the first CERN $(g-2)$ team composed in alphabetic order by G. Charpak, F.J.M. Farley, T. Muller, J.C. Sens and
A. Zichichi (credit by CERN-BUL-PHO-2009-017), formalized the 1st of January 1959 but already operative since
1958. As recall (Combley and Picasso 1974), (Farley and Picasso 1979), (Combley et al. 1981) and (Jegerlehner
2008, Part I, Chapter 1), the breakthrough experiment which made the direct attack on the magnetic moment
anomaly of muons was performed at CERN synchrocyclotron (SC) by the first $(g-2)$ team mentioned above. As a
result of this measurement, the experimental accuracy in the value of the muon anomalous magnetic moment was
reduced to 0.4\% from the level of 15\% at which it had previously stood. Following (Brown and Hoddeson 1983,
Part III, Chapter 8), the CERN experiments performed from 1961 to 1965, have been based on the main idea
according to which, roughly speaking, the muons produced by a beam of pions decaying in flight are
longitudinally polarized; furthermore, in the subsequent decays, the electrons reveal the direction of the muon
spins because they are preferentially emitted along the spin direction at the momentum of decay. Hence, a
$(g-2)$ experiment may be performed trapping the longitudinally polarized muons in a uniform magnetic field and
then measuring the precession frequency of the spins. It has only to be added that, due to the very short muon
lifetime, it was necessary to use high-energy muons in order to lengthen their decay times using the
relativistic time dilation effect. The results reduced the error in the measure of $(g-2)$ from the previous
15\% to 0.4\%.

Following (Jegerlehner 2008, Part I, Chapter 1), surprisingly nothing of special was observed even within 0.4\%
level of accuracy of the experiment; it was the first real evidence that the muon was just a heavy electron, so
reaching to another celebrated experimental evidence of the validity of QED. In particular, this meant that the
muon was point-like and no extra short distance effects could be seen. This latter point was however a matter of
accuracy and therefore the challenge to go further was quite evident; in this regards, see the reviews (Farley
and Semertzidis 2004) and (Garwin 2003). As recalled in (Cabibbo 1994, Part I), G. Bernardini, then research
director responsible for the SC at CERN, remembers as, around the end of 1950s, there were many ideas for the
high precision measurements of the anomalous magnetic moment of the muon, two of them having been that of the
\it screw magnet \rm and that of the \it flat magnet. \rm Gilberto Bernardini consulted the greatest magnet
specialist, Dr. Bent Hedin, who said that would have been necessary some years to fully carried out one of this
project, the flat magnet one, so that it was initially chosen the screw magnet project. In the meanwhile, A.
Zichichi had the ingenious idea to trying a new very simple technique consisting in shaping a flat pole with
very thin iron sheets, glued together by means of the simplest possible method, the scotch tape. In this way,
instead of six years, a few months of hard work allowed Zichichi to built up particular high accuracy magnetic
fields, based on the theoretical notion of \it Garwin-Panofsky-Zichichi polynomial magnetic fields, \rm which
constitute just those experimental tools that needed for attaining high measurements of $a_{\mu}$. The so-called
\it six-meters long flat magnet \rm providing an injection field, followed by two transitions, hence a storage,
then another transition and finally an ejection field, became the core of the first high precision measurement
of the muon $(g-2)$. Likewise, R.L. Garwin, in (Cabibbo 1994, Part I), remembers that, in achieving this, it was
determinant the special responsibility of Zichichi profused by him in producing the bizarre magnetic field in
their storage magnetic system, accomplished with imagination, energy and efficiency. Again, in (Garwin 1986,
1991, 2001) and (Garwin 2003), the author recalls that the 80-ton magnet six-meters long was shimmed in a
wondrous fashion under the responsibility of Nino Zichichi who did a wonderful job in doing this, while the
polarization was measured as the muons emerged from the static magnetic field thanks a system perfected by G.
Charpak; F.J.M. Farley was instead in charge to develop the computer program which would take the individual
counts from the polarization analyzer done by Charpak, while T. Muller played the electronic work with the help
of C. York. Following (Jones 2005), the six-meters magnet came to CERN as the heart of the first $g-2$
experiment, the aim of which was to measure accurately the anomalous magnetic moment, or $g$-factor, of the
muon. This experiment was one of CERN outstanding contributions to fundamental physics and for many years was
unique to the laboratory.

To this point, it is need to retake the equations of motion of a charged particle in a magnetic field $\vec{B}$
from a relativistic viewpoint. Following (Combley et al. 1981), (Picasso 1996) and (Jegerlehner 2008, Part II,
Chapter 6), the cyclotron (or orbital) frequency is given by
\begin{equation}\vec{\omega}_c=\frac{Q}{\gamma mc}\vec{B}\end{equation}where $\gamma=1/\sqrt{1-\beta^2}$ and
$\vec{\beta}=\vec{v}/c$. When a relativistic particle is subject to a circular motion, then it is also need to
take into account the so-called \it Thomas precession, \rm which may be computed as follows. The particle rest
frame of muon rotates around the laboratory frame with angular velocity $\vec{\omega}_T$ given by
\begin{equation}\vec{\omega}_T=\Big(1-\frac{1}{\gamma}\Big)\frac{Q\vec{B}}{mc}\end{equation}and it is different
from the direction of the angular velocity with which the muon's spin rotates in the rest frame, so that the
angular velocity of spin rotation in the laboratory frame is given by
\begin{equation}\vec{\omega}_s\doteq\vec{\omega}_L-\vec{\omega}_T=\Big(a_{\mu}+\frac{1}{\gamma}\Big)
\frac{Q\vec{B}}{mc}\end{equation}which shows that the angular frequency of anomalous magnetic moment is, in
relativistic regime, equal to the angular frequency at very low energies, that is to say
\begin{equation}\vec{\omega}_{a_{\mu}}=\vec{\omega}_s-\vec{\omega}_c=a_{\mu}\frac{Q\vec{B}}{mc}.\end{equation}To
argue upon the electric dipole moment of the muon, we should consider the relativistic equations of the muon in
the laboratory system in presence of an electric field $\vec{E}$ and of a magnetic field $\vec{B}$. In this
case, under the conditions of purely transversal fields $\vec{\beta}\cdot\vec{E}=\vec{\beta}\cdot\vec{B}=0$,
following (Bargmann et al. 1959), the cyclotron angular velocity is given by
\begin{equation}\vec{\omega}_c=\frac{Q}{mc}\Big(\frac{\vec{B}}{\gamma}-\frac{\gamma}{\gamma^2-1}\vec{\beta}
\wedge\vec{E}\Big)\end{equation}while the spin angular velocity is given by
\begin{equation}\vec{\omega}_s=\frac{Q}{mc}\Big(\frac{\vec{B}}{\gamma}-\frac{1}{1+\gamma}\vec{\beta}\wedge\vec{E}
+(\vec{B}-\vec{\beta}\wedge\vec{E})\Big)\end{equation}so that the angular frequency of the muon anomalous
magnetic moment, related to the spin precession, is given by
\begin{equation}\vec{\omega}_{a_{\mu}}=\vec{\omega}_s-\vec{\omega}_c=\frac{Q}{mc}\Bigg(a_{\mu}\vec{B}+
\Big(\frac{1}{\gamma^2-1}-a_{\mu}\Big)\vec{\beta}\wedge\vec{E}\Bigg)\end{equation}which is the key formula for
measuring $a_{\mu}$; $\omega_a=|\vec{\omega}_a|=\omega_s-\omega_c$ is the anomalous frequency difference or
spin-flip transition. If a large enough electric dipole moment given by $(6)_2$ there exists, then either the
applied field $\vec{E}$ (which is zero at the equilibrium beam position) and the motional electric field induced
in the muon rest frame, say $\vec{E}^*=\gamma\vec{\beta}\wedge\vec{B}$, will add an extra precession of the spin
with a component along $\vec{E}$ and one around an axis perpendicular to $\vec{B}$, that is to say
\begin{equation}\vec{\omega}=\vec{\omega}_{a_{\mu}}+\vec{\omega}_{EDM}=\vec{\omega}_{a_{\mu}}+\frac{\eta Q}{2mc}\Big(
{\vec{E}}+\vec{\beta}\wedge\vec{B}\Big)\end{equation}or else
\begin{equation}\Delta\omega_{a_{\mu}}\cong d_e(\vec{E}+\vec{\beta}\wedge\vec{B})\end{equation}which, for
$\beta\sim 1$ and $d_e\vec{E}\sim 0$, yields
\begin{equation}\omega_{a_{\mu}}\cong B\sqrt{\Big(\frac{Q}{mc}a_{\mu}\Big)^2+(d_e)^2}.\end{equation}
The result is that the plane of precession is no longer horizontal but tilted at an angle
\begin{equation}\theta\equiv\arctan\frac{\omega_{EDM}}{\omega_{a_{\mu}}}=\arctan\frac{\eta\beta}{2a_{\mu}}
\cong\frac{\eta}{2a_{\mu}}\end{equation}and the precession frequency is increased by a factor
\begin{equation}\omega'_{a_{\mu}}=\omega_{a_{\mu}}\sqrt{1+\delta^2}.\end{equation}The angle $\theta$ produces a
phase difference in the $(g-2)$ oscillation. It is therefore important to determine whether there is a vertical
component to the precession in order to separate out the effect of an electric dipole moment from the
determination of $\omega_{a_{\mu}}$. The angle of tilt $\theta$ given, in the small angle approximation, by
(70), may be detected by looking for the time variation of the vertical component of the muon polarization with
the same frequency as the $(g-2)$ precession of the horizontal polarization. Therefore, in order to eliminate
the electric dipole moment as a source of any discrepancy which might appear in $(g-2)$ direct measurements of
higher precision is preliminarily required. In any case, the main determination in the electric dipole moment of
the muon is not merely this last clarification of the $(g-2)$ measurements. Indeed, it is also of fundamental
importance in itself since the existence of such a static property for any particle would imply the lack of
invariance for the electromagnetic interaction under both $P$ and $T$, as recalled above. Some of the theories
unifying the weak and electromagnetic interactions predict a small electric dipole moment for some particles
including the muon and a precise measurement of this property would tighten the constrains within which such
theories might operate, so that precise measurements of the electric dipole moment of the muon as of other
particles were and still are highly desirable.

\section*{4. Towards the first exact measurements of the anomalous magnetic moment of the muon}
\addcontentsline{toc}{section}{4. Towards the first exact measurements of the anomalous magnetic moment of the
muon}

In Section 2, we have outlined the first works of A. Zichichi and co-workers on cosmic rays carried out until
the end of 1950s. From this period onwards, A. Zichichi was involved, as briefly said above, in some crucial
experiments concerning the muon $(g-2)$ measurements and carried out at CERN of Geneva. The first work on muon
anomalous magnetic moment in which he was involved is 9. where a precise measurement of the electric dipole
moment of the muon was obtained within the QED context only. The work starts from the above mentioned Michigan
spin precession method used to measure $a_e$ which exploits the possibility to have beams of polarized leptons
underwent to asymmetric decay. With this method, i.e. the spin precession methods (see previous Section), one
can measure $(g-2)$ by storing the particles for some time in a magnetic field and then measuring the relative
precession angle between the spin and the angular momentum which serves as a reference vector. As in the
electron experiments, the primary requirement was in being able in injecting the muons into a magnetic field so
that they could circulate on essentially periodic orbits, hence to trap them in this field for a large number of
orbit periods as possible. Nevertheless, at that time, the available muon beams exhibited, in comparison with
the electron case, very low fluxes, high momenta and large extensions in position and momentum space (hence, low
density in phase space) which implied many other new difficulties besides the above mentioned primary
requirement. On the other hand, the muons did not require the analysis of the spin polarization by scattering
since the asymmetric electron decay reveals the spin deviation; indeed, as said above, the electrons were
emitted along the spin direction at the moment of decay. Starting from the principle of the method of the
experimental apparatus used in (Garwin et al. 1957), the essence of this idea had already been established in
(Berley et al. 1958) where the existence of longitudinally polarized beams of $\mu$ mesons and the availability
of muon decay electron asymmetry as a polarization analyzer suggested this method by means of which one may
search for a muon electric dipole moment. A discussion of the results achieved in (Berley et al. 1958) was then
made in (Garwin and Lederman 1959) from which turns out that several practical methods for overcoming these
difficulties were either experimentally and theoretically undertaken before this work of Charpak, Lederman, Sens
and Zichichi, but without succeed in the enterprize. Instead, this research group was able, for the first time,
to trap 85 MeV/c momentum muons for 28 turns, i.e. orbit periods, with no pulse magnets. Their results clearly
suggested too that minor modifications in their method were enough to enable one in achieving storage for
several hundreds of turns. Well, all this was made possible, as also recalled in the previous section, just
thanks to the ingenious technical and experimental ability of A. Zichichi in building up suitable polynomial
magnetic fields of high precision and thanks to which it was possible to obtain thousand muon turns (see also
(Farley 2005)); in turn, all this was carried out on the basis of the theoretical framework mainly worked out on
previous remarkable studies made by R. Garwin and W.K.H. Panofsky, upon which we shall in-depth return later.
The extreme importance and innovativeness of this experimental technique was successfully carried out later, at
a technical level, in producing the so-called \it six-meters long flat magnet \rm which, in turn, was mainly
built up by A. Zichichi starting from a suitable modification of a previous magnet provided by the University of
Liverpool (see (Zichichi 2010) and (CERN 1960)). Seen the fundamental importance of this event, it is necessary
to outline the early works and ideas which came before the dawning of this experimental apparatus, and mainly
worked out, for the first time, in the paper 9. on whose content we now will briefly argue.

The principle of the method consists in injecting, say along the $Y$ axis, a muon beam into a median $(X,Y)$
plane of a flat magnet gap. A moderator (or absorber) $M$, centered on the origin of the $(X,Y)$ plane, will
contain such a beam through a suitable reduction of the momentum beam $p$ and of the mean vertical (i.e. along
$Z$ direction) field value $B_{z0}$. So, the muons lost energy and consequently follow small and more sharply
orbits which will be contained within the magnetic field region, and to prevent a reabsorption by moderator
after one turn, a small transverse linear gradient of the magnetic field is inserted, causing an orbit drift
along the $X$ axis in the direction opposed to $sign\ a$. The magnetic field configuration is therefore planned
to produce such a drift of the muon orbits along the $X$ axis away from the moderator $M$, focusing the muon
beam in the median $(X,Y)$ plane. The magnetic field therein used has the following polynomial
form\begin{equation}B_z=B_{z0}(1+aY+bY^2)\end{equation}along the median plane, where $a,b\in\mathbb{R}$ have to
be small (Garwin-Panofsky). If $r$ is the distance from the origin and $ar\ll 1$ and $br^2\ll 1$, then the muons
emerging from $M$ will move on nearly circular orbits of radius $r$. A linear gradient alone leads to a
step-size drift of these orbits along the $X$-direction by an amount equal to\begin{equation}s=\pi r^2\langle
grad_Y\frac{B_z}{B_{z0}}\rangle=\pi r^2 a\ \ \mbox{\rm per\ turn}\end{equation}where $\langle\ ,\ \rangle$
denotes average over one orbit loop. This drift will enable some muons to get over $M$ after their first turn,
whereupon they go on along a trochoidal orbit. Moreover, following previous basic and notable studies made by
R.L. Garwin and W.K.H. Panofsky\footnote{See R.L. Garwin, \it Numerical calculations of the stability bands and
solutions of a Hill differential equation, \rm CERN Internal Report (October 1959) and W.K.H. Panofsky, \it
Orbits in the linear magnet, \rm CERN Internal Report (October 1959).}, the linear gradient also produces a weak
vertical focusing with wavelength given by
\begin{equation}\frac{\lambda_{\nu}}{2\pi}\cong\frac{0.76}{a}.\end{equation}Taking into account equation (73),
because we want to be $r/s\gg 1$ in order to store as large as possible a number of turns in a magnet of given
finite size, it follows that this focusing is very weak either because of sensitive variations of the field
index $n$ and since $(r/s\gg 1)\Rightarrow (\lambda_{\nu}/2\pi r\gg 1)$ which implies low frequencies and
consequently a weak focusing, hence a poor storage. Nevertheless, as was pointed out by R.L. Garwin (see his
1959 CERN Internal Report), one can improve the vertical focusing while maintaining a given large value of $r/s$
by the addition of a quadratic term of the type $by^2$ and indeed, for a polynomial magnetic field of the type
(72) with $a$ and $b$ small, one has
\begin{equation}\frac{\lambda_{\nu}}{2\pi}\cong\frac{1}{\sqrt{b+1.74a^2}}\sim\frac{1}{\sqrt{b}}\end{equation}while
the drift step-size is still given by (73), so that we can handle $a$ and $b$ in such a manner to have high
values of the former and low values of the latter. For example, by taking $b=50a^2$, one can, while maintaining
the same $r/s$ of above (for such orbits), improve the focusing to 1 oscillation per 7 turns. Therefore, the
intensity of stored muons is increased by a factor $38/7\sim 5$ by the addition of the quadratic term to the
magnetic field. Thus, to sum up, the term $ay$ produces the $X$ axis drift of an orbit of radius $r$ in
step-sizes of magnitude $a\pi r^2$ per turn\footnote{According to a principle of the method almost similar to
the one proposed by P.S. Farago in (Farago 1958) for the free electron case.}. The next $by^2$ term adds
vertical focusing in such a manner that the wavelength of the vertical oscillations are about $2\pi/\sqrt{b}$;
it has as well the useful function to fix more firmly the magnetic median plane around the center of the magnet
gap because just the median plane begins to touch the poles, then all the particles will go lost. In any case,
it is not allowed to choose $b$ arbitrarily large for vertical defocusings minimizing $\lambda_{\nu}$ because
this would lead to a spread in the drift step-size and hence in storage times. Indeed, orbits emerging at an
angle $\phi$ with respect to the $Y$ axis would have a step-size given by
\begin{equation}s(\phi)=\pi r^2(a-2br\phi)\end{equation}so that the magnitude of $b$ may be chosen in order to
maximize the number of particle stored for a given number of turns.

Once having established these fundamental theoretical points, mainly due, as recalled above, to previous works
of R.L. Garwin and W.K.H. Panofsky, the next step was to practically realize such polynomial magnetic fields,
far from being an easy task. This primary work was masterfully and cleverly accomplished by A. Zichichi starting
from a previous magnet provided by the University of Liverpool for whose technical details we refer to the
Section 2 - Injection and Trapping, of the original work 9. He was very able to set up a complex but efficient
experimental framework that provided suitable polynomial magnetic fields for the magnetic storage of muon beams.
The experimental results are of historical importance and were represented in the Figures 2. and 3.a)-b) of 9.
whose characteristics were adequately theoretically explained in the above mentioned Section 2 of 9. These
results were the first valuable experimental evidence of the fact that particles turning several times inside a
small magnetic arrangement was pursuable, so endorsing that presentiment according to which longer magnetic
systems of this type could give further and more precise measurements. All this was in fact done in the
subsequent experiments made by A. Zichichi and co-workers and that will be described later. The final section of
the work 9. deals then with attempts to measure the electric dipole moment of the muon starting from the
experimental results achieved by the previous works (Berley et al. 1959) and (Garwin and Lederman 1959) and
whose principle of the method was mainly based on the determination of the phase angle given by (59) through the
so-called \it up-down asymmetry parameter\footnote{It is given by $\alpha=(N_{up}-N_{down})/(N_{up}+N_{down})$
respect to the median plane.} $\alpha$, \rm taking into account the original theoretical treatment given by
(Bargmann et al. 1959) and briefly recalled in the previous Section 3. To this end, Charpak, Lederman, Sens and
Zichichi used their innovative experimental arrangement to storage polarized muon beams, just to determine this
EDM of the muon. The related value so found was consistent with time reversal invariance and could be considered
equal to zero within the experimental errors which have been considerably reduced respect to those of the above
mentioned previous works on muon EDM determination. To be precise, their formal treatment is that of (Bargmann
et al. 1959) in which are considered the covariant classical equations of motion of a particle of arbitrary spin
moving in a homogeneous electromagnetic field. As it has already been said, the theoretical considerations made
in (Bargmann et al. 1959) include too the relativistic case because of a remark due to F. Bloch. We consider
longitudinally polarized muons possessing an EDM given by $(6)_2$, which move in a magnetic field $\vec{B}$ in a
plane perpendicular to the latter. In their instantaneous rest frame, they experience an electric field given by
$\vec{E}^*=\gamma\vec{\beta}\wedge\vec{B}$ which causes a precession of the EDM. In the laboratory frame, the
spin precesses around $\vec{v}\wedge\vec{B}$ (hence, out of the orbit plane in which relies
$\vec{v}\wedge\vec{B}$) by an angle $\Theta_s=\omega_st$ when the orbit has gone through an angle
$\Theta_o=\omega_ct$ (or $\Theta_c$) on its orbital plane (see Equation (59)). The polarization (perpendicular
to the orbit) thus produced, is detected by stopping the muons after a known $\Theta_o$ and measuring the
up-down asymmetry of the electrons emerging from the muon decay with respect to the orbit plane (placed in the
median plane of the storage magnet set up in 9. and detected by the scintillator No. 4 of their apparatus). This
determination, successfully achieved by Charpak, Lederman, Sens and Zichichi, was different from the previous
ones only in the magnitude of $\Theta_o$, in which it was assumed to be $\Theta_o\in]0,2\pi[$, whereas they used
the new storage device based on polynomial magnetic fields to get $\Theta_o=2n\pi$ with $n\geq 28$, just thanks
to the multiple turns that their arrangement was able to provide. The principle of the method consisted in
analyzing two range of flight times of particles, a \it group A \rm of \it early \rm particles having made few
turns in the storage magnet and which are used for calibration, and a \it group B \rm of \it late \rm particles
which have made many revolutions. In turn, the measurements were divided into three groups in dependence on the
mean turn index $\langle n\rangle$ of late particles, this being fixed for the early ones and equal to $\langle
n\rangle\approx 1$. The Group $I$ concerns late muons with $\langle n\rangle\approx 11.5$; the Group $II$
concerns late muons with $\langle n\rangle\approx 16.5$, while Group $III$ concerns muons with $\langle
n\rangle\approx 19.5$. For each of these groups, the difference in up-down asymmetry, say
$\Delta^{(i)}=a_{early}^{(i)}-a_{late}^{(i)}\ \ i=I,II,III$, between the early and late ones, is evaluated. The
values so found are reported in the Table I of 9. and from these it is then possible to estimate the angle
$\Theta_s^{(i)}$, through which the spin has rotated out of the median plane, as $\Delta^{(i)}/a_{max}^{(i)}$
where $a_{max}^{(i)}$ is the maximal obtainable value of asymmetry in the given $i$th group. Then
$\Theta_o^{(i)}\approx\omega_c\langle t^{(i)}\rangle$ where $t^{(i)}$ is the beam flight time detected by the
final median plane scintillator. Furthermore, to improve distribution calculations and to reduce systematic
errors, the EDM telescope was also symmetrically displaced at different heights with respect to the magnet
median plane. Finally, combining the three values of $\Theta_s^{(i)}/\Theta_o^{(i)}\ \ i=I,II,III$ (listed in
the above mentioned Table I), it was possible to estimate $\eta$ of $(6)_2$, whence to deduce the upper limit
for the EDM of the muon.

Following (Lee 2004, Chapter 2), the accelerator physics principles involved in the work 9. mainly concern with
transverse particle motion in the sense as first outlined in the 1941 seminal paper (Kerst and Serber 1941) for
the betatron case. In Frenet-Serret coordinates $(x,s,z)$ ($s$ is oriented as the tangent, $x$ as the normal and
$z$ as the binormal respect to the orbit plane) and in zero electric potential, we have a two-dimensional
magnetic field given by $\vec{B}=B_x(x,z)\hat{x}+B_z(x,z)\hat{z}$ where $\hat{z}=\hat{x}\wedge\hat{z}$. In
straight geometries, we have a magnetic flux density given by
\begin{equation}B_z+iB_x=B_0\sum_{n\in\mathbb{N}_0}(b_n+ia_n)(x+iz)^n\end{equation}where $a_n,b_n$ are called
$2(n+1)$th \it multipole coefficients \rm and are given by (Lee 2004, Chapter 2, Section I.3, Equations (2.26)).
The expression (77) is said to be the \it Beth representation \rm (see (Beth 1966, 1967)). For example, in
discussing the focusing of atomic beams, the sextupole terms are show to be able to make high spin focusings
(see (Lee 2004, Chapter 2, Exercise 2.2.18)). In such a case, some historical predecessors of these techniques
to obtain polarized ions may be found in (Haeberli 1967) where, among other things, are discussed too some
previous experiences with separate magnets operating at the quadrupolar or sextupole order, due to H. Friedburg,
W. Paul and H.G. Bennewitz in the early 1950s. In certain sense, looking at the (77), the
Garwin-Panofsky-Zichichi polynomial magnetic fields might be considered as special cases forerunner of such Beth
representations. One of the main aims of this historical paper has just been that pointing out the following
remarkable fact: the first exact measurements of muon AMM will be possible thanks to the use of these
Garwin-Panofsky-Zichichi polynomial magnetic fields which were masterfully used, for the first time, in 9. to
measure the muon EDM; then, the principle of the method there worked out will be gradually improved both
theoretically and experimentally through further pioneering works until the seminal paper 10. in which the first
exact measurement of the muon AMM was finally achieved with success. This marked a milestone of fundamental
physics of the second half of 20th-century, achieved at CERN of Geneva, upon which we shall return later in a
deeper manner. Nevertheless, we must point out as nobody, including the authors themselves of these pioneering
researches, have recognized the right primary role played by polynomial magnetic fields in achieving these,
whose history is utterly neglected. In this regards, the unpublished theoretical work made by Richard L. Garwin
(together to the one made by W.K.H. Panofsky) has been of fundamental importance in setting up the theoretical
bases for these polynomial magnetic fields; later, the genial technical ability of Antonino Zichichi will be
determinant in providing an experimental version of these fields which were very basilar to get the first exact
measurement of the muon AMM. In another place, however, we will deal with this last historical question, also
thanks to precious unpublished bibliographical material which has been kindly provided to me by Professor
Richard L. Garwin to whom I bear my thankful acknowledgements, and that will be historically in-depth analyzed
in another forthcoming paper.

\newpage\section*{List of some publications of A. Zichichi}\addcontentsline{toc}{section}
{List of some publications of A. Zichichi}\begin{description}

\item 1. G. Alexander, J.P. Astbury, G. Ballario, R. Bizzarri, B. Brunelli, A. De Marco, A. Michelini, G.C.
Moneti, E. Zavattini and A. Zichichi, \it A Cloud Chamber Observation of a Singly Charged Unstable Fragment, \rm
Il Nuovo Cimento, Serie X, Vol. 2 (1955) pp. 365-369 [Received on 18 July 1955 and Published in August 1955 -
Registered Preprint No.].\item 2. W.A. Cooper, H. Filthuth, J.A. Newth, G. Petrucci, R.A. Salmeron and A.
Zichichi, \it A Probable Example of the Production and Decay of a Neutral Tau-Meson, \rm Il Nuovo Cimento, Serie
X, Vol. 4 (1956) pp. 1433-1444 [Received on 09 September 1956 and Published in December 1956 - Registered
Preprint No.].\item 3. W.A. Cooper, H. Filthuth, J.A. Newth, G. Petrucci, R.A. Salmeron and A. Zichichi, \it
Example of the Production of $(K^0,\bar{K}^0)$ and $(K^+,\bar{K}^0)$ Pairs of Heavy Mesons, \rm Il Nuovo
Cimento, Serie X, Vol. 5 (1957) pp. 1388-1397 [Received on 14 January 1957 and Published in June 1957 -
Registered Preprint No.].\item 4. C. Ballario, R. Bizzarri, B. Brunelli, A. De Marco, E. Di Capua, A. Michelini,
G.C. Moneti, E. Zavattini and A. Zichichi, \it Life Time Estimate of $\Lambda^0$ and $\theta^0$ Particles, \rm
Il Nuovo Cimento, Serie X, Vol. 6 (1957) pp. 994-996 [Received on 01 August 1957 and Published in October 1957 -
Registered Preprint No.].\item 5. H. Filthuth, J.A. Newth, G. Petrucci, R.A. Salmeron and A. Zichichi, \it
Cosmic Ray Research: Proposal for a New High Energy Experiment, \rm CERN Scientific Policy Committee, Seventh
Meeting, Document No. CERN/SPC/52(A)-3784/e, Geneva, 21-29 October 1957.\item 6. W.A. Cooper, H. Filthuth, L.
Montanet, J.A. Newth, G. Petrucci, R.A. Salmeron and A. Zichichi, \it Neutral $V$-Particle from Copper and
Carbon, \rm Il Nuovo Cimento, Serie X, Vol. 8 (1958) pp. 471-481 [Received on 11 February 1958 and Published on
May 1958 - Registered Preprint No.].\item 7. G. Alexander, C. Ballario, R. Bizzarri, B. Brunelli, E. Di Capua,
A. Michelini, G.C. Moneti and A. Zichichi, \it $\Lambda^0$- and $\theta^0$- Particles Produced in Iron, \rm Il
Nuovo Cimento, Serie X, Vol. 9 (1958) pp. 624-646 [Received on 23 April 1958 and Published in August 1958 -
Registered Preprint No.].\item 8. L. Montanet, J.A. Newth, G. Petrucci, R.A. Salmeron and A. Zichichi, \it A
Cloud Chamber Study of Nuclear Interactions with Energies of about 100 GeV, \rm Il Nuovo Cimento, Serie X, Vol.
17 (1960) pp. 166-188 [Received on 18 March 1960 and Published on 18 July 1960 - Registered Preprint No.].\item
9. G. Charpak, L.M. Lederman, J.C. Sens and A. Zichichi, \it A Method for Trapping Muons in Magnetic Fields, and
Its Application to a Redetermination of the EDM of the Muon, \rm Il Nuovo Cimento, \rm Serie X, Vol. 17 (1960)
pp. 288-303 [Received on 04 April 1960 and Published on 01 August 1960 - Registered Preprint No.
CERN-SC/8431/nc].\item 10. G. Charpak, F.J.M.Farley, R.L. Garwin, T. Muller, J.C. Sens and A. Zichichi, \it The
Anomalous Magnetic Moment of the Muon, \rm Il Nuovo Cimento, Serie X, Vol. 37 (1965) pp. 1241-1363.

\end{description}

\newpage\section*{List of some publications of R.L. Garwin}\addcontentsline{toc}{section}
{List of some publications of R.L. Garwin}

\begin{description}

\item 1'. R.L. Garwin, L.M. Lederman and M. Weinrich, Observations of the Failure of Conservation of Parity and
Charge Conjugation in Meson Decays: the Magnetic Moment of the Free Muon, \it Physical Review, \rm 105, No. 4,
pp. 1415-1417, February 15, 1957. \item 2'. Columbia University Physics Department announcement of parity
experiments by C.S. Wu, E. Ambler, R.L. Garwin, L.M. Lederman, et al., January 15, 1957. \item 3'. D. Berley, T.
Coffin, R.L. Garwin, L. Lederman, and M. Weinrich, Depolarization of Positive Muons in Matter, \it Bulletin of
the American Physical Society, \rm Series II, 2, No. 4, p. 204, April 25, 1957. \item 4'. Berley, T. Coffin,
R.L. Garwin, L. Lederman, and M. Weinrich, Energy Dependence of the Asymmetry in Polarized Muon Decay, \it
Bulletin of the American Physical Society, \rm Series II, 2, No. 4, p. 204, April 25, 1957. \item 5'. T. Coffin,
R.L. Garwin, L.M. Lederman, S. Penman, and A.M. Sachs, Magnetic Resonance Determination of the Magnetic Moment
of the Mu Meson, \it Physical Review, \rm 106, pp. 1108-1110, May 1957. \item 6'. D. Berley, T. Coffin, R.L.
Garwin, L.M. Lederman and M. Weinrich, Energy Dependence of the Asymmetry in the Beta Decay of Polarized Muons,
\it Physical Review, \rm 106, pp. 835-837, May 1957. \item 7'. R.L. Garwin, S. Penman, L.M. Lederman, and A.M.
Sachs, Magnetic Moment of the Free Muon, \it Physical Review, \rm 109, No. 3, pp. 973-979, February 1, 1958.
\item 8'. T. Coffin, R.L. Garwin, S. Penman, L.M. Lederman, and A.M. Sachs, Magnetic Moment of the Free Muon,
\it Bulletin of the American Physical Society, \rm Series II, 3, No. 1, p. 34, January 29, 1958. \item 9'. D.
Berley, R.L. Garwin, G. Gidal and L.M. Lederman, Electric Dipole Moment of the Muon, \it Physical Review
Letters, \rm 1, No. 4, pp. 144-146, August 15, 1958. \item 10'. R.L. Garwin and L.M. Lederman, The Electric
Dipole Moment of Elementary Particles, \it ll Nuovo Cimento, \rm Serie X, 11, pp. 776-780, 1959. \item 11'. D.
Berley, R.L. Garwin, G. Gidal, and L.M. Lederman, Electric Dipole Moment of the Muon, \it Bulletin of the
American Physical Society, \rm Series II, 4, No. 1, Part 1, p. 81, January 28, 1959. \item 12'. D. Berley, R.L.
Garwin, G. Gidal, and L.M. Lederman, Electric Dipole Moment of the Muon, \it Bulletin of the American Physical
Society, \rm Series II, 5, No. 1 Part 2, p. 81, January 27, 1960.\item 13'. Garwin, R.L. (1986, 1991, 2001),
Interviews of Richard L. Garwin by Finn Aaserud and W. Patrick McCray, made on October 23, 1986 in Yorktown
Heights, NY (in three sessions), on June 24, 1991 at the IBM Research Laboratory, Croton-Harmon, NY, and on June
7, 2001 in New York City, \it Oral History Transcripts, Center for History of Physics of the American Institute
of Physics, Niels Bohr Library \& Archives, American Institute of Physics (AIP), \rm College Park (MD), USA.

\end{description}

\newpage\section*{Bibliography}\addcontentsline{toc}{section}{Bibliography}
\begin{description}\item  Alberico, W.M. (1992), \it Introduzione alla Fisica Nucleare, \rm Torino: La
Scientifica Editrice..\item  Araki, H. (1999), \it Mathematical Theory of Quantum Fields, \rm New York: Oxford
University Press, Inc.\item  Archibald, R. C. (1950), Book Reviews: Luigi Berzolari, Enciclopedia delle
matematiche elementari e complementi con estensione alle principali teorie analitiche, geometriche e fisiche,
loro applicazione e notizie storico-bibliografiche, Vol. 3, \it Bulletin of the American Mathematical Society
(N.S.), \rm 56 (6): 517-518. \item  Bacry, H. (1967), \it Le\c{c}ons sur la Théorie des Groupes et les Symétries
des Particules \'{E}lémentaires, \rm Paris: Gordon \& Breach Science Publishers, Inc.\item Bargmann, V., Michel,
L. and Telegdi, V.L. (1959), Precession of the polarization of particle moving in a homogeneous electromagnetic
field, \it Physical Review Letters, \rm 2 (10): 435-436.\item Barone, V. (2004), \it Relatività. Princìpi e
applicazioni, \rm Torino: Bollati Boringhieri editore.\item  Barut, A.O. and R\c{a}czka, R. (1977), \it Theory
of Group Representations and Applications, \rm Warszawa: PWN-Polish Scientific Publishers.\item Bauer, H.H.,
Christian, G.D. and O'Reilly, J.E. (Eds.) (1978), \it Instrumental Analysis, \rm Boston (MA): Allyn \& Bacon,
Inc. (Italian Translation: (1985), \it Analisi Strumentale, \rm Padova: Piccin Nuova Libraria).\item Berezin,
F.A. and Shubin, M.A. (1991) \it The Schr\"{o}dinger Equation, \rm Dordrecht: Kluwer Academic Publishers.\item
Berley, D., Garwin, R.L., Gidal, G. and Lederman, L.M. (1958), Electric Dipole Moment of the Muon, \it Physical
Review Letters, \rm 1 (4): 144-146.\item Beth, R.A. (1966), Complex Representation and Computation of Two
Dimensional Magnetic Fields, \it Journal of Applied Physics, \rm 37 (7): 2568-2571.\item Bertolotti, M. (2005),
\it The History of the Laser, \rm London: IoP - Institute of Physics Publishing, Ltd. (Italian Edition: (1999),
\it Storia del laser, \rm Torino: Bollati Boringhieri editore). \item Beth, R.A. (1967), An Integral Formula for
Two-Dimensional Fields, \it Journal of Applied Physics, \rm 38 (12): 4689-4692.\item Bjorken, J.D. and Drell,
S.D. (1964), \it Relativistic Quantum Mechanics, \rm New York: McGraw-Hill Book Company, Inc.\item Bjorken, J.D.
and  Drell, S.D. (1965), \it Relativistic Quantum Fields, \rm New York: McGraw-Hill Book Company, Inc.\item
Bloch, F. (1946), Nuclear Induction, \it Physical Review, \rm 70 (7-8): 460-474.\item Bogolubov, N.N., Logunov,
A.A. and Todorov, I.T. (1975), \it Introduction to Axiomatic Quantum Field Theory, \rm Reading (MA): W.A.
Benjamin, Inc.\item Bogoliubov, N.N. and Shirkov, D.V. (1980), \it Introduction to the Theory of Quantized
Fields, \rm third edition, New York: A Wiley-Interscience Publication, John Wiley \& Sons, Inc.\item  Bogolubov,
N.N., Logunov, A.A., Oksak, A.I. and Todorov, I.T. (1990), \it General Principles of Quantum Field Theory, \rm
Dordrecht: Kluwer Academic Publishers.\item  Bohr, A. and Mottelson, B.R. (1969, 1975), \it Nuclear Structure,
Volume I, Single-Particle Motion, Volume II, Nuclear Deformations, \rm London: W.A. Benjamin, Inc.\item  Bohm,
A. (1993), \it Quantum Mechanics. Foundations and Applications, \rm third edition, New York:
Springer-Verlag.\item Born, M. (1969), \it Atomic Physics, \rm 8th edition, London-Glasgow: Blackie \& Son, Ltd.
(Italian Translation: (1976), \it Fisica atomica, \rm Torino: Bollati Boringhieri Editore).\item Boudon, R.
(1986) \it L'Ideologie. L'origine des idées re\c{c}ues, \rm Paris: \'{E}ditions Fayard (Italian Translation:
(1991), \it L'ideologia. Origine dei pregiudizi, \rm Torino: Giulio Einaudi editore).\item Breit, G. (1947a),
Relativistic Corrections to Magnetic Moments of Nuclear Particles, \it Physical Review, \rm 71 (7):
400-402.\item Breit, G. (1947b), Does the Electron Have an Intrinsic Magnetic Moment?, \it Physical Review, \rm
72 (10): 984.\item  Brown, L.M. and Hoddeson L. (Eds), \it The Birth of Particle Physics, \rm Based on the
lectures and round-table discussion of the International Symposium on the History of Particle Physics, held at
Fermilab in May, 1980, \rm New York: Cambridge University Press.\item Brown, L.M., Dresden, M. and Hoddeson, L.
(Eds) (1989), \it Pions to Quarks: Particle Physics in the 1950s, \rm Based on the lectures and discussions of
historians and physicists at the Second International Symposium on the History of Particle Physics, held at
Fermilab on May 1-4, 1985, Cambridge (UK): Cambridge University Press.\item Cabibbo, N. (Ed.) (1994), \it Lepton
Physics at Cern and Frascati, \rm Singapore: World Scientific Publishing Company.\item Cahn, R.N. and Goldhaber,
G. (2009), \it The Experimental Foundations of Particle Physics, \rm 2nd Edition, Cambridge (UK): Cambridge
University Press. \item P. Caldirola, \it Dalla microfisica alla macrofisica, \rm Biblioteca EST, Arnoldo
Mondadori Editore, Milano, 1974.\item P. Caldirola, R. Cirelli, G.M. Prosperi, \it Introduzione alla Fisica
Teorica, \rm UTET, Torino, 1982.\item P. Carlson, A. De Angelis, Nationalism and internationalism in science:
the case of the discovery of cosmic rays, \it The European Physical Journal H, \rm 35 (4) (2010) pp.
309-329.\item Carotenuto, A. (1991), \it Trattato di psicologia della personalità, \rm Milano: Raffaello Cortina
Editore.\item P.A. Carruthers, \it Introduction to Unitary Symmetry, \rm Interscience Publishers, a division of
John Wiley \& Sons, Inc., New York, 1966.\item P.A. Carruthers, \it Spin and Isospin in Particle Physics, \rm
Gordon and Breach Science Publishers, New York, 1971.\item C. Castagnoli, \it Lezioni di struttura della
materia, \rm seconda edizione riveduta ed ampliata, Libreria editrice universitaria Levrotto \& Bella, Torino,
1975.\item G. Castelfranchi, \it Fisica moderna, atomica e nucleare, \rm decima edizione completamente
rinnovata, Editore Ulrico Hoepli, Milano, 1959.\item L. Castellani, R. D'Auria, P. Fré, \it Supergravity and
Superstrings. A Geometric Perspective, Volume 1, Mathematical Foundations, Volume 2, Supergravity, Volume 3,
Superstrings, \rm World Scientific Publishing Company, Ltd., 1991.\item A. Cavallucci, E. Lanconelli,
Commemorazione di Bruno Pini, \it La matematica nella Società e nella Cultura. Rivista della Unione Matematica
Italiana, Serie I, \rm Vol. 4, No. 2 (2011) pp. 261-274.\item CERN 1960 Annual Report, Geneva, 1961. \item
Chanowitz, M.S., Furman, M.A. and Hinchliffe, I. (1978), Weak interactions of ultra heavy fermions, \it Physics
Letters B, \rm 78 (2-3): 285-289.\item T-P. Cheng, L-F. Li, \it Gauge theory of elementary particle physics, \rm
Oxford at Clarendon Press, New York, 1984.\item E. Chiavassa, L. Ramello, E. Vercellin, \it Rivelatori di
particelle. Appunti dalle lezioni di Fisica dei Neutroni, \rm La Scientifica Editrice, Torino, 1991.\item
Coffin, T., Garwin, R.L., Penman, S., Lederman, L.M. and Sachs, A.M. (1958), Magnetic Moment of the Free Muon,
\it Physical Review, \rm 109 (3): 973-979.\item Cohen-Tannoudji, C., Diu, B. and Lalo\"{e}, F. (1977), \it
Quantum Mechanics, \rm Volume I, II, Paris-New York: \`{E}ditions Hermann and John Wiley \& Sons, Inc. \item C.
Cohen-Tannoudji, J. Dupont-Roc, G. Grynberg, \it Photons et atomes. Introduction à l'électrodynamique quantique,
\rm Savoirs Actuels, InterEditions/Editions du CNRS, Paris, 1987.\item S. Coleman, \it Aspects of symmetry.
Selected Erice lectures, \rm Cambridge University Press, Cambridge (UK), 1985.\item P.D.B. Collins, A.D. Martin,
E.J. Squires, \it Particle Physics and Cosmology, \rm John Wiley \& Sons, Inc., New York, 1989.\item Combley, F.
and Picasso, E. (1974), The Muon $(g-2)$ Precession Experiments: Past, Present and Future, \it Physics Reports,
\rm 14 (1): 1-58.\item Combley, F., Farley, F.J.M. and Picasso, B. (1981), The CERN Muon $(g-2)$ Experiments,
\it Physics Reports, \rm 68 (2): 93-119.\item Compton, A.H. (1921), The magnetic electron, \it Journal of the
Franklin Institute, \rm 192 (2): 145-155. \item E. Corinaldesi, F. Strocchi, \it Relativistic Wave Mechanics,
\rm North-Holland Publishing Company, Amsterdam, 1963.\item J.F. Cornwell, \it Group Theory in Physics, Volumes
I, II, III, \rm Academic Press, Ltd., London, 1984, 1989.\item Crane, H.R. (1976), $g-2$ techniques: past
evolution and future prospects, \it AIP Conference Proceedings, \rm 35 (1): 306-314.\item Cressant, P. (1970),
\it Lévi Strauss, \rm Paris: \'{E}ditions Universitaires (Italian Translation: (1971), \it Lévi Strauss, \rm
Firenze: C/E Giunti - G. Barbèra).\item Croce, B. (1938), \it La storia come pensiero e come azione, \rm Bari:
Editori Laterza.\item P. Davies (Ed), \it The New Physics, \rm Cambridge University Press, Cambridge (UK), 1989
(Italian Translation: \it La Nuova Fisica, \rm Bollati Boringhieri editore, Torino, 1992).\item A.S. Davydov,
\it Teoria del nucleo atomico, \rm Nicola Zanichelli Editore, Bologna, 1966.\item A.S. Davydov, \it Meccanica
quantistica, \rm Edizioni Mir, Mosca, 1981.\item V. De Alfaro, T. Regge, \it Potential Scattering, \rm
North-Holland Publishing Company, Amsterdam, 1965.\item V. De Alfaro, S. Fubini, G. Furlan, C. Rossetti, \it
Currents in Hadron Physics, \rm North-Holland Publishing Company, Amsterdam and London, 1973.\item V. De Alfaro,
\it Introduzione alla teoria dei campi, \rm Parte I, Lezioni date alla Facoltà di Scienze dell'Università di
Torino, Anno Accademico 1993-1994, Edizioni Cooperativa Libraria Universitaria - CLU, Torino, 1993.\item S.R. de
Groot, L.G. Suttorp, \it Foundations of Electrodynamics, \rm North-Holland Publishing Company, Amsterdam,
1972.\item Dekker, A.J. (1958), \it Solid State Physics, \rm 6th Edition, Englewood Cliffs (NJ): Prentice-Hall,
Inc. (Italian Translation: (1965), \it Fisica dello stato solido, \rm Milano: CEA - Casa Editrice
Ambrosiana).\item P. Deligne, P. Etingof, D.S. Freed, L.C. Jeffrey, D. Kazhdan, J.W. Morgan, D.R. Morrison, E.
Witten (Eds), \it Quantum Fields and Strings: A Course for Mathematicians, \rm Volumes 1, 2, American
Mathematical Society and Institute for Advanced Study, Providence, Rhode Island, 1999.\item A. Derdzinski, \it
Geometry of the Standard Model of Elementary Particles, \rm Springer-Verlag, Berlin and Heidelberg, 1992.\item
A. Di Giacomo, \it Lezioni di Fisica Teorica, \rm Edizioni ETS, Pisa, 1992.\item P.A.M. Dirac, \it The
Principles of Quantum Mechanics, \rm 4th edition, Oxford University Press at Clarendon, Oxford (UK), 1958
(Italian Translation: \it I principi della meccanica quantistica, \rm Editore Boringhieri, Torino, 1959).\item
P.A.M. Dirac, \it Lectures on Quantum Mechanics and Relativistic Field Theory, \rm Tata Institute of Fundamental
Research, Bombay, 1955.\item P.A.M. Dirac, \it Lectures on Quantum Field Theory, \rm Belfer Graduate School of
Science, Yeshiva University, Academic Press, Inc., New York, 1966.\item Ducrot, O., Todorov, T., Sperber, D.,
Safouan, M. and Wahl, F. (1968), \it Qu'est-ce que le structuralisme?, \rm Paris: \'{E}ditions du Seuil (Italian
Translation: (1971), \it Che cos'è lo strutturalismo, \rm Milano: ISEDI).\item L. Eisenbud, E.P. Wigner, \it
Nuclear Structure, \rm Princeton University Press, Princeton (NJ), 1958 (Italian Translation: \it La struttura
del nucleo, \rm Editore Boringhieri, Torino, 1960).\item Farago, P.S. (1958), Proposed Method for Direct
Measurement of the $g$-Factor of Free Electrons, \it Proceedings of the Physical Society (London), \rm 72 (5):
891-894.\item Farago, P.S., Gardiner, R.B. and Rae, A.G.A. (1963), Direct Measurement of the $g$-Factor Anomaly
of Free Electrons, \it Proceedings of the Physical Society (London), \rm 82 (4): 493-500.\item Farley, F.J.M.
and Picasso, E. (1979), The Muon $(g-2)$ Experiments, \it Annual Review of Nuclear and Particle Science, \rm 29
(1): 243-282 [Registered Preprint: The Muon $(g-2)$ Experiments at CERN, Report No. CERN-EP/79-20, 13 March
1979].\item Farley, F.J.M. and Semertzidis, Y.K. (2004), The 47 years of muon $g-2$, \it Progress in Particle
and Nuclear Physics, \rm 52 (1): 1-83.\item Farley, F.J.M. (2005), A new method of measuring the muon $g-2$,
talk presented in the Section: ''Role of innovative detectors'' of the Symposium \it The Golden Age of Particle
Physics and its Legacy: A Festschrift in honour of Larry Sulak, \rm Boston University, October 21-22, 2005.
\item E. Fermi, \it Notes on Quantum Mechanics. A Course Given by Enrico Fermi at the University of Chicago, \rm
Phoenix Books, The University of Chicago Press, Chicago \& London, 1962.\item E. Fermi, \it Elementary
Particles, \rm Yale University Press, Inc., New Haven, 1951 (Italian Translation: \it Particelle elementari, \rm
Editore Boringhieri, Torino, 1963).\item P. Fleury, J.P. Mathieu, \it Cours de physique générale et
expérimentale, 8. Atomes, Molécules, Noyaux, \rm \'{E}ditions Eyrolles, Paris, 1963 (Italian Translation: \it
Trattato di fisica generale e sperimentale, 8. Atomi, Molecole, Nuclei, \rm Nicola Zanichelli editore, Bologna,
1965).\item Foldy, L.L. and Wouthuysen, S.A. (1950), On the Dirac Theory of Spin 1/2 Particles and its
Non-Relativistic Limit, \it Physical Review, \rm 78 (1): 29-36.\item L. Fonda, G.C. Ghirardi, \it Symmetry
Principles in Quantum Physics, \rm Marcel Dekker, Inc., New York, 1970.\item Ford, J.W., Luxon, J.L., Rich, A.,
Wesley, J.C. and Telegdi, V.L. (1972), Resonant Spin Rotation - a New Lepton $g-2$ Technique, \it Physical
Review Letters, \rm 29 (25): 1691-1695.\item K.W. Ford, \it The World of Elementary Particles, \rm Blaisdell
Publishing Company, 1963 (Italian Translation: \it La fisica delle particelle, \rm Biblioteca EST, Arnoldo
Mondadori Editore, Milano, 1965).\item A. Frank, P. Van Isacker, \it Algebraic Methods in Molecular and Nuclear
Structure Physics, \rm John Wiley \& Sons, Inc., New York, 1994.\item G. Friedlander, J.W. Kennedy, \it Nuclear
and Radiochemistry, \rm John Wiley \& Sons, Inc., New York, 1960 (Italian Translation: \it Chimica nucleare e
radiochimica, \rm Carlo Manfredi Editore, Milano, 1965).\item Friedman, J.I. and Telegdi, V.L. (1957), Nuclear
Emulsion Evidence for Parity Nonconservation in the Decay Chain $p^+\rightarrow\mu^+\rightarrow e^+$, \it
Physical Review, \rm 106 (6): 1290-1293.\item K.O. Friedrichs, \it Mathematical Aspects of the Quantum Theory of
Fields, \rm Interscience Publishers, Inc., New York, 1953.\item U. Galimberti, \it Dizionario di Psicologia, \rm
UTET Libreria, Torino, 2006.\item G. Gamow, \it Biography of Physics, \rm Harper \& Row, New York, 1961 (Italian
Translation: \it Biografia della fisica, \rm Biblioteca EST, Arnoldo Mondadori Editore, Milano, 1963).\item G.
Gamow, \it Thirty Years That Shook Physics. The Story of Quantum Theory, \rm Anchor Books Doubleday \& Company,
Inc., Garden City, New York, 1966 (Italian Translation: \it Trent'anni che sconvolsero la fisica. La storia
della Teoria dei Quanti, \rm Nicola Zanichelli editore, Bologna, 1966).\item Gardner, J.H. and Purcell, E.M.
(1949), A Precise Determination of the Proton Magnetic Moment in Bohr Magnetons, \it Physical Review, \rm 76
(8): 1262-1263.\item Gardner, J.H. (1951), Measurement of the Magnetic Moment of the Proton in Bohr Magnetons,
\it Physical Review, \rm 83 (5): 996-1004.\item Garwin, R.L., Lederman, L.M. and Weinrich, M. (1957),
Observation of the Failure of Conservation of Parity and Charge Conjugation in Meson Decays: the Magnetic Moment
of the Free Muon, \it Physical Review, \rm 105 (4): 1415-1417.\item Garwin, R.L., Hutchinson, D.P., Penman, S.
and Shapiro, G. (1960), Accurate Determination of the $\mu^+$ Magnetic Moment, \it Physical Review, \rm 118 (1):
271-283.\item Garwin, R.L. and Lederman, L.M. (1959), The Electric Dipole Moment of Elementary Particles, \it Il
Nuovo Cimento, \rm 11 (6): 776-780.\item Garwin, R.L. (2003), The first muon spin rotation experiment, \it
Physica B, \rm 326 (1): 1-10. \item H. Georgi, \it Lie Algebras in Particle Physics. From Isospin to Unified
Theories, \rm Addison-Wesley Publishing Company, Inc., Reading, Massachusetts, 1982.\item Germain, P. (1989),
\it Introduction aux accélérateurs de particules, \rm Cours édité par D. Dekkers et D. Manglunki, CERN 89-07, 7
July 1989 - Rév. 15/28.02.2005, Genève.\item V.L. Ginzburg, \it Questioni di fisica e astrofisica, \rm Editori
Riuniti-Edizioni Mir, Roma-Mosca, 1983.\item J. Glimm, A. Jaffe, \it Quantum Physics. A Functional Integral
Point of View, \rm Springer-Verlag, New York, 1981.\item M. Gliozzi, Storia del pensiero fisico, Articolo LX in:
L. Berzolari (Ed), \it Enciclopedia delle Matematiche Elementari e Complementi, con estensione alle principali
teorie analitiche, geometriche e fisiche, loro applicazioni e notizie storico-bibliografiche, \rm Volume III,
Parte 2$^{a}$, Editore Ulrico Hoepli, Milano, 1949 (ristampa anastatica 1972).\item M. Gliozzi, \it Storia della
fisica, \rm a cura di Alessandra e Ferdinando Gliozzi, Bollati Boringhieri Editore, Torino, 2005.\item
Gottfried, K. (1966), \it Quantum Mechanics, Volume I, Foundations, \rm New York: W.A. Benjamin, Inc.\item
Gr\"{a}ff, G. (1971), Methods for Lepton $g$-Factor Anomaly Measurement, in: Langenberg, D.N. and Taylor, B.N.
(Eds.) (1971), \it Precision Measurement and Fundamental Constants. Proceedings of the International Conference
held at the National Bureau of Standards in Gaithersburg, Maryland, August 3-7, 1970, \rm Washington (D.C.):
U.S. Government Printing Office, U.S. Department of Commerce, National Bureau of Standards Special Publication
No. 343, CODEN: XNBSA.\item M.B. Green, J.H. Schwarz, E. Witten, \it Superstring Theory, Volume 1, Introduction,
Volume 2, Loop amplitudes, anomalies and phenomenology, \rm Cambridge University Press, Cambridge (UK),
1987.\item Grodzins, L. (1959), Measurements of helicity, in: Frisch, R.O. (Ed.) (1959), \it Progress in Nuclear
Physics, \rm Volume 7, London: Pergamon Press, Ltd., pp. 163-241.\item E. Guadagnini, \it Fisica Teorica, \rm
Lezioni per il Corso di Dottorato di Ricerca, Edizioni ETS, Pisa, 1999.\item R. Haag, \it Local Quantum Physics.
Fields, Particles, Algebras, \rm Springer-Verlag, Berlin and Heidelberg, 1992.\item Haeberli, W. (1967), Sources
of Polarized Ions, \rm Annual Review of Nuclear Science, \rm 17: 373-426.\item Hahn, E.L. (1950), Spin Echoes,
\it Physical Review, \rm 80 (4): 580-594. \item Haken, H. and Wolf, H.C. (2005), \it The Physics of Atoms and
Quanta. Introduction to Experiments and Theory, \rm 7th Edition, Berlin and Heidelberg: Springer-Verlag (Italian
Translation of the 1987 2nd Edition: (1990), \it Fisica atomica e quantistica. Introduzione ai fondamenti
sperimentali e teorici, \rm Torino: Bollati Boringhieri Editore).\item S. Hartmann, \it Models and stories in
hadron physics, \rm in: M.S. Morgan, M. Morrison (Eds), \it Models as Mediators. Perspectives on Natural and
Social Science, \rm Cambridge University Press, Cambridge (UK), 1999, Chapter 11, pp. 326-346.\item W.P. Healy,
\it Non-Relativistic Quantum Electrodynamics, \rm Academic Press, Ltd., London, 1982.\item W. Heisenberg, \it
Die Physikalischen Prinzipien der Quantentheorie, \rm S. Hirzel Verlag, Lipsia, 1930 (Italian Translation: \it I
principi fisici della teoria dei quanti, \rm Editore Boringhieri, Torino, 1976).\item W. Heisenberg, \it
Wandlungen in den Grundlagen der Naturwissenshaft, \rm S. Hirzel Verlag, Stuttgart (FRG), 1959 (Italian
Translation: \it Mutamenti nelle basi della scienza, \rm Paolo Boringhieri editore, Torino, 1978).\item W.
Heitler, \it The Quantum Theory of Radiation, \rm Third Edition, Oxford University Press at Clarendon, Oxford
(UK), 1954.\item Holzscheiter, M.H. (1995), A brief history in time of ion traps and their achievements in
science, \it Physica Scripta, \rm Supplement T59: 69-76, Special volume devoted to the \it Proceedings of the
Nobel Symposium 91: Trapped Charged Particles and Related Fundamental Physics, \rm 19-26 August 1994, Lysekil,
Sweden.\item K. Huang, \it Quarks, Leptons \& Gauge Fields, \rm 2nd edition, World Scientific Publishing Co.,
Singapore, 1992.\item D.J. Hughes, \it The Neutron Story, \rm Doubleday \& Company, Inc., Garden City, New York,
1959 (Italian Translation: \it Fisica del Neutrone, \rm Piccola Biblioteca Einaudi, Torino, Giulio Einaudi
editore, Torino, 1960).\item Hughes, V.W. and Schultz, H.L. (Eds) (1967), \it Methods of Experimental Physics,
\rm Vol. 4, Part B: Atomic and Electron Physics. Atomic Sources and Detectors, \rm New York: Academic Press,
Inc.\item Hughes, V.W. and Wu, C.S. (Eds.) (1977), \it Muon Physics, Vol. I. Electromagnetic Interactions, Vol.
II, Weak Interactions, Vol. III, Chemistry and Solids, \rm New York: Academic Press, Inc.\item Hughes, V.W. and
Sichtermann, E.P. (2003), The Anomalous Magnetic Moment of the Muon, \it International Journal of Modern Physics
A, \rm 18 (S1): 215-272.\item Humphries, S. Jr. (1999), \it Principles of Charged Particle Acceleration, \rm New
York: John Wiley \& Sons, Inc. \item F. Iachello, A. Arima, \it The interacting boson model, \rm Cambridge
University Press, Cambridge (UK), 1987.\item F. Iachello, P. Van Isacker, \it The interacting boson-fermion
model, \rm Cambridge University Press, Cambridge (UK), 1991.\item Iurato, G. (2013), On Collingwood's
historicism, \it hal.archives-ouvertes.fr/hal-00921948-version 1.\item \rm Jackson, J.D. (1975), \it Classical
Electrodynamics, 2nd Edition, \rm New York: John Wiley \& Sons, Inc. (Italian Translation: (1984), \it
Elettrodinamica Classica, \rm Bologna: Nicola Zanichelli Editore). \item M. Jacob, Pas d'exclusion pour Wolfgang
Pauli, \it CERN Courier - International Journal of High-Energy Physics, \rm Volume 40, No. 9, September 2000,
pp. 30-32.\item J.M. Jauch, F. Rohrlich, \it The Theory of Photons and Electrons. The Relativistic Quantum Field
Theory of Charged Particles with Spin One-half, \rm Second Expanded Edition, Springer-Verlag, New York, Berlin
and Heidelberg, 1976.\item C.J. Joachain, \it Quantum Collision Theory, \rm North-Holland Publishing Company,
Amsterdam, 1975.\item R. Jost, \it The General Theory of Quantized Fields, \rm American Mathematical Society,
Providence, Rhode Island, 1965.\item M. Kaku, \it Introduction to Superstrings, \rm Springer-Verlag, New York,
1988.\item M. Kaku, \it Quantum Field Theory. A Modern Introduction, \rm Oxford University Press, New York,
1993.\item D. Kastler, \it Introduction a l'électrodynamique quantique, \rm Dunod, Paris, 1961.\item Kastler, A.
(1976), \it Cette \'{E}strange Matière, \rm Paris: Editions Stock (Italian Translation: (1977), \it Questa
strana materia, \rm Milano: Arnoldo Mondandori Editore).\item E.C. Kemble, \it The Fundamental Principles of
Quantum Mechanics with Elementary Applications, \rm Dover Publications, Inc., New York, 1958.\item Kerst, D.W.
and Serber, R. (1941), Electronic Orbits in the Induction Accelerators, \it Physical Review, \rm 60 (1):
53-58.\item Kinoshita, T. (Ed.) (1990), \it Quantum Electrodynamics, \rm Singapore: World Scientific Publishing
Company. \item E. Kiritsis, \it Introduction to Superstring Theory, \rm preprint CERN-TH/97-218
(hep-th/9709062), March 1997.\item Kittel, C. (1966), \it Introduction to Solid State Physics, \rm 3rd Edition,
New York: John Wiley \& Sons, Inc. (Italian Translation: (1971), \it Introduzione alla fisica dello stato
solido, \rm Torino: Editore Boringhieri).\item Koenig, S.H., Prodell, A.G. and Kusch, P. (1952), The Anomalous
Magnetic Moment of the Electron, \it Physical Review, \rm 88 (2): 191-199.\item H.S. Kragh, An Introduction to
Historiography of Science, \rm Cambridge University Press, Cambridge (UK), 1987 (Italian Translation: \it
Introduzione alla storiografia della scienza, \rm Nicola Zanichelli Editore, Bologna, 1990).\item H.S. Kragh,
(1990), \it Dirac. A Scientific Biography, \rm Cambridge (UK): Cambridge University Press.\item H.S. Kragh,
(2002), \it Quantum Generations. A History of Physics in the Twentieth Century, \rm Princeton (NJ): Princeton
University Press.\item Kusch, P. (1956), Magnetic Moment of the Electron, \it Science, \rm 123 (3189): 207-211.
\item E. Lanconelli, Commemorazione di Bruno Pini, necrologio tenuto all'Accademia Nazionale dei Lincei, Roma,
2007.\item E. Lanconelli, Bruno Pini and the Parabolic Harnack Inequality: The Dawning of Parabolic Potential
Theory, in: S. Coen (Ed), \it Mathematicians in Bologna: 1861-1960, \rm Springer Basel AG, Basel, 2012, pp.
317-332.\item Landau, L.D. (1957), On the conservation laws for weak interactions, \it Nuclear Physics, \rm 3
(1): 127-131.\item L.D. Landau, E.M. Lif\v{s}its, \it Fisica Teorica, 3. Meccanica Quantistica. Teoria non
relativistica, \rm Editori Riuniti-Edizioni Mir, Roma-Mosca, 1982.\item L.D. Landau, E.M. Lif\v{s}its, \it
Fisica teorica, 4. Teoria quantistica relativistica, \rm a cura di V.B. Berestetskij, E.M. Lif\v{s}its, L.P.
Pitaevskij, Editori Riuniti-Edizioni Mir, Roma-Mosca, 1978.\item Lederman, L.M. (1992), \it Observations in
Particle Physics from Two Neutrinos to the Standard Model, \rm Fermilab Golden Book Collection, Batavia,
Illinois: Fermi National Accelerator Laboratory.\item Lee, T.D. and Yang, C.N. (1956), Question of Parity
Conservation in Weak Interactions, \it Physical Review, \rm 104 (1): 254-258.\item Lee, T.D., Oehme, R. and
Yang, C.N. (1957), Remarks on Possible Noninvariance under Time Reversal and Charge Conjugation, \it Physical
Review, \rm 106 (2): 340-345.\item T.D. Lee, \it Particle Physics and Introduction to Field Theory, \rm Harwood
Academic Publishers, New York, 1981.\item Lee, S.J. (2004), \it Accelerator Physics, \rm Second Edition,
Singapore: World Scientific Publishing Company, Ltd. \item B. Lehnert, \it Dynamics of Charged Particles, \rm
North-Holland Publishing Company, Amsterdam, 1964.\item J. Leite Lopes, \it Gauge Field Theories. An
Introduction, \rm Pergamon Press, Ltd., Oxford (UK), 1981.\item B.G. Levich, V.A. Myamlin, Yu.A. Vdovin, \it
Theoretical Physics. An Advanced Text, Volume 3, Quantum Mechanics, \rm North-Holland Publishing Company,
Amsterdam, 1973.\item J. \L opuszànski, \it An Introduction to Symmetry and Supersymmetry in Quantum Field
Theory, \rm World Scientific Publishing Company, Ltd., Singapore, 1991.\item Louisell, W.H., Pidd, R.W. and
Crane, H.R. (1954), An Experimental Measurement of the Gyromagnetic Ratio of the Free Electron, \it Physical
Review, \rm 94 (1): 7-16.\item Luttinger, J.M. (1948), A Note on the Magnetic Moment of the Electron, \it
Physical Review, \rm 74 (8): 893-898.\item L. Maiani and R.A. Ricci (Eds), \it Symposium in honour of Antonino
Zichichi to celebrate the 30th anniversary of The Discovery of Nuclear Antimatter, \rm Conference Proceedings of
the Italian Physical Society, Volume 53, Bologna, 18 December 1995, Published for the Italian Physical Society
by Editrice Compositori, Bologna, 1996.\item Mandl, F. and Shaw, G. (1984), \it Quantum Field Theory, \rm
Chichester (WS-UK): John Wiley and Sons, Ltd. \item K.B. Marathe, G. Martucci, \it The Mathematical Foundations
of Gauge Theories, \rm North-Holland Elsevier Science Publishers B.V., Amsterdam, 1992.\item P.T. Matthews, \it
The Nuclear Apple. Recent discoveries in fundamental physics, \rm Chatto and Windus, London, 1971 (Italian
Translation: \it Nel nucleo dell'atomo. Le più recenti scoperte della fisica fondamentale, \rm Biblioteca EST,
Arnoldo Mondadori Editore, Milano, 1972).\item P.T. Matthews, \it Introduction to Quantum Mechanics, \rm
McGraw-Hill Publishing Company, Ltd., Maidenhead, Berkshire (UK), 1974 (Italian Translation: \it Introduzione
alla meccanica quantistica, \rm Nicola Zanichelli Editore, Bologna, 1978.\item Melnikov, K. and Vainshtein, A.
(2006), \it Theory of the Muon Anomalous Magnetic Moment, \rm Berlin and Heidelberg: Springer-Verlag. \item G.A.
Miller, Book Reviews: Enciclopedia delle Matematiche Elementari, Vol. 1, \it Bulletin of the American
Mathematical Society, \rm 38 (3) (1932) pp. 157-158.\item Miller, J.P., de Rafael, E. and Roberts, L.B. (2007),
Muon $(g-2)$: experiment and theory, \it Reports on Progress in Physics, \rm 70 (5): 795-881.\item G. Morpurgo,
\it Lezioni sulle forze nucleari, \rm A.A. 1954-1955, Scuola di Perfezionamento in Fisica Nucleare, Istituto di
Fisica dell'Università di Roma, Litografia Marves, Roma, 1955.\item G. Morpurgo, \it Introduzione alla fisica
delle particelle, \rm Nicola Zanichelli Editore, Bologna, 1987.\item H. Muirhead, \it The Physics of Elementary
Particles, \rm Pergamon Press, Ltd., Oxford (UK), 1965.\item H.J.W. M\"{u}ller-Kirsten, A. Wiedemann, \it
Supersymmetry. An Introduction with Conceptual and Calculational Details, \rm World Scientific Publishing
Company, Ltd., Singapore, 1987. \item Y. Ne'eman, \it Algebraic Theory of Particle Physics. Hadron Dynamics in
terms of Unitary Spin Currents, \rm W.A. Benjamin, Inc., New York, 1967.\item R.G. Newton, \it Scattering Theory
of Waves and Particles, \rm Second Edition, Springer-Verlag, New York, Heidelberg and Berlin, 1982.\item R.G.
Newton, \it The Complex j-Plane. Complex Angular Momentum in Nonrelativistic Quantum Scattering Theory, \rm W.A.
Benjamin, Inc., New York, 1964.\item Ohanian, H.C. (1988), \it Classical Electrodynamics, \rm Boston (MA): Allyn
\& Bacon, Inc.\item L.B. Okun, \it Leptons and Quarks, \rm Elsevier Science Publishers, Ltd., Amsterdam and New
York, 1982 (Italian Edition: \it Leptoni e quark, \rm Editori Riuniti-Edizioni Mir, Roma-Mosca, 1986).\item L.
O'Raifeartaigh, \it Group Structure of Gauge Theories, \rm Cambridge University Press, Cambridge (UK),
1986.\item Panofsky, W.K.H. and Phillips, M. (1962), \it Classical Electricity and Magnetism, \rm Reading (MA):
Addison-Wesley Publishing Company, Inc. (Italian Translation: (1966), \it Elettricità e magnetismo, \rm Milano:
CEA - Casa Editrice Ambrosiana). \item J.C. Parikh, \it Group Symmetries in Nuclear Structure, \rm Plenum Press,
New York and London, 1978.\item Pauli, W. (1941), Relativistic Field Theory of Elementary Particles, \it Review
of Modern Physics, \rm 13 (3): 203-232.\item W. Pauli, \it Wellenmechanik, \rm Verlag des Vereins der
Mathematiker und Physiker an der ETH, Zürich, 1959 (Italian Translation: \it Meccanica ondulatoria, \rm Editore
Boringhieri, Torino, 1962).\item W. Pauli, \it Pauli Lectures on Physics, 6. Selected Topics in Field
Quantization, \rm Edited by C. Enz, The MIT Press, Cambridge, Massachusetts, 1973.\item W. Pauli, \it Lectures
on continuous groups and reflections in quantum mechanics, \rm Notes by R.J. Riddell jr., Radiation Laboratory
UCRL-8213, University of California, Berkeley, Printed for the U.S. Atomic Energy Commission, 1958.\item
Pedulli, G.F., Alberti, A. and Lucarini, M. (1996), \it Metodi Fisici in Chimica Organica. Princìpi ed
applicazioni di tecniche spettroscopiche, \rm Padova: Piccin Nuova Libraria.\item E. Persico, \it Fondamenti
della Meccanica Atomica, \rm Casa Editrice Nicola Zanichelli, Bologna, 1936.\item E. Persico, \it Gli atomi e la
loro energia, \rm Nicola Zanichelli Editore, Bologna, 1959.\item M. Piattelli-Palmarini, \it Scienza come
Cultura. Protagonisti, Luoghi e Idee delle Scienze Contemporanee, \rm edizione paperback a cura di Simone
Piattelli, Saggi Oscar Mondadori Editore, Milano, 1992.\item Picasso, E. (1985), Le misure del momento magnetico
del muone, \it Il Nuovo Saggiatore. Bollettino della Società Italiana di Fisica, \rm 4: 22-30.\item Picasso, E.
(1996), The anomalous magnetic moment of the muon and related topics, \it Atti dell'Accademia Nazionale dei
Lincei, Rendiconti della classe di scienze fisiche, matematiche e naturali, \rm 7 (9): 209-241.\item A.
Pignedoli, \it Alcune teorie meccaniche ''superiori'', \rm CEDAM, Padova, 1969.\item M. Planck, \it La
conoscenza del mondo fisico, \rm Paolo Boringhieri Editore, Torino, 1964.\item J. Polchinski, \it String Theory,
Volume I, An Introduction to the Bosonic String, Volume II, Superstring Theory and Beyond, \rm Cambridge
University Press, Cambridge (UK), 1998.\item S. Pokorski, \it Gauge Field Theories, \rm Cambridge University
Press, Cambridge, 1987.\item V. Polara, \it L'atomo e il suo nucleo. Struttura dell'atomo e disintegrazione
spontanee ed artificiali del nucleo, \rm Perrella Editore, Roma, 1949.\item A.M. Polyakov, \it Gauge Fields and
Strings, \rm Harwood Academic Publishers GmbH, Chur, Switzerland, 1987.\item Povh, B., Rith, K., Scholz, C. and
Zetsche, F. (1995), \it Particles and Nuclei. An Introduction to the Physical Concepts, \rm Berlin and
Heidelberg: Springer-Verlag (Italian Translation: (1998), \it Particelle e nuclei. Un'introduzione ai concetti
fisici, \rm Torino: Bollati Boringhieri editore).\item Rabi, I.I., Zacharias, J.R., Millman, S. and Kusch, P.
(1938), A New Method of Measuring Nuclear Magnetic Moment, \it Physical Review, \rm 53 (4): 318(L). \item Rabi,
I.I., Millman, S., Kusch, P. and Zacharias, J.R. (1939), The Molecular Beam Resonance Method for Measuring
Nuclear Magnetic Moments. The Magnetic Moments of $_3Li^6, _3Li^7$ and $_9Fe^{19}$, \it Physical Review, \rm 55
(6): 526-535. \item M. Reed, B. Simon, \it Methods of Modern Mathematical Physics, Volume I, Functional
Analysis, Volume II, Fourier Analysis, Self-Adjointness, Volume III, Scattering Theory, Volume IV, Analysis of
Operators, \rm Academic Press, Inc., New York, 1980, 1975, 1979, 1978.\item Rich, A. and Wesley, J. (1972), The
Current Status of the Lepton $g$ Factors, \it Reviews of Modern Physics, \rm 44 (2): 250-283.\item P. Ring, P.
Schuck, \it The Nuclear Many-Body Problem, \rm Springer-Verlag, New York, 1980.\item R.G. Roberts, \it The
Structure of the Proton, \rm Cambridge University Press, Cambridge (UK), 1990.\item Roberts, B.L. and Marciano,
W.J. (Eds.) (2010), \it Lepton Dipole Moments, \rm Singapore: World Scientific Publishing Company, Ltd. \item P.
Roman, \it Theory of Elementary Particles, \rm North-Holland Publishing Company, Amsterdam, 1960.\item P. Roman,
\it Advanced Quantum Theory. An outline of the fundamental ideas, \rm Addison-Wesley Publishing Company, Inc.,
Reading, Massachusetts, 1965.\item C. Rossetti, \it Elementi di teoria dell'urto, \rm Libreria editrice
universitaria Levrotto \& Bella, Torino, 1985.\item P.A. Rowlatt, \it Group Theory and Elementary Particles, \rm
Longmans, Green and Company, Ltd., London, 1966.\item Rossi, B. (1964), \it Cosmic Rays, \rm New York:
McGraw-Hill Book Company, Inc. (Italian Translation: (1971), \it I raggi cosmici, \rm Torino: Giulio Einaudi
editore).\item J.J. Sakurai, \it Invariance Principles and Elementary Particles, \rm Princeton University Press,
Princeton (NJ), 1964.\item J.J. Sakurai, \it Advanced Quantum Mechanics, \rm Addison-Wesley Publishing Company,
Inc., Reading, Massachusetts, 1967.\item J.J. Sakurai, S.F. Tuan, \it Modern Quantum Mechanics, \rm The
Benjamin/Cummings Publishing Company, Inc., Menlo Park (CA), 1985 (Italian Translation: \it Meccanica
quantistica moderna, \rm Nicola Zanichelli Editore, Bologna, 1990).\item F. Scheck, H. Upmeier, W. Werner (Eds),
\it Noncommutative Geometry and the Standard Model of Elementary Particle Physics, \rm Springer-Verlag, Berlin
and Heidelberg, 2002.\item L.I. Schiff, \it Quantum Mechanics, \rm McGraw-Hill Book Company, Inc., New York,
1952 (Italian Translation: \it Meccanica quantistica, \rm Edizioni Scientifiche Einaudi, Torino, 1952).\item H.
Schlaepfer, Cosmic Rays, \it Spatium, Published by International Space Science Institute of Bern, \rm 11 (2003)
pp. 3-15.\item Schultz, D.P. (1969), \it A History of Modern Psychology, \rm New York: Academic Press, Inc.
(Italian Translation: (1974), \it Storia della psicologia moderna, \rm Firenze: Giunti-Barbèra).\item L.
Schwartz, \it Application of Distributions to the Theory of Elementary Particles in Quantum Mechanics, \rm
Gordon \& Breach Science Publishers, Inc., New York, 1968.\item Schwartz, M. (1972), \it Principles of
Electrodynamics, \rm New York: McGraw-Hill Book Company, Inc.\item A.S. Schwarz, \it Quantum Field Theory and
Topology, \rm Springer-Verlag, Berlin and Heidelberg, 1993.\item S.S. Schweber, \it An Introduction to
Relativistic Quantum Field Theory, \rm Row, Peterson and Company, Evanston, Illinois and Elmsford, New York,
1961.\item Schweber, S.S. (1983), Some chapters for a history of quantum field theory: 1938-1952, in: DeWitt, B.
and Stora, R. (Eds) (1984), \it Relativity, Groups and Topology, II, \rm Parts 1, 2, 3, Les Houches, Session XL,
27 June - 4 August 1983, Amsterdam (The Netherlands): North-Holland Physics Publishing Company, Inc., pp.
37-220.\item S.S. Schweber, \it QED And The Men Who Made It: Dyson, Feynman, Schwinger, and Tomonaga, \rm
Princeton University Press, Princeton (NJ), 1994.\item J. Schwinger, On Quantum-Electrodynamics and the Magnetic
Moment of the Electron, \it Physical Review, \rm 73(4): 416-417. \item J. Schwinger, \it Quantum Mechanics.
Symbolism of Atomic Measurements, \rm edited by B-G. Englert, Springer-Verlag, Berlin and Heidelberg, 2001.\item
I.E. Segal, \it Mathematical Problems of Relativistic Physics, \rm American Mathematical Society, Providence,
Rhode Island, 1963.\item E. Segrè, \it Nuclei and Particles. An Introdduction to Nuclear and Subnuclear Physics,
\rm W.A. Benjamin, Inc., New York, 1964 (Italian Translation: \it Nuclei e particelle. Introduzione alla fisica
nucleare e subnucleare, \rm Nicola Zanichelli Editore, Bologna, 1966; seconda edizione, 1999).\item E. Segrè,
\it Personaggi e scoperte della fisica contemporanea, \rm Biblioteca EST, Arnoldo Mondadori Editore, Milano,
1976.\item Shankar, R. (1994), \it Principles of Quantum Mechanics, \rm 2nd Edition, New York: Kluwer
Academic/Plenum Publishers. \item A.G. Sitenko, V.K. Tartakovskij, \it Lezioni di teoria del nucleo, \rm
Edizioni Mir, Mosca, 1981.\item Slater, J.C. (1968), \it Quantum Theory of Matter, \rm New York: McGraw-Hill
Book Company, Inc. (Italian Translation: (1980), \it Teoria quantistica della materia, \rm Bologna: Nicola
Zanichelli Editore).\item \u{S}polskij, E.D. (1986), \it Fisica atomica, Volume I, Introduzione alla fisica
atomica, Volume II, Fondamenti della meccanica quantistica e struttura del guscio elettronico dell'atomo, \rm
Mosca-Roma: Edizioni Mir.\item Fl. Stancu, \it Group Theory in Subnuclear Physics, \rm Oxford University Press
at Clarendon, Oxford (UK), 1996.\item S. Sternberg, \it Group Theory and Physics, \rm Cambridge University
Press, Cambridge (UK), 1994.\item P. Straneo, Materia, irraggiamento e fisica quantica, Articolo LII in: L.
Berzolari (Ed), \it Enciclopedia delle Matematiche Elementari e Complementi, con estensione alle principali
teorie analitiche, geometriche e fisiche, loro applicazioni e notizie storico-bibliografiche, \rm Volume III,
Parte 1$^{a}$, Editore Ulrico Hoepli, Milano, 1947 (ristampa anastatica 1975).\item R.F. Streater, A.S.
Wightman, \it PCT, Spin and Statistics, and all that, \rm W.A. Benjamin, Inc., New York, 1964.\item Sundermeyer,
K. (1982) \it Contrained Dynamics, with Applications to Yang-Mills Theory, General Relativity, Classical Spin
and Dual String Model, \rm Berlin and Heidelberg: Springer-Verlag.\item Tagliagambe, S. and Malinconico, A.
(2011), \it Pauli e Jung. Un confronto su materia e psiche, \rm Milano: Raffaello Cortina Editore.\item J.R.
Taylor, \it Scattering Theory. The Quantum Theory on Nonrelativistic Collisions, \rm John Wiley \& Sons, Inc.,
New York, 1972.\item W. Thirring, \it A Course in Mathematical Physics, Volume 1, Classical Dynamical Systems,
Volume 2, Classical Field Theory, Volume 3, Quantum Mechanics of Atoms and Molecules, Volume 4, Quantum
Mechanics of Large Systems, \rm Springer-Verlag New York, Inc., 1978, 1979, 1981, 1983.\item H. Thomä, H.
Kächele, \it Psychoanalytic Practice, Volume 1, Principles, Volume 2, Praxis, \rm Springer-Verlag, Berlin and
Heidelberg, 1987, 1992 (Italian Translation: \it Trattato di terapia psicoanalitica, Volume 1, Fondamenti
teorici, Volume 2, Pratica clinica, \rm Bollati Boringhieri editore, Torino, 1990, 1993).\item I.T. Todorov,
M.C. Mintchev, V.B. Petrova, \it Conformal Invariance in Quantum Field Theory, \rm Pubblicazioni della Classe di
Scienze della Scuola Normale Superiore, SNS, Pisa, 1978.\item S. Tolansky, \it Introduction to Atomic Physics,
\rm 5th edition, Longmans, Green and Company, Inc., London, 1963 (Italian Translation: \it Introduzione alla
fisica atomica, \rm 2 voll., Editore Boringhieri Torino, 1966).\item Tomonaga, S.I. (1997), \it The Story of
Spin, \rm Chicago: The University of Chicago Press.\item F.G. Tricomi, \it La mia vita di matematico attraverso
la cronistoria dei miei lavori (bibliografia commentata 1916-1967), \rm Casa Editrice Dott. Antonio Milani -
CEDAM, Padova, 1967.\item Tsai, S.Y. (1981), Universal weak interactions involving heavy leptons, \it Lettere al
Nuovo Cimento, \rm 2 (18): 949-952.\item C. Villi, G. Pisent, V. Vanzani, \it Lezioni di Istituzioni di Fisica
Nucleare, \rm Anno Accademico 1970-1971, Istituto di Fisica dell'Università di Padova, Padova, 1971.\item C.
Villi, F. Zardi, \it Appunti di lezioni di fisica nucleare, \rm Anno Accademico 1974-1975, Istituto di Fisica
dell'Università di Padova, Padova, 1975.\item A. Visconti, \it Théorie quantique des champs, \rm Tomes I, II,
Gauthier-Villars, Paris, 1961, 1965.\item A.H. V\"{o}lkel, \it Field, Particles and Currents, \rm
Springer-Verlag, Berlin and Heidelberg, 1977. \item J.D. Walecka, \it Theoretical Nuclear and Subnuclear
Physics, \rm Oxford University Press, New York, 1995.\item S. Weinberg, \it The Quantum Theory of Fields, Volume
I, Foundations, Volume II, Modern Applications, Volume III, Supersymmetry, \rm Cambridge University Press,
Cambridge (UK), 1995, 1996, 2000 (Italian Translation of Volume I: \it La teoria quantistica dei campi, \rm
Nicola Zanichelli editore, Bologna, 1999).\item Weisskopf, V.F. (1949), Recent Developments in the Theory of the
Electron, \it Reviews of Modern Physics, \rm 21 (2): 305-315.\item G. Wentzel, \it Quantum Theory of Fields, \rm
Interscience Publishers, Inc., New York, 1949.\item Wertheimer, M. (1979), \it A Brief History of Psychology,
\rm New York: Holt, Rinehart and Wilson, Inc. (Italian Translation: (1983), \it Breve storia della psicologia,
\rm Bologna: Nicola Zanichelli editore).\item H. Weyl, \it Gruppentheorie und quantenmechanik, \rm Verlag Von S.
Hirzel, Leipzig, 1928; second edition, 1931 (English Translation: \it The Theory of Groups and Quantum
Mechanics, \rm translated from the second revised German edition by H.P. Robertson, A. Methuen \& Co., Ltd.,
London, 1931; reprinted by Dover Publications, Inc., New York, 1950).\item E.V.H. Wichmann, \it Quantum Physics,
\rm McGraw-Hill Book Company, Inc., New York, 1971 (Italian Translation: \it La Fisica di Berkeley, 4. Fisica
Quantistica, \rm Nicola Zanichelli Editore, Bologna, 1973).\item Wick, G.C. (1945), \it Appunti di Fisica
Nucleare, parte I, \rm Anno Accademico 1944-1945, redatti da E. Amaldi, Roma: Tipo-litografia Romolo Piola.\item
Wick, G.C. (1946), \it Appunti di Fisica Nucleare, parte II, \rm Anno Accademico 1945-1946, redatti a cura di E.
Amaldi, Roma: Tipo-litografia Romolo Piola.\item E.P. Wigner, \it Gruppentheorie und ihre Anwendung auf die
Quantenmechanik der Atomspektren, \rm Friedrich Vieweg und Sohn Akt. Ges., Braunschweig, 1931 (English
Translation: \it Group Theory and Its Application To The Quantum Mechanics of Atomic Spectra, \rm expanded and
improved edition of the first one translated from the German by J.J. Griffin, Academic Press, Inc., New York,
1959).\item R.R. Wilson, R. Littauer, \it Accelerators. Machines of Nuclear Physics, \rm Anchor Books Doubleday
\& Company, Inc., Garden City, New York, 1960 (Italian Translation: \it Acceleratori di particelle. Macchine
della fisica nucleare, \rm Nicola Zanichelli editore, Bologna, 1965).\item C.N. Yang, \it Elementary Particles.
A Short History of Some Discoveries in Atomic Physics - 1959 Vanuxem Lectures, \rm Princeton University Press,
Princeton (NJ), 1961 (Italian Translation: \it La scoperta delle particelle elementari, \rm Editore Boringhieri,
Torino, 1969).\item Yang, C.N. (2005), \it Selected Papers (1945-1980) With Commentary, \rm World Scientific
Series in 20th Century Physics, Vol. 36, Singapore: World Scientific Publishing Company, Ltd. \item H. Yukawa,
On the Interaction of Elementary Particles. I, \it Proceedings of the Physico-Mathematical Society of Japan, \rm
17 (1935) pp. 48-57 (reprinted in \it Progress of Theoretical Physics, Supplement, \rm 1 (1955) pp. 1-10).\item
Zel'dovich, Ya.B. (1961), Dipole moment of unstable elementary particles, \it Soviet Physics - JETP (Journal of
Experimental and Theoretical Physics), \rm 12: 1030-1031. \item Zichichi, A. (1981), Struttura delle particelle,
in: \it Enciclopedia del Novecento, \rm Vol. V, Roma: Istituto della Enciclopedia Italiana fondata da G.
Treccani, pp. 125-215.\item A. Zichichi, \it Scienza ed emergenze planetarie, \rm 6$^{a}$ edizione Superbur
saggi, Biblioteca Universale Rizzoli, RCS Libri, Milano, 1997.\item A. Zichichi, \it Creativity in Science, \rm
6th International Zermatt Symposium on Creativity in Economics, Arts and Science, Zermatt, Switzerland, 12-16
January 1996, World Scientific Publishing Company, Ltd., Singapore, 1999.\item A. Zichichi, \it Subnuclear
Physics. The first 50 years: highlights from Erice to ELN, \rm edited by O. Bernabei, P. Pupillo, F. Roversi
Monaco, World Scientific Publishing Company, Ltd., Singapore, 2001.\item Zichichi, A. (2008), The 40th
Anniversary of EPS: Gilberto Bernardini's contributions to the Physics of the XX Century, \it Il Nuovo
Saggiatore. Bollettino della Società Italiana di Fisica, \rm 24 (5-6): 77-94.\item Zichichi, A. (2010), In
ricordo di George Charpak (1924-2010), \it Il Nuovo Saggiatore. Bollettino della Società Italiana di Fisica, \rm
26 (5-6): 109.\item J. Zinn-Justin, \it Quantum Field Theory and Critical Phenomena, \rm Oxford University Press
at Clarendon, New York, 1989.

\end{description}

\end{document}